\documentclass[10pt]{article} 
\usepackage[accepted]{tmlr}

\usepackage{amsmath,amsfonts,bm}
\usepackage{amsthm}

\def\eqref#1{equation~\ref{#1}}

\def\1{\bm{1}}

\DeclareMathAlphabet{\mathsfit}{\encodingdefault}{\sfdefault}{m}{sl}
\SetMathAlphabet{\mathsfit}{bold}{\encodingdefault}{\sfdefault}{bx}{n}

\newcommand{\E}{\mathbb{E}}
\newcommand{\Ls}{\mathcal{L}}

\usepackage{hyperref}
\usepackage{url}
\usepackage{comment}
\usepackage{enumitem}
\usepackage{ulem}

\usepackage{multirow}
\usepackage{threeparttable}
\usepackage{booktabs}
\usepackage{pdflscape}
\usepackage{tabularx}
\newcolumntype{C}[1]{>{\centering\arraybackslash}p{#1}}
\newcolumntype{L}[1]{>{\raggedright\arraybackslash}p{#1}}
\newcolumntype{R}[1]{>{\raggedleft\arraybackslash}p{#1}}

\usepackage[super]{nth}

\usepackage{tablefootnote}

\usepackage{xspace}

\usepackage{graphicx}
\usepackage{subfig}
\usepackage[section]{placeins}
\newcounter{mysfig}
\counterwithin{mysfig}{figure}

\renewcommand\themysfig{(\alph{mysfig})}
\makeatletter
\newcommand\Scaption[1]{%
\refstepcounter{mysfig}%
\vskip.5\abovecaptionskip
  \sbox\@tempboxa{\small\themysfig~#1}%
  \ifdim \wd\@tempboxa >\hsize
    \small\themysfig~#1\par
  \else
    \global \@minipagefalse
    \hb@xt@\hsize{\hfil\box\@tempboxa\hfil}%
  \fi
  \vskip\belowcaptionskip}
\makeatother

\newcommand{\mfm}{\normalfont\textsc{SoniDo}\xspace}
\newcommand{\musicgen}{\normalfont\textsc{MusicGen}\xspace}
\newcommand{\musicgens}{\normalfont\textsc{MusicGen Small}\xspace}
\newcommand{\musicgenl}{\normalfont\textsc{MusicGen Large}\xspace}
\newcommand{\blockcomment}[1]{}

\title{Music Foundation Model as Generic Booster for Music Downstream Tasks}

\author{\name WeiHsiang Liao$^{1*}$, \quad\name Yuhta Takida$^{1*}$, \quad\name Yukara Ikemiya$^{1*}$, \quad\name Zhi Zhong$^{2*}$, \\ 
\name Chieh\nobreakdash-Hsin Lai$^{1}$, \quad\name Giorgio Fabbro$^{3}$, 
\quad\name Kazuki Shimada$^{1}$,
\quad\name Keisuke Toyama$^{2}$,\\
\name Kin Wai Cheuk$^{1}$,
\quad\name Marco A. Mart\'{i}nez-Ram\'{i}rez$^{1}$, 
\quad\name Shusuke Takahashi$^{2}$, \\
\name Stefan Uhlich$^{3}$, \quad\name Taketo Akama$^{4}$,
\quad\name Woosung Choi$^{1}$,
\quad\name Yuichiro Koyama$^{2}$,
\quad\name Yuki Mitsufuji$^{1,2}$\\
\\
\addr $^1$SonyAI, Tokyo, Japan\\
\addr $^2$Sony Group Corporation, Tokyo, Japan\\
\addr $^3$Sony Europe B.V., Stuttgart, Germany\\
\addr $^4$Sony CSL, Tokyo, Japan\\
$^*${\normalfont Equal contribution}\\
}

\def\month{05}  
\def\year{2025} 
\def\openreview{\url{https://openreview.net/forum?id=kHl4JzyNzF}} 

\definecolor{c_revise}{rgb}{0.0,0.0,0.0}
\definecolor{c_yuhta}{rgb}{0.831,0.184,0.494}
\definecolor{c_kinwai}{rgb}{0.0, 0.87, 0.87}
\definecolor{c_hidden}{rgb}{0.8, 0.8, 0.8}

\begin{document}

\maketitle

\begin{abstract}
We demonstrate the efficacy of using intermediate representations from a single foundation model to enhance various music downstream tasks. We introduce \mfm, a music foundation model (MFM) designed to extract hierarchical features from target music samples. By leveraging hierarchical intermediate features, \mfm constrains the information granularity, leading to improved performance across various downstream tasks including both understanding and generative tasks. We specifically evaluated this approach on representative tasks such as music tagging, music transcription, music source separation, and music mixing. Our results reveal that the features extracted from foundation models provide valuable enhancements in training downstream task models. This highlights the capability of using features extracted from music foundation models as a booster for downstream tasks. Our approach not only benefits existing task-specific models but also supports music downstream tasks constrained by data scarcity. This paves the way for more effective and accessible music processing solutions.
\end{abstract}

\begin{figure*}[th]
    \centering
    \includegraphics[width=0.95\textwidth,keepaspectratio,clip]{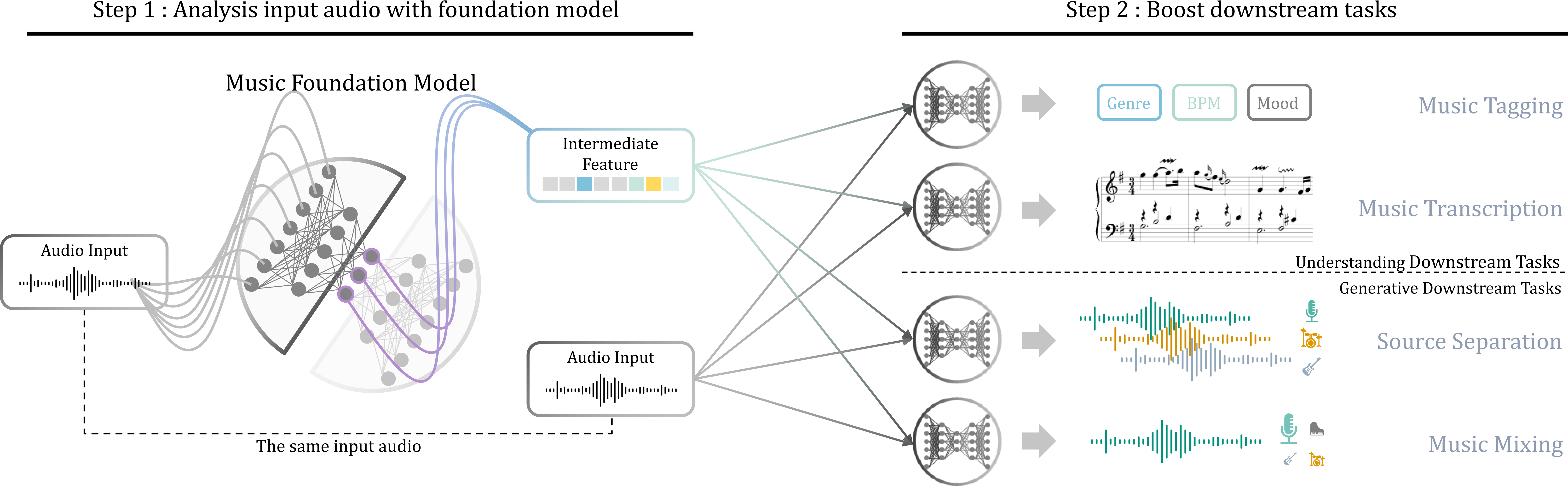}
    \caption{\mfm extracts hierarchical features of target music samples, which are useful for solving music downstream tasks including understanding and generative tasks.}
    \label{fig:user_workflow}
\end{figure*}

\section{Introduction}
\label{sec:intro}

A foundation model is a pre-trained model developed on a large-scale dataset that can be adapted for a variety of downstream tasks~\citep{Bommasani2021FoundationModels}. Several language processing models~\citep{radford2021learning,Brown2020gpt3,Delvin2018bert,Geminiteam2024gemini} are considered foundation models due to their ability to unify all language tasks as sequence prediction tasks, effectively addressing multiple tasks with a single model. These foundation models have gained significant attraction and are widely used in everyday applications. 
\textcolor{c_revise}{In contrast, a powerful \textit{music foundation model} capable of handling various \textit{music downstream tasks} for music production is lacking~\citep{ma2024foundation}.} We categorize the tasks that a music foundation model primarily addresses into two types: \textit{understanding tasks}, such as tagging and transcription, and \textit{generative tasks}, such as mixing and mastering.

Several multi-task models have been proposed as potential music foundation models~\citep{li2023jen1, yang2023uniaudio, copet2023musicgen, agostinelli2023musiclm}. However, this approach necessitates the inclusion of the desired tasks during the training phase. A notable strategy to overcome this limitation is to inject features extracted from a pre-trained large-scale model into smaller back-end models for downstream tasks that were not seen during training. This ensemble approach, which combines a large-scale model with various smaller models, can effectively function as a music foundation model. The codified audio language modeling (CALM) framework proposed by~\citet{CastellonDL21calm} is the first work in this direction, utilizing the intermediate representations from Jukebox~\citep{dhariwal2020jukebox} to tackle music information retrieval (MIR) tasks, covering most music understanding tasks. Beyond MIR,\citet{donahue2022} leveraged representations from Jukebox for melody transcription. Other studies have followed this approach to address time-invariant MIR tasks using the latest generative models based on residual quantized variational Autoencoders (RQ-VAEs)~\citep{zeghidour2022soundstream, defossez2023encodec}, enhancing the state-of-the-art (SOTA). However, these applications remain limited to music understanding tasks. \citet{li2023mert} expanded the focus to include music source separation, a generative task, but encountered instability issues during training. The performance of this extension does not yet match that of the baselines mentioned by~\citet{mitsufuji2022music}. An extensive overview of related work can be found in Appendix~\ref{sec:app_related_works}.

\textcolor{c_revise}{We extended the methodology from MIR tasks to generic music downstream tasks. To address both understanding and generative tasks, we focus on a tokenization-based generative modeling approach, which is directly analogous to foundation models in the natural language processing field. Furthermore, We hypothesize that the representation structure of foundation models is crucial in this context.} Specifically, we propose that hierarchical representations, which divide information of varying granularity into different levels of embedding, are expected to provide efficient information hierarchy for all downstream tasks including both understanding and generative tasks. We empirically verify this hypothesis in Section~\ref{sec:experiments}. In contrast, music foundation models that have been applied to boost music downstream tasks do not have such a hierarchical structure. For example, Jukebox~\citep{CastellonDL21calm,donahue2022} is trained to have multi-level representation inspired from hierarchical latent representation~\citep{razavi2019generating}; however, each level is independently trained. RQ-VAEs~\citep{yang2023uniaudio,li2023mert} learn factorized representation that has a self-organized coarse-to-fine structure, however, they are not hierarchical.

In accordance with the aforementioned hypothesis, we outline this study as follows. We propose and train our music foundation model, \mfm (meaning \textit{sound} in Spanish), on a high-fidelity internal dataset
\footnote{The rights of this internal dataset are trained on licensed content only. Except for as specifically authorized by the rights owner, the rights owner expressly prohibits and has opted out of any text or data mining, web scraping or similar, reproductions, extractions or uses, of its content for any purposes, including in relation to training, developing, or commercializing any Al System.}
to establish a task-agnostic feature extraction pipeline. \mfm is a generative model consisting of a multi-level transformer with a multi-level hierarchical encoder. With proper pre-processing, we infuse its intermediate representation as features to task-specific models on various music downstream tasks with data augmentation. Moreover, for understanding tasks, we proposed an on-the-fly data augmentation called \textit{token-out} to avoid overfitting. Performance evaluation was done by benchmarking with representative tasks from understanding to generative tasks: music tagging, music transcription, music source separation, and music mixing.

\newcolumntype{H}{>{\setbox0=\hbox\bgroup}c<{\egroup}@{}}
\begin{table}[t]
\centering
\caption{Performance overview of applying extracted features to various downstream tasks. Bold: best, underline: second best. 
The result marked with * is obtained with a different evaluation protocol. The results marked with \textdagger{} are numbers reported in~\citet{CastellonDL21calm}.
}
\resizebox{\linewidth}{!}{
\begin{tabular}{ccc||ccccHcH|cc}
\hline
Downstream Task & Dataset & Metric & \mfm & \begin{tabular}[c]{@{}c@{}} \musicgen \\ \textsc{Small}\end{tabular} & \begin{tabular}[c]{@{}c@{}} \musicgen \\ \textsc{Large}\end{tabular} & Jukebox-5B & MULE & MERT & LLark & \multicolumn{2}{c}{Task-Specific SOTA}\\
\hline
\multirow{2}{*}{Multi-task Music Tagging} & \multirow{2}{*}{MusicTagATune} & ROC-AUC & \underline{91.7} & 90.4 & 90.5 & 91.5\textsuperscript{\textdagger} & 91.4 & 91.3 & - & \textbf{92.0} & \multirow{2}{*}{\citep{HuangJLGLE22mulan}}\\ 
 &  & mAP & \textbf{41.5} & 38.8 & 39.0 & \underline{41.4}\textsuperscript{\textdagger} & 40.4 & 40.2 & - & 38.4 \\
Pitch Estimation & \multirow{2}{*}{Nsynth} & Acc. & \underline{93.8} & 93.3 & 92.8 & 91.6\textsuperscript{\textdagger} & 89.2 & \textbf{94.4} & - & 89.2 & \citep{McCallumKOGE22mule}\\
Instrument Classification &  & Acc. & \underline{78.0} & 71.9 & 74.2 & 70.4\textsuperscript{\textdagger} & 74.0 & 72.6 & - & \textbf{78.2} & \citep{wang2022slowfast}\\

Emotion Regression & EmoMusic & Averaged $R^2$ & 64.7 & 45.6 & 46.2 & \underline{66.9}\textsuperscript{\textdagger} & 64.6 & \textbf{68.0} & - & \textbf{63.0}* & \citep{CastellonDL21calm} \\
Key Detection & GiantSteps & Weighted Acc. & 63.5 & 65.2 & 62.4 & 66.7\textsuperscript{\textdagger} & 66.7 & 65.6 & \underline{70.0}* & \textbf{79.6} & \citep{CastellonDL21calm}\\
Genre Classification & GTZAN & Acc. & \underline{80.7} & 75.2 & 70.3 & 79.7\textsuperscript{\textdagger} & 73.5 & 79.3 & 56.0* & \textbf{83.5} & \citep{McCallumKOGE22mule} \\
Singer Identification & \multirow{2}{*}{VocalSet} & Acc. & \underline{87.0} & 82.3 & 83.3 & 82.6\textsuperscript{\textdagger} & 87.5 & \textbf{87.1} & - & 80.3 & \citep{modrzejewski2023} \\
Technique Identification &  & Acc. & 74.4 & 66.1 & 63.9 & \underline{76.7}\textsuperscript{\textdagger} & 75.5 & \textbf{76.9} & - & 65.6 & \citep{YamamotoNT22}\\
\hline

\multirow{5}{*}{Music Transcription} & \multirow{5}{*}{MAPS} & Frame F1 & \underline{83.92} & 82.94 & 81.53 & \textbf{84.23} & - & - & - & 82.89 & \multirow{5}{*}{\citep{toyama2023}}\\
 &  & Note F1 & \underline{86.45} & 85.97 & 85.14 & \textbf{86.54} & - & - & - & 85.14 &\\
 &  & Note w/ Offset F1 & \underline{68.27} & \textbf{68.27} & 66.28 & 68.26 & - & - & - & 66.34 &\\
 &  & \begin{tabular}[c]{@{}c@{}}Note w/ \\Offset \& Velocity F1\end{tabular} & \textbf{51.34} & \underline{50.42} & 48.69 & 50.46 & - & - & - & 48.20 &\\
\hline

\multirow{8}{*}{Source Separation} & \multirow{4}{*}{MUSDB18}  & SDR (bass) & \underline{9.50} & 8.86 & 8.17 & 7.12 & - & 5.6 & - & \textbf{11.31} & \multirow{4}{*}{\citep{lu2024bsroformer}}\\ 
 &  & SDR (drums) & \underline{8.65} & 8.03 & 7.50 & 6.65 & - & 3.6 & - & \textbf{9.49} \\
 &  & SDR (other) & \underline{5.91} & 5.59 & 5.54 & 4.77 & - & 3.0 & - & \textbf{7.73} \\
 &  & SDR (vocals) & \underline{8.07} & 7.57 & 7.66 & 6.84 & - & 5.3 & - & \textbf{10.66} \\
\cmidrule(lr){2-12}

 & \multirow{4}{*}{\begin{tabular}[c]{@{}c@{}}MDXDB21\\ hidden\end{tabular}} & SDR (bass) & \textbf{8.14} & 7.44 & 7.40 & 6.58 & - & - & - & \underline{7.86} & \multirow{4}{*}{\citep{rouard2023hybrid}}\\
 &  & SDR (drums) & \underline{8.16} & \textbf{8.31} & 7.37 & 6.58 & - & - & - & 7.89 \\
 &  & SDR (other) & \underline{5.21} & \textbf{5.26} & 4.93 & 4.59 & - & - & - & 5.09 \\
 &  & SDR (vocals) & \textbf{8.04} & \underline{7.81} & 7.73 & 7.12 & - & - & - & 7.70\\
 \hline
 \multirow{5}{*}{Music Mixing} 
 & \multirow{5}{*}{\begin{tabular}[c]{@{}c@{}}MDXDB21-dry\\ hidden\end{tabular}} & Stereo-Invariant & \textbf{79.86} & 87.27 & 87.32 & 87.97 & - & - & - & \underline{82.09} & \multirow{5}{*}{\citep{martinez2022automatic}}\\
 &  & Spectral$_{\textrm{mape}}$ & \underline{0.221} & 0.229 & 0.228          & 0.231 & - & - & - & \textbf{0.193}\\ 
 &  & Panning$_{\textrm{mape}}$ & \textbf{0.175} & 0.244 & 0.219              & 0.249 & - & - & - & \underline{0.179}\\
 &  & Dynamic$_{\textrm{mape}}$ & \textbf{0.064} & 0.072 & 0.073              & 0.075 & - & - & - & \underline{0.070}\\
 &  & Loudness$_{\textrm{mape}}$ & 0.171 & 0.148 & \textbf{0.132} & \underline{0.144} & - & - & - & 0.152\\ 
 \hline
 \end{tabular}}
\label{tab:summary_of_mdt}
\end{table}

The encoder design of \mfm is inspired by Jukebox but makes the representation hierarchical by enforcing the fine level to be conditioned by the coarse levels using a hierarchical autoencoder framework called hierarchically quantized VAE (HQ-VAE)~\citep{takida2023hq-vae}. We then use a transformer-based multi-level auto-regressive model to characterize the probability mass of learnt HQ-VAE embeddings. 
We extract features from the intermediate representation of \mfm by first converting input audio with the encoder into tokens, feeding them into the transformers, and extracting the intermediate output from the midst layer. We refer to these extracted features as \emph{\mfm features}.

As shown in Table~\ref{tab:summary_of_mdt}, we test \textcolor{c_revise}{\mfm's feature injection for selected downstream tasks along with several baselines}. To the best of our knowledge, this is the first study on enhancing both understanding and generative tasks with the intermediate representation from a single model. We briefly list our major findings: 
\begin{enumerate}[itemsep=0pt, topsep=0pt, partopsep=0pt]
\item We empirically show that, with an auto-regressive generative model that is established on hierarchical representation, its intermediate representation can serve as generic booster of various music downstream tasks. 

\item We verify that the extracted intermediate representation is beneficial for music understanding tasks even with only an extra shallow back-end network. The extension of the shallow network with attention layers leads to further improvement.

\item We show that the extracted intermediate representation is beneficial for enhancing task-specific models, through the applications to both understanding and generative tasks.

\item Several of the above improvements in each task category result in new SOTA scores. The summary of our results is shown in Table~\ref{tab:summary_of_mdt}.
\end{enumerate}

\section{Proposed Two-stage Hierarchical Model: \mfm}
\label{sec:foundation_model}
To explore the effectiveness of hierarchical modeling in boosting downstream tasks, we adopt a typical two-stage generative modeling~\citep{dhariwal2020jukebox, copet2023musicgen, li2023jen1}. In  stage-1, we use an HQ-VAE for hierarchical representation learning, dividing information into different levels based on granularity.
In stage-2, we use auto-regressive modeling to learn the multi-level token streams extracted from the stage-1 model. Finally, features from stage-2 model are extracted as described in Section~\ref{ssec:dt_fe}. 

\subsection{Stage-1 Model: HQ-VAE}
\label{ssec:stage1}
We construct the architecture of \mfm to learn a hierarchical representation of the target dataset.
Consider a music sample $\bm{x}$ with length $T$, where $\bm{x}\in\mathcal{X}\subset\mathbb{R}^T$. 
A set of codebooks $\{\bm{B}_1,\bm{B}_2,\bm{B}_3\}$ is used for learning a three-layer hierarchical representation on $\bm{x}$.
For $l\in\{1,2,3\}$, the $l$th codebook is denoted as $\bm{B}_l=\{\bm{b}_{l,k}\}_{k=1}^{K_l}$, which consists of $K_l$ $d_l$-dimensional trainable vectors $\bm{b}_{l,k}\in\mathbb{R}^{d_l}$. 
The architecture is designed to extract a hierarchical latent representation of music samples, which is denoted as $\bm{Z}_{1,2,3}:=\bm{Z}_1\otimes\bm{Z}_2\otimes\bm{Z}_3$ with $\bm{Z}_l\in\bm{B}_l^{t_l}$ ($l=1,2,3$), where $t_l$ is the latent sequence length at the $l$th layer. The discrete tensors $\bm{Z}_1$, $\bm{Z}_2$, and $\bm{Z}_3$ are expected to convey the coarse, medium, and fine-grained information. The reconstruction can be done with a well-optimized neural function $\bm{f}:\bm{B}_1^{t_1}\otimes\bm{B}_2^{t_2}\otimes\bm{B}_3^{t_3}\to\mathcal{X}$, i.e., $\bm{x}\approx \bm{f}(\bm{Z}_{1,2,3})$.

\begin{figure*}[t]
    \centering
    \subfloat[Stage 1 model architecture]{\includegraphics[width=0.9\textwidth,keepaspectratio,clip]{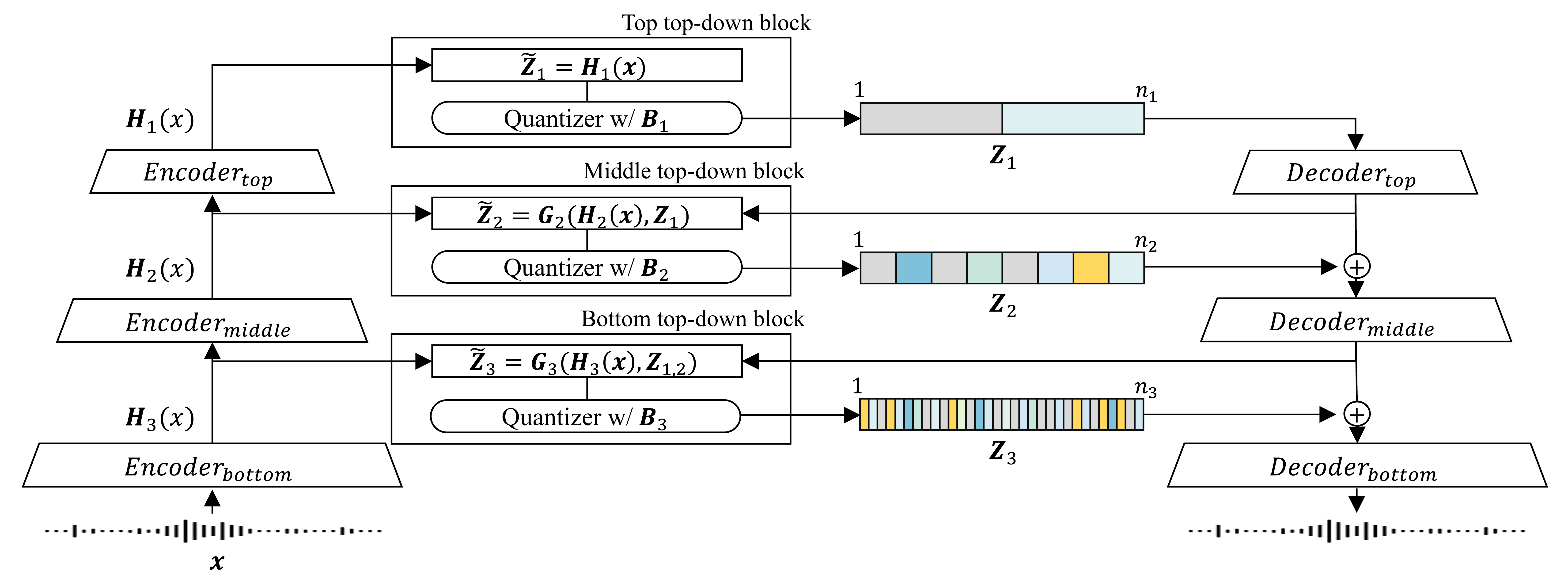}} \\
    \centering
    \subfloat[Stage-2 model architecture]    {\includegraphics[width=0.9\textwidth,keepaspectratio,clip]{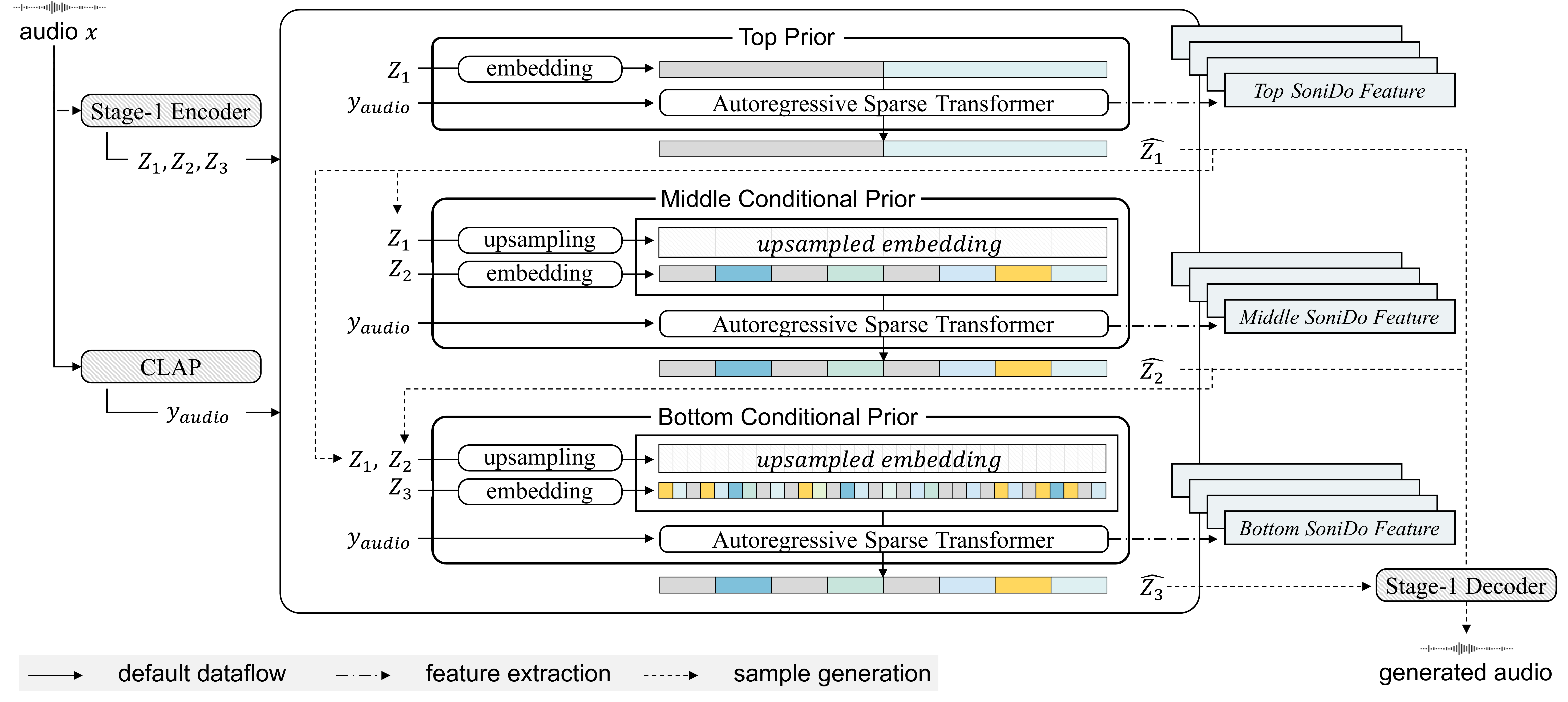}}
    \caption{The two stages of \mfm.} 
    \label{fig:mfm_overall}    
\end{figure*}

The architecture is composed of bottom-up and top-down paths, as illustrated in Figure~\ref{fig:mfm_overall}(a), the inference process of which is as follows. A series of encoders in the bottom-up path extracts feature tensors for three different information resolutions, which are denoted as $\bm{H}_{l}(\bm{x})$ $(l=1,2,3)$, from sample $\bm{x}$. The feature $\bm{H}_{l}(\bm{x})$ is used for the top-down path to process the data in a hierarchical manner. The top-down path has three (top, middle, and bottom) top-down blocks to model hierarchical discrete latent representations. The top block first quantizes $\tilde{\bm{Z}}_1:=\bm{H}_1(\bm{x})$, which has the most global (coarse) information amongst the encoded features, into discrete tensor $\bm{Z}_1$ by the nearest neighbor search in codebook $\bm{B}_1$. At the next step, the middle latent tensor is conditioned on the top $\bm{Z}_1$ to focus more on local details, with the injection of $\bm{H}_2(\bm{x})$. Therefore, the block takes both tensors processed in the top block and bottom-up paths, i.e., 
$\bm{Z}_1$ and $\bm{H}_2(\bm{x})$, generating a raw continuous feature $\tilde{\bm{Z}}_2:=\bm{G}_2(\bm{H}_2(\bm{x}),\bm{Z}_1)$. The raw feature is then quantized into $\bm{Z}_2$ in the same manner as with codebook $\bm{B}_2$. The bottom block repeats a similar process with $\bm{Z}_2$ and $\bm{H}_3(\bm{x})$ to further refine the representation with the additional discrete feature $\bm{Z}_3$. Finally, the set of $\bm{Z}_1$, $\bm{Z}_2$, and $\bm{Z}_3$ is decoded to the data space to reconstruct $\bm{x}$.

We train the architecture including the codebooks within the variational Bayes framework, as an instance of HQ-VAE, stochastically quantized VAE-2 (SQ-VAE-2)~\citep{takida2023hq-vae}. To establish a generative process in this VAE, we first define the prior probability distribution on $\bm{Z}_{1,2,3}$ as $P(\bm{Z}_{1,2,3})=P_1(\bm{Z}_1)P_2(\bm{Z}_{2}|\bm{Z}_{1}) P_3(\bm{Z}_{3}|\bm{Z}_{1,2})$.
Given a chunk of latent variables $\bm{Z}_1$, $\bm{Z}_2$, and $\bm{Z}_3$, a data sample can be generated under a conditional probability distribution $p(\bm{x}|\bm{Z}_{1,2,3})$. Concretely, we parameterize the conditional distribution as a normal distribution with function $\bm{f}$ and a trainable isotropic covariance matrix as $p(\bm{x}|\bm{Z}_{1,2,3})=\mathcal{N}(\bm{f}(\bm{Z}_{1,2,3}),\sigma^2\bm{I})$.
To summarize, the generative process consists of two steps: sampling $\bm{Z}_{1,2,3}$ from the prior distribution and decoding it with the conditional distribution. Note that, in practice, $\bm{Z}_{1,2,3}$ is sampled from an estimated posterior distribution instead of the prior distribution, as presented in Section~\ref{ssec:stage2}.
Next, the approximated posterior distribution for $p(\bm{Z}_{1,2,3}|\bm{x})$ is set as $Q(\bm{Z}_{1,2,3}|\bm{x})=Q_1(\bm{Z}_1|\bm{x})Q_2(\bm{Z}_2|\bm{Z}_1,\bm{x})Q_3(\bm{Z}_3|\bm{Z}_{1,2},\bm{x})$. We connect each $Q_1$, $Q_2$, and $Q_3$ with the components in Figure~\ref{fig:mfm_overall}(a). Specifically, the categorical distribution at the $l$th layer, $Q_l(\bm{Z}_l|\bm{Z}_{<l},\bm{x})$, is defined as a stochastic quantization that is $\hat{P}_{s_l^2}(\bm{z}_{l,n}=\bm{b}_{l,k}|\tilde{\bm{z}}_{l,n})\propto\exp(-\|\tilde{\bm{z}}_{l,n}-\bm{b}_{l,k}\|^2/2s_l^2)$ with a trainable positive scalar $s_l^2$, where $\bm{z}_{l,n}$ and $\tilde{\bm{z}}_{l,n}$ indicate the $n$th vectors in $\bm{Z}_l$ and $\tilde{\bm{Z}}_l$, respectively.
Finally, the resulting training objective consists of terms for reconstruction and latent regularization:
\begin{align}
\Ls_1(\bm{x})=
\frac{T}{2}\log\sigma^2+\mathbb{E}_{Q(\bm{Z}_{1,2,3}|\bm{x})}
\left[\frac{\|\bm{x}-\bm{f}(\bm{Z}_{1,2,3})\|_2^2}{2\sigma^2}
+\sum_{l=1}^3\left(\frac{\|\tilde{\bm{Z}}_l-\bm{Z}_l\|_F^2}{2s_l^2}
-H(\hat{P}_{s_l^2}(\bm{Z}_l|\tilde{\bm{Z}}_{l}))\right)
\right],
\label{eq:sqvae2_gaussian}
\end{align}
where $H(\cdot)$ is the entropy of a probability mass function. 
Progressive coding~\citep{takida2023hq-vae} is applied to ensure the amount of information is balanced across the three layers.

\subsection{Stage-2 Model: Sparse Transformers}
\label{ssec:stage2}
The stage-2 model addresses the gap between the pre-set prior distribution (i.e., $P(\bm{Z}_{1,2,3})$) and marginalized posterior distribution (i.e., $Q(\bm{Z}_{1,2,3}):=\E_{p(\bm{x})}[Q(\bm{Z}_{1,2,3}|\bm{x})]$) by directly learning the posterior distribution.
We incorporated the contrastive language-audio pretraining (CLAP) model proposed by LAION~\citep{laionclap2023} into the stage-2 model. To include the CLAP conditioning, we approximate $Q_{\bm{\phi}}(\bm{Z}_{1,2,3})$ with a conditioned decomposition as
\begin{align}
    P_{\bm{\Pi}}(\bm{Z}_{1,2,3}|\bm{y}_{\text{audio}})=
    P_{\bm{\pi}_1}(\bm{Z}_1|\bm{y}_{\text{audio}})
    P_{\bm{\pi}_2}(\bm{Z}_{2}|\bm{Z}_{1},\bm{y}_{\text{audio}})
    P_{\bm{\pi}_3}(\bm{Z}_{3}|\bm{Z}_{1,2},\bm{y}_{\text{audio}}),
\end{align}
where $\bm{\Pi}:=\{\bm{\pi}_1,\bm{\pi}_2,\bm{\pi}_3\}$ is a set of neural networks for the stage-2 model, and $\bm{y}_{\text{audio}}\in\mathbb{R}^{512}$ denotes the feature produced from the CLAP encoder. Thanks to the alignment between the audio and text embeddings of CLAP, even if the audio dataset has no text caption, we can still feed audio in the training phase, whereas it allows either audio or text input in the inference stage. The use of a pre-trained encoder is common in modern generative models. For example, MusicGen~\citep{copet2023musicgen} uses the pre-trained T5 encoder~\citep{Raffel2019T5textencoder} to model the text conditions. 
The training objective is negative log-likelihood:
\begin{align}
    \Ls_2(\bm{x})=\E_{Q_{\bm{\phi}}(\bm{Z}_{1,2,3}|\bm{x})p_{\text{CLAP}}(\bm{y}_{\text{audio}}|\bm{x})}[-\log P_{\bm{\Pi}}(\bm{Z}_{1,2,3}|\bm{y}_{\text{audio}})].
\end{align}

We follow Jukebox~\citep{dhariwal2020jukebox} to construct the networks $\bm{\Pi}$ with sparse transformers~\citep{vaswani2017attention, child2019generating}. As illustrated in Figure~\ref{fig:mfm_overall}(b), we train three auto-regressive sparse transformers to model $P(\bm{Z}_1|\bm{y}_{\text{audio}})$, $P(\bm{Z}_2|\bm{Z}_1, \bm{y}_{\text{audio}})$, and $P(\bm{Z}_3|\bm{Z}_{1,2}, \bm{y}_{\text{audio}})$, which we refer to as top prior, middle conditional prior, and bottom conditional prior, respectively. 
The middle and bottom priors use the token sequences from the upper levels, with up-sampling achieved through \textit{upsampling modules}, corresponding to the \textit{conditioners} of Jukebox.
We additionally condition each prior on $\bm{y}_{\text{audio}}$. 
Appendix~\ref{sec:app_details_of_mfm_stage2} provides further details of the stage-2 model.
Appendix~\ref{sec:app_eval_music_generation} evaluates the common objective metrics on \mfm.  

\subsection{{\mfm} vs. Other Music Foundation Models}
\label{sec:mfm_compare}

\begin{figure*}[th]
  \centering
  \begin{minipage}{0.32\textwidth}
    \centering
    \includegraphics[height=3cm]{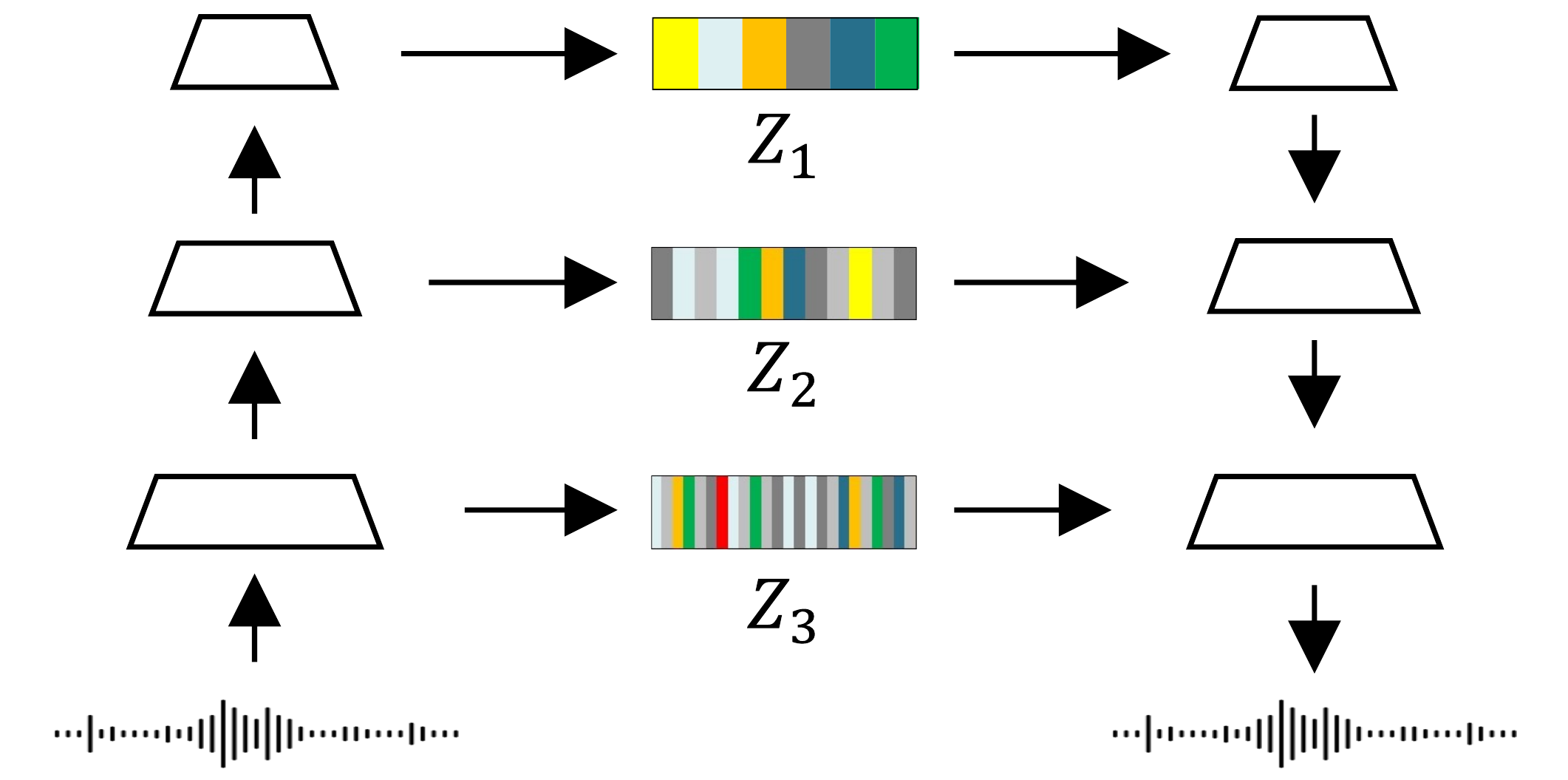}
    \Scaption{SQ-VAE-2}
    \label{fig:mfm_comparison_sub_sqvae2}
  \end{minipage}%
  \hfill
  \begin{minipage}{0.32\textwidth}
    \centering
    \includegraphics[height=3cm]{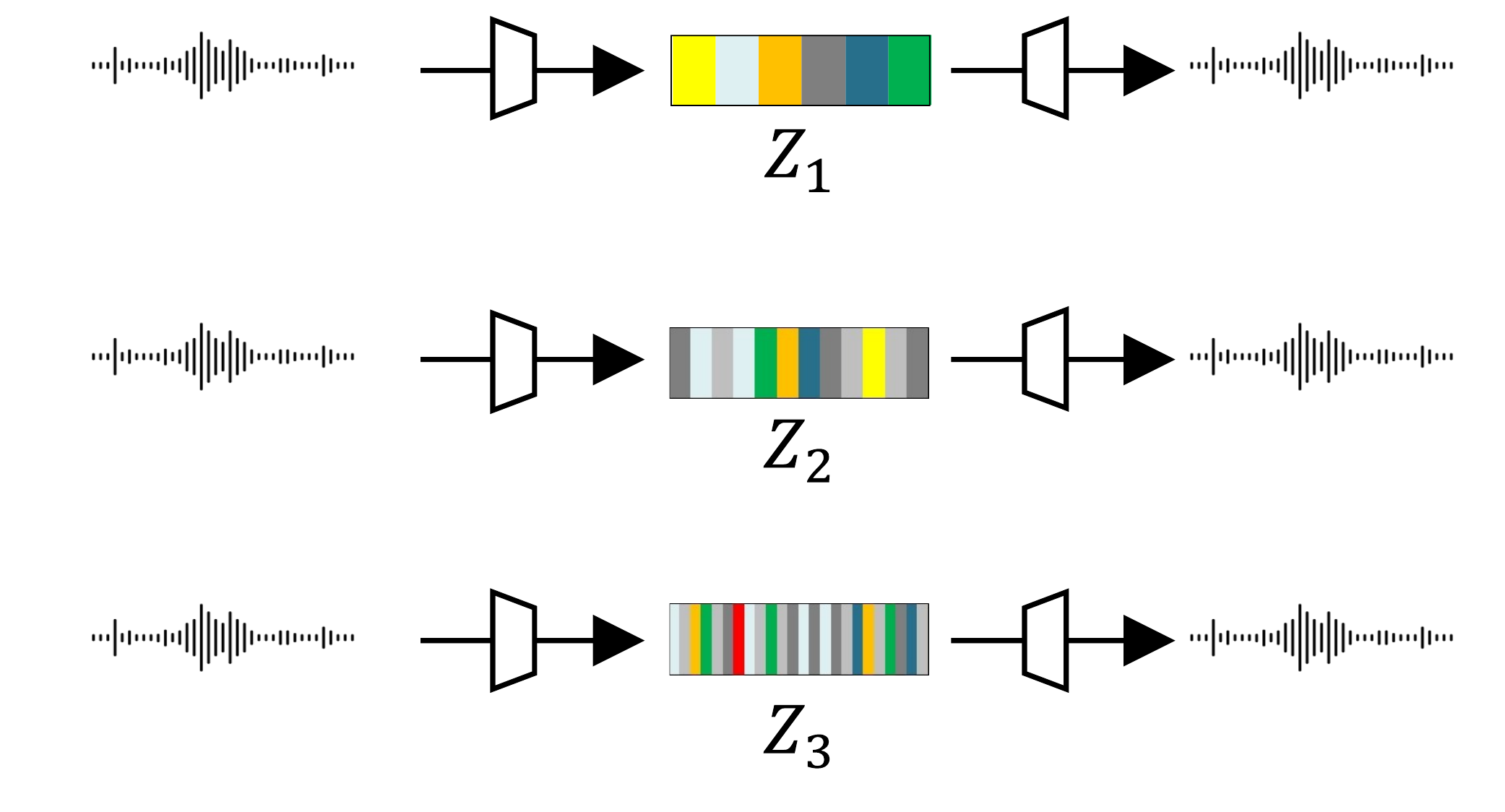}
    \Scaption{multi-resolution VQ-VAEs}
    \label{fig:mfm_comparison_sub_jukebox}
  \end{minipage}%
  \hfill
  \begin{minipage}{0.32\textwidth}
    \centering
    \includegraphics[height=3cm]{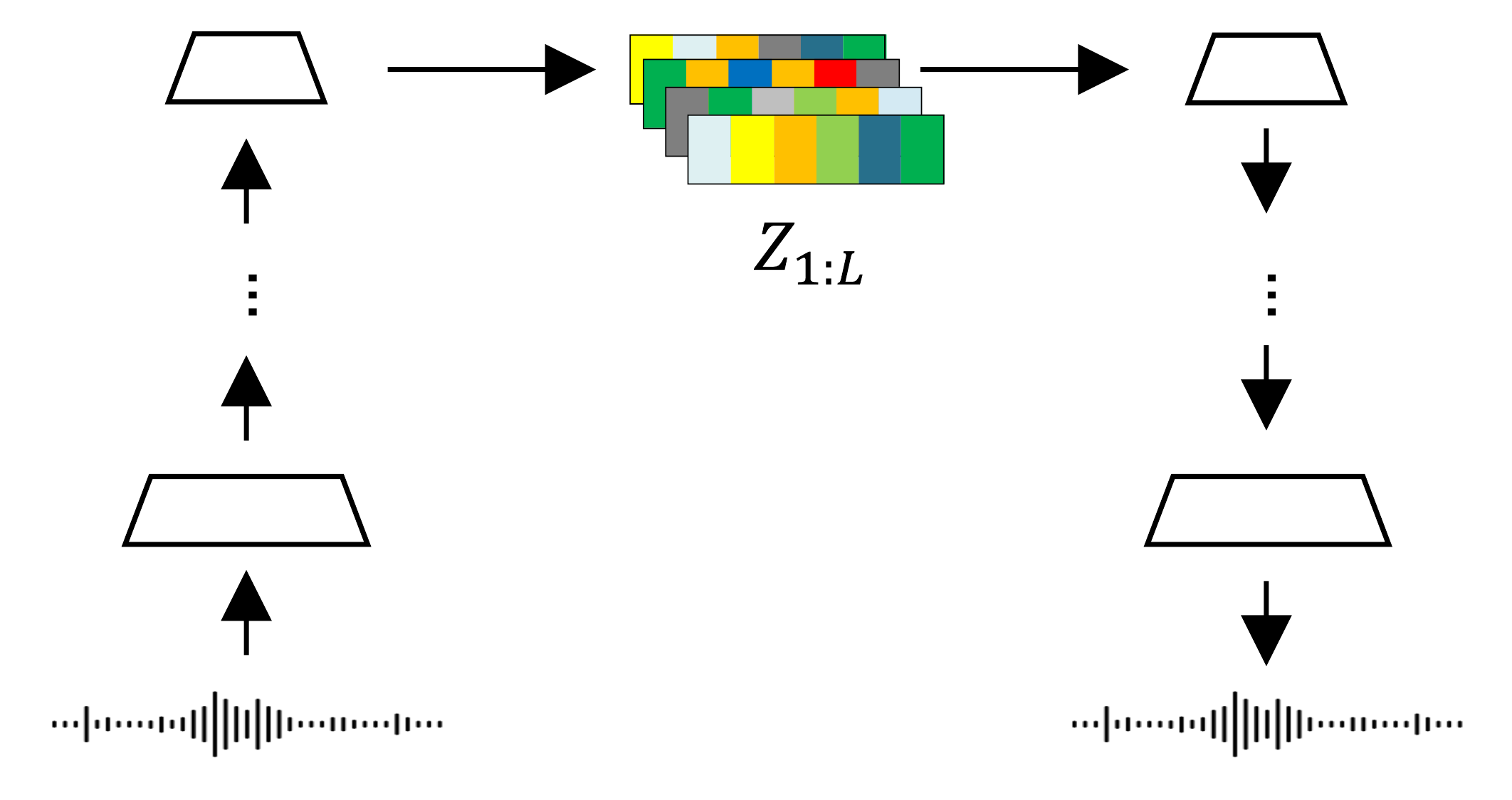}
    \Scaption{RVQ}
    \label{fig:mfm_comparison_sub_rvq}
  \end{minipage}
  \caption{Stage-1 model comparison.}
  \label{fig:mfm_comparison}
\end{figure*}

This section compares the architecture of \mfm with those of other well-known music foundation models, i.e., Jukebox~\citep{dhariwal2020jukebox}, MusicLM~\citep{agostinelli2023musiclm}, and MusicGen~\citep{copet2023musicgen}. These models are categorized on the basis of how their stage-1 models are constructed; (a) SQ-VAE-2, (b) multi-resolution VQ-VAEs, and (c) residual vector quantization (RVQ), as illustrated in Figure \ref{fig:mfm_comparison}.

While \mfm and Jukebox exhibit some shared characteristics, such as a three-level architecture in the stage-1 model, \mfm is based on SQ-VAE-2, whereas Jukebox used multi-resolution VQ-VAEs.
In Jukebox, token streams $\bm{Z}_1$, $\bm{Z}_2$, and $\bm{Z}_3$ were independently and separately trained for different sampling rates. Consequently, the $l$th transformer was designed to generate $\bm{Z}_l$ by \textit{upsampling} the previous token sequence $\bm{Z}_{l-1}$ for $l=2, 3$. In contrast, \mfm's token streams from the stage-1 model are jointly trained and collaboratively contribute to the comprehensive modeling of the waveform at the original sampling rate from scratch. Given the tight interrelation between token streams from different levels, \mfm's $l$th transformer is conditioned on all the upper token streams. 

Recent approaches such as MusicLM and MusicGen used RVQ in a bottleneck feature space instead of applying these hierarchical quantization methods (e.g., SQ-VAE-2).
These approaches also use transformers to model the prior of the music-token streams $P(Z_{1:L})$.
In the context of token-sequence length for generating 1~s of audio at a target sampling rate $sr$, the bottom-most token-sequence length in \mfm and Jukebox is $sr/8$, while MusicGen requires a token-sequence length of $sr/640$.
RVQ-based models excute quantization in highly compressed latent spaces using a series of vector quantization layers, effectively shortening the token sequence to be learned by transformers in the stage-2 model.

\section{{\mfm} on Music Downstream Tasks}
We first obtain \mfm features from input audio with a task-agnostic feature extraction process. Depending on whether the downstream task is time-invariant or time-varying, we then apply different pre-processing steps. Finally, we inject the pre-processed \mfm features into a proper location of a target task-specific model. The selection of such a location is explained in Section~\ref{sec:experiments}.

\subsection{Task-agnostic Feature Extraction}
\label{ssec:dt_fe}
We follow the feature extraction pipeline in \citet{CastellonDL21calm, ntt2022mae, huang2022investigating} based on the pre-trained frozen \mfm. 
The music waveform is first converted to multi-level token sequences via the stage-1 encoder of \mfm. The tokens are then fed into the top prior, middle conditional, and bottom conditional priors of \mfm without auto-regressive iteration. 
The middle and bottom priors are conditioned by the ground-truth tokens produced with the stage-1 encoders. 
We extract the output of the $N$-th ($N=36$ as in \citet{CastellonDL21calm}) transformer layer in those prior models as \mfm features. 
If the CLAP audio embedding is not used as the condition of priors, we call the feature extraction unconditional extraction; otherwise, CLAP-conditional extraction.

The maximum sequence length of the priors is $8192$. 
Since the down-sampling rates in the stage-1 model are $128\times$ (top), $32\times$ (middle), and $8\times$ (bottom), 
the same amount of $8192$ tokens in different priors correspond to 24~s, 6~s and 1.5~s in the time domain, 
forming a set of hierarchical multi-rate features. 
To save computational resources, the \mfm features are pre-computed for most downstream tasks, except for HTDemucs mentioned in Section~\ref{sssec:music_editing_mss}, where a clip of music is randomly selected on-the-fly during training. 
To compute features for a long audio input, we treat the input as overlapping segments with the ratio $N_\mathrm{ovlp}$. 
If $N_\mathrm{ovlp}$ is sufficiently large such that the overlap is longer than the perception field of the stage-2 model, it is guaranteed that the feature extraction result is not affected by the segmentation.

\subsection{Feature Pre-processing for Time-invariant Downstream Tasks}
\label{ssec:dt_pp_1}
If the downstream task is time-invariant, we first divide input audio into non-overlapping segments of 24 (top), 6 (middle), and 1.5 (bottom)~s. For each prior, the \mfm features of the segment are reduced to a single token via average pooling, forming $3$ \mfm token sequences in the end.

The common practice~\citep{CastellonDL21calm, li2023mert} suggests using a multi-layer perceptron (MLP) with a single hidden layer of $512$ dimensions to probe the features. 
However, \mfm token sequences originate from priors with different time resolutions, which is different from prior studies. To effectively use these hierarchical features, a sequence aggregation is required.  
We thus propose to aggregate the sequences via a standard attention block, which is an attention layer followed by a feed-forward layer. This is inspired by the attention-based feature aggregation in instrument classification tasks~\citep{gururani2019attention, zhong2023tagging}. 
We first concatenate the hierarchical \mfm features into a single token sequence then attach a learnable class token at the front. The attention block is trained to aggregate all features into the class token, which is then converted to music tags or emotion scores by the aforementioned MLP. Hyperparameters as well as an ablation study on the sequence aggregation are provided in Appendix~\ref{sec:app_mir}.

\begin{figure*}[tb]
    \centering
    \includegraphics[width=0.6\textwidth,keepaspectratio,clip]{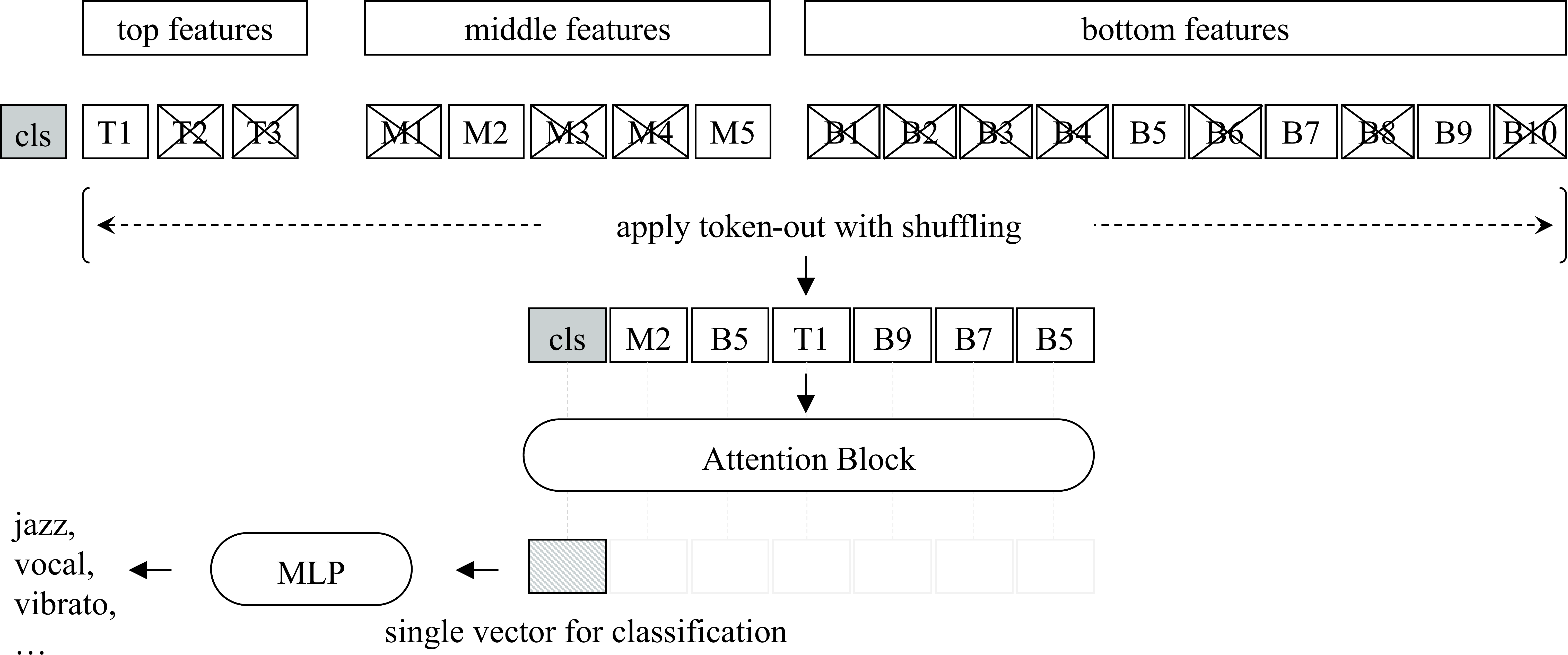}
    \caption{Attention-based feature aggregation and token-out data augmentation. ``T'', ``M'', ``B'' mean top, middle, and bottom priors, respectively. Token-out augmentation deletes masked tokens from input sequence. Attention block aggregates sequence into single vector and is followed by MLP to predict tags.}
    \label{fig:mir_tokenout}
\end{figure*}

To prevent overfitting when using the concatenated token sequence to train the attention block, we propose an on-the-fly data augmentation method called \textit{token-out}. This is inspired by SpecAugment~\citep{specaugment2019} and masked Transformers~\citep{koutini2022patchout,zhong2023extending,comunita2024specmaskgit}, in which a part of the input is masked before feeding into deep neural networks. Unlike prior arts, 
 \textcolor{c_revise}{token-out is applied to the whole \textit{token} sequence extracted with the multiple layers in \mfm,} as illustrated in Figure~\ref{fig:mir_tokenout}.The masking ratio is sampled between 0 and 100\% uniformly. As shown in Appendix~\ref{sec:app_mir}, aggregating the \mfm features with the shallow attention layer and token-out augmentation led to performance improvement.

\subsection{Feature Pre-processing for Time-varying Downstream Tasks}
\label{ssec:dt_pp_2}
When applying the \mfm features to a task-specific model on a time-varying downstream task, we face several challenges, such as temporal alignment, proper amount of information compression, and sufficient feature adaptation before injecting features into the task-specific models. 

The temporal alignment between the \mfm features and target model can be achieved by either pooling the \mfm features or using linear layers. 
Examples of these two cases are provided in Appendices~\ref{sec:app_mss_umx_ablation} and~\ref{sec:app_amt}, respectively. 
Compared with average pooling or max-pooling, we found that using linear layers can yield better performance, as described in Appendix~\ref{sec:app_mss_umx_ablation}. 

Both information compression and feature adaptation can be done with linear layers. The output dimension is simply set to match the feature dimension of the target task-specific model.
Empirically, one layer is sufficient for most of the models we have tested, except for music transcription with hFT-Transformer~\cite{toyama2023}, which requires four layers, as described in Appendix \ref{sec:app_amt_hft_transformer}.

\section{Experiments}
\label{sec:experiments}
We conducted experiments to examine the usefulness of features extracted from music foundation models for understanding and generative downstream tasks by addressing two questions: \textbf{Q1:} Do extracted features have useful information for music understanding? \textbf{Q2:} Can extracted features boost current task-specific models for both understanding and generative tasks?
To verify the generalizability of the results, we test not only the \mfm features, but also features extracted from \textcolor{c_revise}{Jukebox and} the two public versions of \musicgen~\citep{copet2023musicgen}, namely \musicgens and \musicgenl. \textcolor{c_revise}{As a major focus of this work is to extend the applicable downstream tasks, we evaluate Jukebox's features specifically for time-varying tasks and report the results from~\citet{CastellonDL21calm} for time-invariant tasks in Table~\ref{tab:summary_of_mdt}}.

To address the first question, we selected eight music tagging tasks and music transcription as representatives for understanding tasks.We verified that the extracted features encompass both time-invariant information of overall musical properties and time-varying information of specific musical events.
For the second question, we tested the injection of extracted features into several task-specific models for understanding and generative tasks. They consist of one transcription model~\citep{toyama2023}, two  separation models~\citep{mitsufuji2022music,fabbro2023sound}, and two mixing models from~\citet{martinez2022automatic}. 

The feature extraction described in Section~\ref{ssec:dt_fe} is used for all experiments. Before applying the features into downstream tasks, task-dependent pre-processing is applied (see Sections~\ref{ssec:dt_pp_1} and~\ref{ssec:dt_pp_2}). Details of the experimental setup, results, and further ablation studies are provided in Appendices~\ref{sec:app_mir},~\ref{sec:app_amt},~\ref{sec:app_mss_ds}, and~\ref{sec:app_mixing}, respectively. We only use features extracted from the top and middle layers of \mfm. In the preliminary experiments, we found that including the bottom-layer features does not always improve the performance of understanding tasks. We assume this is due to the bottom layer mostly containing only the fine-grained information irrelevant to the tasks, thus degrading performance. This is discussed further in each subsection.

\subsection{Usefulness of Extracted Features for Music Understanding}
\label{ssec:exp_understanding}

The following feature probing experiments demonstrate that features extracted from \mfm, \musicgen and Jukebox all contain valuable information, which is consistent with~\citet{CastellonDL21calm}. However, for music transcription, the shallow network remains insufficient to match SOTA models. In Section~\ref{ssec:exp_gen}, we show that injecting extracted features into task-specific models can boost their performance beyond SOTA.

\subsubsection{Music Tagging}
\label{sssec:exp_mir}
We test a wide range of music tagging tasks as well as the emotion regression task: MagnaTagATune (MTAT)~\citep{LawWMBD09mtat} for auto tagging, Nsynth~\citep{nsynth2017} for pitch and instrument recognition, EmoMusic~\citep{soleymani2013emomusic} for emotion regression, GTZAN~\citep{tzanetakis2002gtzan} for genre classification, GiantSteps~\citep{knees2015giantsteps, widmer2017mtg-giantsteps} for musical key estimation, and VocalSet~\citep{wilkins2018vocalset} for singer and singing technique identification. A summary of the datasets is shown in Table~\ref{tab:mir_datasets} in Appendix \ref{sec:app_mir}. We followed the pre-processing in previous studies~\citep{li2023mert, yuan2023marble} and used scikit-learn~\citep{scikit-learn} and mir\_eval~\citep{raffel2014mir_eval} for metric computation. The average R$^2$ of arousal and valence axis is reported for EmoMusic.
The feature pre-processing for time-invariant downstream tasks in Section~\ref{ssec:dt_pp_1} including the feature aggregation and token-out augmentation is applied for all tasks. Following common practice~\citep{CastellonDL21calm,li2023mert}, an MLP is then used to probe the aggregated features.

We conduct a preliminery study for \mfm with top prior features to compare CLAP-conditional extraction, unconditional extraction and features from the CLAP encoder. We use the tagging task for coarse-grained concepts on MTAT, and the classification task for fine-grained concepts (pitch) on Nsynth. We found that CLAP performs well for coarse concepts, while unconditional extraction results in better accuracy for pitch estimation. CLAP-conditional extraction achieves better scores in both tasks. Details can be found in Appendix~\ref{sec:app_mir} and Table~\ref{tab:mir_exp_preliminary}. Consequently, we report \mfm's scores with the CLAP-conditional feature extraction for time-invariant understanding tasks.

The test results on various datasets and benchmarks with prior studies are listed in Table~\ref{tab:mir_exp_benchmark}.
Probing the \mfm and \musicgen features both shown competitive scores in most tasks. The \mfm's features reached the top-2 performance in auto tagging, pitch estimation, instrument classification, and genre classification. They also performs well for emotion regression and singer identification. While these are prompt-conditioned generative models, feature probing using these models reached comparable performance compared with SOTA encoder-only models specialized for understanding tasks.   

\begin{table}[t] 
\centering
\caption{Music tagging. Benchmark results of \mfm features in music tagging tasks (\textbf{bold}: top-2 score).}
    \resizebox{\linewidth}{!}{
    \begin{tabular}{cccccccccc}
\toprule
    Dataset & \multicolumn{2}{c}{MTAT} & Nsynth & Nsynth & EmoMusic & GiantSteps & GTZAN &VocalSet & VocalSet\\
    Task & \multicolumn{2}{c}{Auto tagging}& Pitch & Instrument & Emotion regression & Key & Genre & Singer & Vocal techniques\\
\midrule
    Metrics & ROC-AUC & mAP & Acc. & Acc. & Average R$^2$ & Weighted acc. & Acc. & Acc. & Acc.\\
\hline
    \textit{Supervised} \\
    MusiCNN \cite{pons2019musicnn}  & 90.6 & 38.3 & 64.1 & 72.6 & 58.5 & 12.8 & 79.0 & 57.0 & 70.3\\
    MULE-supervised \cite{McCallumKOGE22mule} & \textbf{91.7} & 41.3 & 79.3 & 73.1 & 64.6 & 28.6 & \textbf{83.5} & - & - \\
\hline
    \textit{Auto-regression} \\
    Jukebox \cite{dhariwal2020jukebox, CastellonDL21calm} & 91.5 & \textbf{41.4} & 
91.6 & 70.4 & \textbf{66.9} & \textbf{66.7} & 79.7 & 82.6 & \textbf{76.7} \\
    MusicGen-small \cite{copet2023musicgen} & 90.4 & 38.8 & 93.3 & 71.9 & 45.6 & 65.2 & 75.2 & 82.3 & 66.1\\
    MusicGen-large \cite{copet2023musicgen} & 90.5 & 39.0 & 92.8 & 74.2 & 46.2 & 62.4 & 70.3 & 83.3 & 63.9\\
\hline
    \textit{Contrastive} \\
    CLMR \cite{janne2021clmr} & 89.4 & 36.1 & 47.0 & 67.9 & 56.8 & 14.9 & 68.6 & 49.9 & 58.1 \\
    Slowfast-NFNet-F0 \cite{wang2022slowfast} & - & 39.5 & 88.0 & \textbf{78.2} & - & - & - & - & - \\
    MULE-contrastive \cite{McCallumKOGE22mule} & 91.4 & 40.4 & 89.2 & 74.0 & 63.9 & \textbf{66.7} & 73.5 & \textbf{87.5} & 75.5 \\
\hline
    \textit{Mask reconstruction} \\
    HuBERT music \cite{hsu2021hubert, li2023mert} & 90.2 & 37.7 & 77.4 & 69.3 & 54.3 & 14.7 & 70.0 & 75.3 & 65.9 \\
    data2vec music \cite{baevski2022data2vec, li2023mert} & 90.0 & 36.2 & 93.1 & 69.4 & 61.6 & 50.6 & 74.1 & 81.4 & 71.1 \\
    MERT-330M \cite{li2023mert} & 91.3 & 40.2 & \textbf{94.4} & 72.6 & \textbf{68.0} & 65.6 & 79.3 & \textbf{87.1} & \textbf{76.9} \\
\hline
    \textit{Hierarchical auto-regression (ours)} \\
    \mfm & \textbf{91.7} & \textbf{41.5} & \textbf{93.8} & \textbf{78.0} & 64.7 & 63.5 & \textbf{80.7} & 87.0 & 74.4\\
\bottomrule
    \end{tabular}}

\label{tab:mir_exp_benchmark}    
\end{table}

\begin{table}[tb]
    \centering
    \caption{Results of feature probing using shallow back-end on MAPS (\textbf{bold}: best, \underline{underline}: second-best).
    }
    \label{tab:amt_linear_piano}
    \small
    \begin{tabular}{lc}
    \toprule
    Input&Note F1(\%)\\
    \midrule
    Spectrogram&18.83\\
    Spectrogram + \mfm Top&57.20\\
    Spectrogram + \mfm Middle&\underline{64.98}\\
    Spectrogram + \mfm Top + \mfm Middle&\textbf{66.02}\\
    Spectrogram + \musicgens&53.18\\
    Spectrogram + \musicgenl&49.16\\
    \textcolor{c_revise}{Spectrogram + Jukebox}&\textcolor{c_revise}{57.13}\\
    \bottomrule
    \end{tabular}
\end{table}

\subsubsection{Music Transcription}

Beyond time-invariant understanding tasks, we continue the test on music transcription, which is a time-varying understanding task. \textcolor{c_revise}{All of \mfm, \musicgen, and Jukebox} features are obtained with unconditional extraction described in Section~\ref{ssec:dt_fe}. 
The dimension and time resolution of extracted features are aligned to those of the spectrogram with linear layers.
Following~\citet{CastellonDL21calm}, the features are concatenated with the spectrogram. A single-layer shallow back-end network is used to probe these features. 

We show the transcription performance of the feature probing mentioned above on the MAPS dataset~\citep{emiya2010} in Table~\ref{tab:amt_linear_piano}. All the features greatly improved the note-wise F1 score compared with using the spectrogram only. This suggests that \textcolor{c_revise}{all of \mfm, \musicgen, and Jukebox} features contain useful information for time-varying understanding tasks.

\subsection{Using Extracted Features to Boost Existing Task-specific Models}
\label{ssec:exp_gen}
In Section~\ref{ssec:exp_understanding}, we showed that the extracted features contain useful knowledge for music understanding. 
In this section, we test \textcolor{c_revise}{all of \mfm, \musicgen, and Jukebox} features on several SOTA task-specific models, the tasks include music transcription, music source separation, and music mixing, which covered both music understanding and generative tasks. 
The experimental results indicate that the extracted features consistently boost the performances of task-specific models. We also observed that injecting the \mfm features accelerated the decrement of training loss in early epochs.

\subsubsection{Music Transcription: hFT-Transformer}\label{sssec:exp_amt}

We applied the extracted features to hFT-Transformer~\citep{toyama2023}, a SOTA music transcription model for piano on MAPS~\citep{emiya2010}, to assess whether it surpasses existing models that rely solely on the spectrogram. On the basis of the input spectrogram, hFT-Transformer estimates the frame-based note activation, along with the onset, offset, and velocity of a note (\textit{frame}, \textit{onset}, \textit{offset}, and \textit{velocity}). It is a transformer-based model consisting of two transformer encoders that work on different axes of the input and a transformer decoder in the middle of these two encoders. Following the processing pipeline in Section~\ref{sec:experiments}, we attempted injecting the \textcolor{c_revise}{\mfm, \musicgen, and Jukebox} features before the 1st encoder, 2nd encoder, and decoder. We found that feature injection before the decoder yields the best result, thus we adopt this injection method in the following experiments.  
All the training hyperparameters were kept the same as in a previous study~\citep{toyama2023}. Further details are provided in Appendix~\ref{sec:app_amt_hft_transformer}.

\begin{table}[tb]
    \centering
    \caption{
    Music transcription results of F1 scores on MAPS (\textbf{bold}: best score, \underline{underline}: second best). ``Note'' refers to note-wise estimation. First row corresponds to hFT-Transformer~\citep{toyama2023}.
    }
    \small
    \tabcolsep=3pt
    \resizebox{0.95\linewidth}{!}{
    \begin{tabular}{clcccc}
    \toprule
    Training data&Input&Frame&Note&Note w/ Offset&Note w/ Offset\&Velocity\\
    \midrule
    \multirow{6}{*}{100[\%]}&Spectrogram&82.89&85.14&66.34&48.20\\
    &Spectrogram + \mfm Top&83.92&\underline{86.45}&\underline{68.27}&\textbf{51.34}\\
    &Spectrogram + \mfm Top + \mfm Middle&\underline{84.16}&85.96&67.37&\underline{50.98}\\
    &Spectrogram + \musicgens&82.94&85.97&\textbf{68.27}&50.42\\    
    &Spectrogram + \musicgenl&81.53&85.14&66.28&48.69\\
    &\textcolor{c_revise}{Spectrogram + Jukebox}&\textcolor{c_revise}{\textbf{84.23}}&\textcolor{c_revise}{\textbf{86.54}}&\textcolor{c_revise}{68.26}&\textcolor{c_revise}{50.46}\\
    \midrule
    \multirow{6}{*}{10[\%]}&Spectrogram&9.83&0.59&0.17&0.46\\
    &Spectrogram + \mfm Top&65.91&66.64&39.88&25.87\\
    &Spectrogram + \mfm Top + \mfm Middle&\textbf{71.57}&\textbf{75.00}&\textbf{46.18}&\textbf{30.63}\\
    &Spectrogram + \musicgens&63.73&65.90&39.00&24.94\\    
    &Spectrogram + \musicgenl&61.81&63.27&37.03&24.01\\
    &\textcolor{c_revise}{Spectrogram + Jukebox}&\textcolor{c_revise}{\underline{70.43}}&\textcolor{c_revise}{\underline{73.76}}&\textcolor{c_revise}{\underline{45.80}}&\textcolor{c_revise}{\underline{30.42}}\\
    \bottomrule
    \end{tabular}
    }
    \label{tab:amt_result_hft}
\end{table}

Following the common evaluation practice~\citep{gardner2022,toyama2023}, we report four F1 scores: frame-wise, note-wise, note-wise with offset, and note-wise with offset and velocity using the checkpoint with the best validation F1 score.
As shown in Table~\ref{tab:amt_result_hft}, injecting either \textcolor{c_revise}{\mfm, \musicgens, or Jukebox} features improves the performance of hFT-Transformer. The performance gap is especially huge when the model is trained with a small subset of MAPS. This demonstrates the usefulness of injecting music foundation model features into downstream task models when training data are scarce. We also observed that the decrement of training loss is faster when either \textcolor{c_revise}{\mfm, \musicgen, or Jukebox} features are injected, as shown in Figure \ref{fig:amt_loss_curve_hft} in Appendix~\ref{sec:app_amt_hft_transformer}. 

In the experiment involving the full MAPS, injecting top and middle \mfm features yields performance improvement. However, no improvement is observed when all the features from three layers are injected. A similar trend can be observed from the results of \musicgenl. We assume that the network capacity required to interpret all the information contained in the features could exceeded that of hFT-Transformer, negatively impacting the model. This suggests that, disentangling feature information on the basis of information granularity to filter out irrelevant information is crucial for such injection.

\subsubsection{Music Source Separation: UMX, HTDemucs}
\label{sssec:music_editing_mss}
We select Open-Unmix (UMX)~\citep{stoter19umx} and Demucs (HTDemucs)~\citep{rouard2023hybrid} for feature injection. UMX estimates the time-frequency mask of the target source using recurrent neural network (RNN) blocks. HTDemucs is a hybrid model with waveform U-Net branch and spectral U-Net branches. We inject the extracted features into the encoder block for UMX using a down-sampling block and in each HTDemucs branch using a cross-domain Transformer (details in Appendices~\ref{sec:app_mss_umx_detail} and ~\ref{appendix:HTDemucsMFM}). Based on the observation in Section~\ref{sssec:exp_amt} and for simplicity, we inject only top-level features from \mfm.
 
Table~\ref{tab:mss_result} lists the SDR scores on the test split of MUSDB18~\cite{rafii2017musdb18} and the hidden split of MDXDB21~\cite{mitsufuji2022music,fabbro2023sound}.
The details of the experiments are provided in Appendix~\ref{sec:app_mss_ds}. Similar to the experiment discussed in Section~\ref{sssec:exp_amt}, a faster loss decrement is observed, as shown in Figures~\ref{fig:umx_trainvalid_curves} and~\ref{fig:htdemucs_trainvalid_curves} in Appendix~\ref{sec:app_mss_ds}. Injecting the \mfm features into both UMX and HTDemucs greatly boosts the separation performance for both models, even on the unseen dataset MDXDB21. It also improves the separation performance of HTDemucs when training on data corrupted by bleeding errors (\emph{SDXDB23\_Bleeding})~\cite{fabbro2023sound}.
However, the \musicgen features do not always improve the results. Injecting the \musicgens features improved UMX, but not for the other cases. 
According to the ablation study results, injecting short-term Fourier transform (STFT) signal, CLAP features, \musicgen features, \textcolor{c_revise}{or Jukebox} features leads to unstable training. However, no such behavior is observed when injecting the \mfm features into HTDemucs. We assume that performance can be improved if instability during training is avoided. As mentioned in Section~\ref{sssec:exp_amt}, interpreting information contained in \musicgenl could cost too much capacity of the downstream model and result in performance degradation. 

In summary, we observed that injecting the \mfm features into separation models not only yields faster training and better performance but also improves the robustness to dataset corruption. 

\begin{table}[t]
\centering
\caption{Music source separation. Evaluation results on MUSDB18 and MDXDB21.}
\renewcommand{\arraystretch}{1.1}
\resizebox{\textwidth}{!}{
\begin{tabular}{@{}lccccclccccc@{}}
\toprule
\multirow{2}{*}{\textbf{Model}}  & \multicolumn{5}{c}{\textbf{MUSDB18 (BSSEval v4 SDR} (dB)\textbf{)}} & \phantom{a} & \multicolumn{5}{c}{\textbf{MDXDB21 (global SDR} (dB)\textbf{)}} \\
\cline{2-6}\cline{8-12}
& Bass & Drums & Other & Vocals & Average & & Bass & Drums & Other & Vocals & Average \\
\midrule
Open-Unmix (UMX) & 4.01 & 4.35 & 2.79 & 5.66 & 4.20 & 
& 4.50 & \textbf{4.46} & 2.66 & 5.55 & 4.29 
\\
UMX + MusicGen Small & 4.25 & \textbf{4.55}  & \textbf{3.18}  & 5.66  & 
 \textbf{4.41}  & 
& 4.61  & 4.29 & 2.92 & 5.42  & 4.31
\\
UMX + MusicGen Large & 3.97 & 4.25  & 3.13  & 5.28  & 
 4.16  & 
& 4.55  & 4.04 & \textbf{2.95} & 5.35  & 4.22
\\
UMX + \mfm & \textbf{4.37} & 4.16 & 3.00 & \textbf{5.91} & 
 4.36 & 
& \textbf{4.71} & 4.43 & 2.64 & \textbf{5.69} & \textbf{4.37}
\\
\midrule
HTDemucs (default) & 8.94 & 8.22 & 5.55 & 7.56 & 7.57 & & 7.86 & 7.89 & 5.09 & 7.70 & 7.13 \\
HTDemucs (ablation 1) & 8.81 & 8.20 & 5.70 & 7.69 & 7.60 & & 7.94 & 7.97 & 5.16 & 7.91 & 7.24 \\
HTDemucs (ablation 2) & 8.75 & 8.64 & 5.78 & 7.85 & 7.76 & & 7.96 & 7.69 & 5.12 & 7.89 & 7.17 \\
HTDemucs + STFT-2048 & 5.65 & 6.22 & 4.45 & 6.56 & 5.72 & & 5.84 & 6.13 & 4.40 & 6.85 & 5.80 \\
HTDemucs + STFT-4096 & 6.44 & 6.25 & 4.29 & 6.28 & 5.81 & & 6.18 & 6.19 & 4.43 & 6.83 & 5.91 \\
HTDemucs + CLAP & 8.25 & 7.37 & 5.21 & 7.21 & 7.01 & & 7.37 & 7.51 & 4.82 & 7.47 & 6.79 \\
HTDemucs + MusicGen Small & 8.86 & 8.03 & 5.59 & 7.57 & 7.51 & & 7.44 & \textbf{8.31} & \textbf{5.26} & 7.81 & 7.21 \\
HTDemucs + MusicGen Large & 8.17 & 7.50 & 5.54 & 7.66 & 7.22 & & 7.40 & 7.37 & 4.93 & 7.73 & 6.86 \\  
\textcolor{c_revise}{HTDemucs + Jukebox} & 7.12 & 6.65 & 4.77 & 6.84 & 6.35 & & 6.58 & 6.58 & 4.59 & 7.12 & 6.22 \\
\\
HTDemucs + \mfm & \textbf{9.50} & \textbf{8.65} & \textbf{5.91} & \textbf{8.07} & \textbf{8.03} & & \textbf{8.14} & 8.16 & 5.21 & \textbf{8.04} & \textbf{7.39} \\
\midrule
HTDemucs (trained on \emph{SDXDB23\_Bleeding}) & 3.86 & 5.52 & 3.53 & 5.70 & 4.65 & & 6.20 & 5.98 & 4.53 & 6.69 & 5.85 \\
HTDemucs + \mfm (trained on \emph{SDXDB23\_Bleeding}) & \textbf{5.50} & \textbf{6.06} & \textbf{3.97} & \textbf{5.82} & \textbf{5.43} & & \textbf{6.41} & \textbf{6.40} & \textbf{4.64} & \textbf{7.19} & \textbf{6.16} \\
\bottomrule
\end{tabular}}
\label{tab:mss_result}
\end{table}

\subsubsection{Music Mixing: Mix-Wave-U-Net, CRAFx2}
\label{sssec:exp_mix}

Mix-Wave-U-Net~\citep{steinmetz2022automix} along with a modified CRAFx~\citep{martinez2020deep}, henceforth referred to as CRAFx2, are used as the baselines. 
The input to both networks is the stereo stems pre-processed by Fx-normalization~\citep{martinez2022automatic}, and the output is the stereo mixture. 
These models do not handle high-level information relevant to mixing, such as genre, instrumentation, or mood. Conditioning these models with extracted features, which implicitly contain such information, is expected to improve mixing performance. The features are computed from the monaural downmix of the mixture, which corresponds to the summation of the Fx-normalized input stems. To incorporate these features, we condition both networks using Feature-wise Linear Modulation (FiLM) layers~\citep{perez2018film}. For Mix-Wave-U-Net, we inject features into the up-sampling and bottleneck one-dimensional (1D) convolutional blocks. For CRAFx2, we use FiLM layers to condition both the latent-space mixer and synthesis back-end (see Appendix~\ref{appendix:mixing_waveunet}).

We train all models on MUSDB18, and the stereo-invariant loss, along with all training hyperparameters, remains the same, as in a previous study~\citep{martinez2022automatic}. 
Due to the inherent subjectivity of the task, identifying the best model is challenging. Thus, as shown in Table~\ref{tab:summary_of_mdt}, an objective evaluation is conducted by measuring the proximity between the output mixes and target mixes on the same test sets as in~\citet{martinez2022automatic}. The proximity measurement is based on objective metrics~\citep{steinmetz2022automix, martinez2022automatic}. These metrics consist of spectral, panning, dynamic, and loudness low-level audio features, which are the key audio characteristics often manipulated during the mixing process. Further details and experiments are provided in Appendices~\ref{appendix:mixing_experiments} and~\ref{appendix:mixing_results}, respectively.

As shown in Table~\ref{tab:mixing}, conditioning both architectures with the \mfm features improved objective performance. The training and validation curves in Figure~\ref{fig:mixing_trainvalid_curves} of Appendix~\ref{appendix:mixing_results} show a faster loss decrease in early epochs and better generalization, respectively. Although there is no standardized objective evaluation due to the subjective nature of the task~\citep{steinmetz2022automix}, the presented metrics suggest that the best-performing model closely aligns with the target mixes, resembling professional human-made mixes.

Using the \mfm features leads to the best performance. \textcolor{c_revise}{The Jukebox and \musicgen features provide improvements but not as effective as those of \mfm, particularly for CRAFx2, where the default model outperforms both Jukebox and \musicgen in various metrics}. For \musicgen, we assume this performance gap may be attributed to its training on 32~kHz audio compared with the 44.1~kHz used for \mfm, which limit its effectiveness for full-band tasks, such as music mixing, that require higher sampling rates.
There is no data-driven approach that used task-agnostic features of the input stems for music mixing improvement. Thus, we can conclude that incorporating the \mfm features benefits both the training and performance of automatic music mixing models. This aligns with recent design studies~\citep{lefford2021context, vanka2023adoption}, advocating for the incorporation of contextual inputs.

\begin{table}[t]
\centering
\caption{Music mixing. Evaluation results on the MDXDB21-dry and MUSDB18 test sets include mean absolute percentage error for audio effect-related features, their average, and stereo-invariant loss. More details are provided in Table~\ref{tab:features_mixing_combined}.}
\renewcommand{\arraystretch}{1.2}
\resizebox{\textwidth}{!}{
\begin{tabular}{@{}p{5.6125cm}ccccc|cccccc|c@{}}
\toprule
\multirow{2}{*}{\textbf{Model}}  & \multicolumn{6}{c}{\textbf{MDXDB21-dry test set}} & \multicolumn{6}{c}{\textbf{MUSDB18 test set}} \\
\cmidrule(r){2-7}\cmidrule(r){8-13}

& \multirow{2}{*}{Spectral} & \multirow{2}{*}{Panning} & \multirow{2}{*}{Dynamic} & \multirow{2}{*}{Loudness} & \multirow{2}{*}{Average} & Stereo  &  \multirow{2}{*}{Spectral} & \multirow{2}{*}{Panning} & \multirow{2}{*}{Dynamic} & \multirow{2}{*}{Loudness} & \multirow{2}{*}{Avg} & Stereo  \\

&  &  &  &  &  & Invariant  &   &  & & & & Invariant\\
\midrule
Mix-Wave-U-Net (default)    & 0.234 & 0.215 & 0.073 & 0.168 & 0.173 &  89.631 &  0.201 & 0.164 & 0.085 & 0.167 & 0.154 & 34.253 \\

\textcolor{c_revise}{Mix-Wave-U-Net + Jukebox} & \textcolor{c_revise}{0.240} & \textcolor{c_revise}{0.231} & \textcolor{c_revise}{0.075} & \textcolor{c_revise}{0.154} & \textcolor{c_revise}{0.175} & \textcolor{c_revise}{83.717} & \textcolor{c_revise}{0.206} & \textcolor{c_revise}{0.187} & \textcolor{c_revise}{0.082} & \textcolor{c_revise}{\textbf{0.157}} & \textcolor{c_revise}{0.158} & \textcolor{c_revise}{32.573} \\

Mix-Wave-U-Net + MusicGen Small       & 0.240 & 0.197 & \textbf{0.064} & 0.147 & 0.162 &  80.161 &  0.214 & \textbf{0.158} & 0.079 & 0.163 & 0.153 & 32.151\\

Mix-Wave-U-Net + MusicGen Large       & 0.241 & 0.231 & 0.066 & 0.145 & 0.171 &  81.161 &  0.205 & 0.192 & 0.075 & 0.167 & 0.160 & 32.649\\

Mix-Wave-U-Net + \mfm       & \textbf{0.226} & \textbf{0.180} & 0.067 & \textbf{0.131} & \textbf{0.151} &  \textbf{78.180} &  \textbf{0.186} & 0.175 & \textbf{0.063} & 0.179 & \textbf{0.151} & \textbf{30.116}\\
\\

CRAFx2 (default)        & \textbf{0.193} & 0.179 & 0.070 & 0.152 & \textbf{0.148} & 82.095 &  0.193 & \textbf{0.154} & 0.081 & \textbf{0.165} & 0.148 & 32.856\\

\textcolor{c_revise}{CRAFx2 + Jukebox} & \textcolor{c_revise}{0.231} & \textcolor{c_revise}{0.249} & \textcolor{c_revise}{0.075} & \textcolor{c_revise}{0.144} & \textcolor{c_revise}{0.175} & \textcolor{c_revise}{87.973} & \textcolor{c_revise}{0.216} & \textcolor{c_revise}{0.220} & \textcolor{c_revise}{0.082} & \textcolor{c_revise}{\textbf{0.165}} & \textcolor{c_revise}{0.171} & \textcolor{c_revise}{36.172} \\

CRAFX2 + MusicGen Small          & 0.229 & 0.244 & 0.072 & 0.148 & 0.173 & 87.273 & 0.211 & 0.204 & 0.083 & 0.178 & 0.169 & 36.418\\

CRAFX2 + MusicGen Large          & 0.228 & 0.219 & 0.073 & \textbf{0.132} & 0.163 & 87.318 & 0.224 & 0.206 & 0.080 & 0.175 & 0.171 & 36.519\\

CRAFX2 + \mfm           & 0.221 & \textbf{0.175} & \textbf{0.064} & 0.171 & 0.158 & \textbf{79.861} &  \textbf{0.187} & \textbf{0.154} & \textbf{0.076} & 0.169 & \textbf{0.146} & \textbf{30.155}\\
\bottomrule
\end{tabular}}

\label{tab:mixing}
\end{table}

\section{Ethical Concerns}
\label{sec:ai_ethics}
To train \mfm, we acquired an internal dataset of library music with licensing explicitly allowing machine learning training.
The dataset is mostly non-vocal, biased toward orchestral and western music. A model trained on this dataset is unlikely to characterize equally well for all types of music. The learnt intermediate embedding may reflect the bias.
When using such a biased music foundation model as a performance booster, thorough verification is required before using such a model for practical use or the decision process.

\section{Conclusions}
\label{sec:conclusions}
We extended the use of music foundation models from MIR to generic music downstream tasks. The task-agnostic intermediate representation extracted using proposed \mfm model has been applied to task-specific models of music tagging, music transcription, music source separation, and music mixing. On the basis of the evaluation results, performance improvement is observed for all selected music downstream tasks. This suggests that incorporating the intermediate features extracted from a pre-trained auto-regressive music foundation model should be considered as a generic booster in future development of task-specific models. This is especially helpful when it is difficult to acquire a sufficient dataset or the computation resource does not allow large-scale training. A study on the bias propagation of a pre-trained music foundation model to a downstream task model should be conducted in another future work.

\section*{Acknowledgement}

We would like to thank Takashi Shibuya, Naoya Takahashi, Marc Ferras, and Masato Ishii for many helpful comments during the preparation of this manuscript. Besides, we thank anonymous reviewers for their valuable suggestions and comments.

\clearpage

\bibliography{str_def_abrv,refs_ml,refs_dgm, refs_music_amt, refs_music_ass, refs_music_dgm, refs_music_mir, refs_music_mixing}

\begin{thebibliography}{104}
\providecommand{\natexlab}[1]{#1}
\providecommand{\url}[1]{\texttt{#1}}
\expandafter\ifx\csname urlstyle\endcsname\relax
  \providecommand{\doi}[1]{doi: #1}\else
  \providecommand{\doi}{doi: \begingroup \urlstyle{rm}\Url}\fi

\bibitem[Agostinelli et~al.(2023)Agostinelli, Denk, Borsos, Engel, Verzetti,
  Caillon, Huang, Jansen, Roberts, Tagliasacchi, Sharifi, Zeghidour, and
  Frank]{agostinelli2023musiclm}
Andrea Agostinelli, Timo~I. Denk, Zalán Borsos, Jesse Engel, Mauro Verzetti,
  Antoine Caillon, Qingqing Huang, Aren Jansen, Adam Roberts, Marco
  Tagliasacchi, Matt Sharifi, Neil Zeghidour, and Christian Frank.
\newblock Musiclm: Generating music from text, 2023.

\bibitem[Baevski et~al.(2022)Baevski, Hsu, Xu, Babu, Gu, and
  Auli]{baevski2022data2vec}
Alexei Baevski, Wei-Ning Hsu, Qiantong Xu, Arun Babu, Jiatao Gu, and Michael
  Auli.
\newblock Data2vec: A general framework for self-supervised learning in speech,
  vision and language.
\newblock In \emph{International Conference on Machine Learning}, pp.\
  1298--1312. PMLR, 2022.

\bibitem[Bommasani et~al.(2021)Bommasani, Hudson, Adeli, Altman, Arora, von
  Arx, Bernstein, Bohg, Bosselut, Brunskill, Brynjolfsson, Buch, Card,
  Castellon, Chatterji, Chen, Creel, Davis, Demszky, Donahue, Doumbouya,
  Durmus, Ermon, Etchemendy, Ethayarajh, Fei-Fei, Finn, Gale, Gillespie, Goel,
  Goodman, Grossman, Guha, Hashimoto, Henderson, Hewitt, Ho, Hong, Hsu, Huang,
  Icard, Jain, Jurafsky, Kalluri, Karamcheti, Keeling, Khani, Khattab, Koh,
  Krass, Krishna, Kuditipudi, Kumar, Ladhak, Lee, Lee, Leskovec, Levent, Li,
  Li, Ma, Malik, Manning, Mirchandani, Mitchell, Munyikwa, Nair, Narayan,
  Narayanan, Newman, Nie, Niebles, Nilforoshan, Nyarko, Ogut, Orr,
  Papadimitriou, Park, Piech, Portelance, Potts, Raghunathan, Reich, Ren, Rong,
  Roohani, Ruiz, Ryan, R'e, Sadigh, Sagawa, Santhanam, Shih, Srinivasan,
  Tamkin, Taori, Thomas, Tram{\`e}r, Wang, Wang, Wu, Wu, Wu, Xie, Yasunaga,
  You, Zaharia, Zhang, Zhang, Zhang, Zhang, Zheng, Zhou, and
  Liang]{Bommasani2021FoundationModels}
Rishi Bommasani, Drew~A. Hudson, Ehsan Adeli, Russ Altman, Simran Arora, Sydney
  von Arx, Michael~S. Bernstein, Jeannette Bohg, Antoine Bosselut, Emma
  Brunskill, Erik Brynjolfsson, S.~Buch, Dallas Card, Rodrigo Castellon,
  Niladri~S. Chatterji, Annie~S. Chen, Kathleen~A. Creel, Jared Davis, Dora
  Demszky, Chris Donahue, Moussa Doumbouya, Esin Durmus, Stefano Ermon, John
  Etchemendy, Kawin Ethayarajh, Li~Fei-Fei, Chelsea Finn, Trevor Gale,
  Lauren~E. Gillespie, Karan Goel, Noah~D. Goodman, Shelby Grossman, Neel Guha,
  Tatsunori Hashimoto, Peter Henderson, John Hewitt, Daniel~E. Ho, Jenny Hong,
  Kyle Hsu, Jing Huang, Thomas~F. Icard, Saahil Jain, Dan Jurafsky, Pratyusha
  Kalluri, Siddharth Karamcheti, Geoff Keeling, Fereshte Khani, O.~Khattab,
  Pang~Wei Koh, Mark~S. Krass, Ranjay Krishna, Rohith Kuditipudi, Ananya Kumar,
  Faisal Ladhak, Mina Lee, Tony Lee, Jure Leskovec, Isabelle Levent, Xiang~Lisa
  Li, Xuechen Li, Tengyu Ma, Ali Malik, Christopher~D. Manning, Suvir~P.
  Mirchandani, Eric Mitchell, Zanele Munyikwa, Suraj Nair, Avanika Narayan,
  Deepak Narayanan, Benjamin Newman, Allen Nie, Juan~Carlos Niebles, Hamed
  Nilforoshan, J.~F. Nyarko, Giray Ogut, Laurel Orr, Isabel Papadimitriou,
  Joon~Sung Park, Chris Piech, Eva Portelance, Christopher Potts, Aditi
  Raghunathan, Robert Reich, Hongyu Ren, Frieda Rong, Yusuf~H. Roohani, Camilo
  Ruiz, Jack Ryan, Christopher R'e, Dorsa Sadigh, Shiori Sagawa, Keshav
  Santhanam, Andy Shih, Krishna~Parasuram Srinivasan, Alex Tamkin, Rohan Taori,
  Armin~W. Thomas, Florian Tram{\`e}r, Rose~E. Wang, William Wang, Bohan Wu,
  Jiajun Wu, Yuhuai Wu, Sang~Michael Xie, Michihiro Yasunaga, Jiaxuan You,
  Matei~A. Zaharia, Michael Zhang, Tianyi Zhang, Xikun Zhang, Yuhui Zhang,
  Lucia Zheng, Kaitlyn Zhou, and Percy Liang.
\newblock On the opportunities and risks of foundation models.
\newblock \emph{ArXiv}, 2021.
\newblock URL \url{https://crfm.stanford.edu/assets/report.pdf}.

\bibitem[Brown et~al.(2020)Brown, Mann, Ryder, Subbiah, Kaplan, Dhariwal,
  Neelakantan, Shyam, Sastry, Askell, Agarwal, Herbert{-}Voss, Krueger,
  Henighan, Child, Ramesh, Ziegler, Wu, Winter, Hesse, Chen, Sigler, Litwin,
  Gray, Chess, Clark, Berner, McCandlish, Radford, Sutskever, and
  Amodei]{Brown2020gpt3}
Tom~B. Brown, Benjamin Mann, Nick Ryder, Melanie Subbiah, Jared Kaplan,
  Prafulla Dhariwal, Arvind Neelakantan, Pranav Shyam, Girish Sastry, Amanda
  Askell, Sandhini Agarwal, Ariel Herbert{-}Voss, Gretchen Krueger, Tom
  Henighan, Rewon Child, Aditya Ramesh, Daniel~M. Ziegler, Jeffrey Wu, Clemens
  Winter, Christopher Hesse, Mark Chen, Eric Sigler, Mateusz Litwin, Scott
  Gray, Benjamin Chess, Jack Clark, Christopher Berner, Sam McCandlish, Alec
  Radford, Ilya Sutskever, and Dario Amodei.
\newblock Language models are few-shot learners.
\newblock \emph{CoRR}, abs/2005.14165, 2020.
\newblock URL \url{https://arxiv.org/abs/2005.14165}.

\bibitem[Castellon et~al.(2021)Castellon, Donahue, and
  Liang]{CastellonDL21calm}
Rodrigo Castellon, Chris Donahue, and Percy Liang.
\newblock Codified audio language modeling learns useful representations for
  music information retrieval.
\newblock In Jin~Ha Lee, Alexander Lerch, Zhiyao Duan, Juhan Nam, Preeti Rao,
  Peter van Kranenburg, and Ajay Srinivasamurthy (eds.), \emph{Proceedings of
  the 22nd International Society for Music Information Retrieval Conference,
  {ISMIR} 2021, Online, November 7-12, 2021}, pp.\  88--96, 2021.
\newblock URL \url{https://archives.ismir.net/ismir2021/paper/000010.pdf}.

\bibitem[Chen et~al.(2023)Chen, Wichern, Germain, and Roux]{chen2023pac}
Ke~Chen, Gordon Wichern, Fran{\c{c}}ois~G Germain, and Jonathan~Le Roux.
\newblock Pac-hubert: Self-supervised music source separation via primitive
  auditory clustering and hidden-unit bert.
\newblock \emph{arXiv preprint arXiv:2304.02160}, 2023.

\bibitem[Chen et~al.(2020)Chen, Radford, Child, Wu, Jun, Luan, and
  Sutskever]{openai2020imagegpt}
Mark Chen, Alec Radford, Rewon Child, Jeffrey Wu, Heewoo Jun, David Luan, and
  Ilya Sutskever.
\newblock Generative pretraining from pixels.
\newblock In \emph{International conference on machine learning}, pp.\
  1691--1703. PMLR, 2020.

\bibitem[Cheuk et~al.(2021{\natexlab{a}})Cheuk, Herremans, and
  Su]{cheuk2021reconvat}
Kin~Wai Cheuk, Dorien Herremans, and Li~Su.
\newblock Reconvat: A semi-supervised automatic music transcription framework
  for low-resource real-world data.
\newblock In \emph{Proceedings of the 29th ACM International Conference on
  Multimedia}, pp.\  3918--3926, 2021{\natexlab{a}}.

\bibitem[Cheuk et~al.(2021{\natexlab{b}})Cheuk, Luo, Benetos, and
  Herremans]{Cheuk_IJCNN2021}
Kin~Wai Cheuk, Yin-Jyun Luo, Emmanouil Benetos, and Dorien Herremans.
\newblock Revisiting the onsets and frames model with additive attention.
\newblock In \emph{Proceedings of the International Joint Conference on Neural
  Networks}, pp.\  In press. IEEE, 2021{\natexlab{b}}.
\newblock \doi{10.1109/SPW.2018.00014}.

\bibitem[Child et~al.(2019)Child, Gray, Radford, and
  Sutskever]{child2019generating}
Rewon Child, Scott Gray, Alec Radford, and Ilya Sutskever.
\newblock Generating long sequences with sparse transformers.
\newblock \emph{CoRR}, abs/1904.10509, 2019.
\newblock URL \url{http://arxiv.org/abs/1904.10509}.

\bibitem[Colonel \& Reiss(2021)Colonel and Reiss]{colonel2021reverse}
Joseph~T Colonel and Joshua Reiss.
\newblock Reverse engineering of a recording mix with differentiable digital
  signal processing.
\newblock \emph{The Journal of the Acoustical Society of America}, 150\penalty0
  (1):\penalty0 608--619, 2021.

\bibitem[Comunit{\`a} et~al.(2024)Comunit{\`a}, Zhong, Takahashi, Yang, Zhao,
  Saito, Ikemiya, Shibuya, Takahashi, and Mitsufuji]{comunita2024specmaskgit}
Marco Comunit{\`a}, Zhi Zhong, Akira Takahashi, Shiqi Yang, Mengjie Zhao,
  Koichi Saito, Yukara Ikemiya, Takashi Shibuya, Shusuke Takahashi, and Yuki
  Mitsufuji.
\newblock Specmaskgit: Masked generative modeling of audio spectrograms for
  efficient audio synthesis and beyond.
\newblock \emph{arXiv preprint arXiv:2406.17672}, 2024.

\bibitem[Copet et~al.(2023)Copet, Kreuk, Gat, Remez, Kant, Synnaeve, Adi, and
  Défossez]{copet2023musicgen}
Jade Copet, Felix Kreuk, Itai Gat, Tal Remez, David Kant, Gabriel Synnaeve,
  Yossi Adi, and Alexandre Défossez.
\newblock Simple and controllable music generation, 2023.

\bibitem[D{\'e}fossez et~al.(2023)D{\'e}fossez, Copet, Synnaeve, and
  Adi]{defossez2023encodec}
Alexandre D{\'e}fossez, Jade Copet, Gabriel Synnaeve, and Yossi Adi.
\newblock High fidelity neural audio compression.
\newblock \emph{Transactions on Machine Learning Research}, 2023.
\newblock ISSN 2835-8856.
\newblock URL \url{https://openreview.net/forum?id=ivCd8z8zR2}.
\newblock Featured Certification, Reproducibility Certification.

\bibitem[Devlin et~al.(2018)Devlin, Chang, Lee, and Toutanova]{Delvin2018bert}
Jacob Devlin, Ming{-}Wei Chang, Kenton Lee, and Kristina Toutanova.
\newblock {BERT:} pre-training of deep bidirectional transformers for language
  understanding.
\newblock \emph{CoRR}, abs/1810.04805, 2018.
\newblock URL \url{http://arxiv.org/abs/1810.04805}.

\bibitem[Dhariwal et~al.(2020)Dhariwal, Jun, Payne, Kim, Radford, and
  Sutskever]{dhariwal2020jukebox}
Prafulla Dhariwal, Heewoo Jun, Christine Payne, Jong~Wook Kim, Alec Radford,
  and Ilya Sutskever.
\newblock Jukebox: A generative model for music, 2020.

\bibitem[Donahue et~al.(2022)Donahue, Thickstun, and Liang]{donahue2022}
Chris Donahue, John Thickstun, and Percy Liang.
\newblock Melody transcription via generative pre-training.
\newblock In \emph{Proceedings of the 23rd International Society for Music
  Information Retrieval Conference}, 2022.

\bibitem[Duan et~al.(2010)Duan, Pardo, and Zhang]{Bach10}
Zhiyao Duan, Bryan Pardo, and Changshui Zhang.
\newblock Multiple fundamental frequency estimation by modeling spectral peaks
  and non-peak regions.
\newblock \emph{IEEE Transactions on Audio, Speech, and Language Processing},
  18\penalty0 (8):\penalty0 2121--2133, 2010.
\newblock \doi{10.1109/TASL.2010.2042119}.

\bibitem[Emiya et~al.(2010)Emiya, Badeau, and David]{emiya2010}
Valentin Emiya, Roland Badeau, and Bertrand David.
\newblock Multipitch estimation of piano sounds using a new probabilistic
  spectral smoothness principle.
\newblock \emph{IEEE Transactions on Audio, Speech, and Language Processing},
  18\penalty0 (6):\penalty0 1643--1654, 2010.

\bibitem[Engel et~al.(2017)Engel, Resnick, Roberts, Dieleman, Eck, Simonyan,
  and Norouzi]{nsynth2017}
Jesse Engel, Cinjon Resnick, Adam Roberts, Sander Dieleman, Douglas Eck, Karen
  Simonyan, and Mohammad Norouzi.
\newblock Neural audio synthesis of musical notes with wavenet autoencoders,
  2017.

\bibitem[Fabbro et~al.(2023)Fabbro, Uhlich, Lai, Choi,
  Mart{\'\i}nez-Ram{\'\i}rez, Liao, Gadelha, Ramos, Hsu, Rodrigues,
  et~al.]{fabbro2023sound}
Giorgio Fabbro, Stefan Uhlich, Chieh-Hsin Lai, Woosung Choi, Marco
  Mart{\'\i}nez-Ram{\'\i}rez, Weihsiang Liao, Igor Gadelha, Geraldo Ramos,
  Eddie Hsu, Hugo Rodrigues, et~al.
\newblock The {Sound Demixing Challenge} 2023 -- music demixing track.
\newblock \emph{arXiv preprint arXiv:2308.06979}, 2023.

\bibitem[Forsgren \& Martiros(2022)Forsgren and
  Martiros]{Forsgren_Martiros_2022riffusion}
Seth* Forsgren and Hayk* Martiros.
\newblock {Riffusion - Stable diffusion for real-time music generation}.
\newblock 2022.
\newblock URL \url{https://riffusion.com/about}.

\bibitem[Fritsch(2012)]{TRIOS2014}
Joachim Fritsch.
\newblock High quality musical audio source separation.
\newblock \emph{Master’s thesis}, 2012.

\bibitem[Gardner et~al.(2022)Gardner, Simon, Manilow, Hawthorne, and
  Engel]{gardner2022}
Josh Gardner, Ian Simon, Ethan Manilow, Curtis Hawthorne, and Jesse Engel.
\newblock Mt3: Multi-task multitrack music transcription.
\newblock In \emph{Proceedings of the 10th International Conference on Learning
  Representations}, 2022.

\bibitem[Gururani et~al.(2019)Gururani, Sharma, and
  Lerch]{gururani2019attention}
Siddharth Gururani, Mohit Sharma, and Alexander Lerch.
\newblock An attention mechanism for musical instrument recognition.
\newblock In \emph{ISMIR 2019}, 2019.

\bibitem[Hawthorne et~al.(2018)Hawthorne, Elsen, Song, Roberts, Simon, Raffel,
  Engel, Oore, and Eck]{hawthorne2018}
Curtis Hawthorne, Erich Elsen, Jialin Song, Adam Roberts, Ian Simon, Colin
  Raffel, Jesse Engel, Sageev Oore, and Douglas Eck.
\newblock Onsets and frames: Dual-objective piano transcription.
\newblock In \emph{Proceedings of the 19th International Society for Music
  Information Retrieval Conference, ISMIR 2018, Paris, France, 2018}, 2018.
\newblock URL \url{https://arxiv.org/abs/1710.11153}.

\bibitem[Hsu et~al.(2021)Hsu, Bolte, Tsai, Lakhotia, Salakhutdinov, and
  Mohamed]{hsu2021hubert}
Wei-Ning Hsu, Benjamin Bolte, Yao-Hung~Hubert Tsai, Kushal Lakhotia, Ruslan
  Salakhutdinov, and Abdelrahman Mohamed.
\newblock Hubert: Self-supervised speech representation learning by masked
  prediction of hidden units.
\newblock \emph{IEEE/ACM TASLP}, 29:\penalty0 3451--3460, 2021.

\bibitem[Hu et~al.(2018)Hu, Shen, and Sun]{hu2018squeeze}
Jie Hu, Li~Shen, and Gang Sun.
\newblock Squeeze-and-excitation networks.
\newblock In \emph{Proceedings of the IEEE conference on computer vision and
  pattern recognition}, 2018.

\bibitem[Huang et~al.(2022{\natexlab{a}})Huang, Jansen, Lee, Ganti, Li, and
  Ellis]{HuangJLGLE22mulan}
Qingqing Huang, Aren Jansen, Joonseok Lee, Ravi Ganti, Judith~Yue Li, and
  Daniel P.~W. Ellis.
\newblock Mulan: {A} joint embedding of music audio and natural language.
\newblock In Preeti Rao, Hema~A. Murthy, Ajay Srinivasamurthy, Rachel~M.
  Bittner, Rafael~Caro Repetto, Masataka Goto, Xavier Serra, and Marius Miron
  (eds.), \emph{Proceedings of the 23rd International Society for Music
  Information Retrieval Conference, {ISMIR} 2022, Bengaluru, India, December
  4-8, 2022}, pp.\  559--566, 2022{\natexlab{a}}.
\newblock URL \url{https://archives.ismir.net/ismir2022/paper/000067.pdf}.

\bibitem[Huang et~al.(2023)Huang, Park, Wang, Denk, Ly, Chen, Zhang, Zhang, Yu,
  Frank, Engel, Le, Chan, Chen, and Han]{huang2023noise2music}
Qingqing Huang, Daniel~S. Park, Tao Wang, Timo~I. Denk, Andy Ly, Nanxin Chen,
  Zhengdong Zhang, Zhishuai Zhang, Jiahui Yu, Christian Frank, Jesse Engel,
  Quoc~V. Le, William Chan, Zhifeng Chen, and Wei Han.
\newblock Noise2music: Text-conditioned music generation with diffusion models,
  2023.

\bibitem[Huang et~al.(2022{\natexlab{b}})Huang, Watanabe, Yang, García, and
  Khudanpur]{huang2022investigating}
Zili Huang, Shinji Watanabe, Shu-wen Yang, Paola García, and Sanjeev
  Khudanpur.
\newblock Investigating self-supervised learning for speech enhancement and
  separation.
\newblock In \emph{ICASSP 2022 - 2022 IEEE International Conference on
  Acoustics, Speech and Signal Processing (ICASSP)}, pp.\  6837--6841,
  2022{\natexlab{b}}.
\newblock \doi{10.1109/ICASSP43922.2022.9746303}.

\bibitem[Hung et~al.(2022)Hung, wei Fu, Tseng, Chiang, Tsao, and
  Lin]{hung22_interspeech}
Kuo-Hsuan Hung, Szu wei Fu, Huan-Hsin Tseng, Hsin-Tien Chiang, Yu~Tsao, and
  Chii-Wann Lin.
\newblock {Boosting Self-Supervised Embeddings for Speech Enhancement}.
\newblock In \emph{Proc. Interspeech 2022}, pp.\  186--190, 2022.
\newblock \doi{10.21437/Interspeech.2022-10002}.

\bibitem[ITU-R(2011)]{itu2011itu}
Rec ITU-R.
\newblock Itu-r bs. 1770-2, algorithms to measure audio programme loudness and
  true-peak audio level.
\newblock \emph{International Telecommunications Union, Geneva}, 2011.

\bibitem[Kelz et~al.(2019)Kelz, B{\"o}ck, and Widmer]{kelz2019deep}
Rainer Kelz, Sebastian B{\"o}ck, and Gerhard Widmer.
\newblock Deep polyphonic adsr piano note transcription.
\newblock In \emph{ICASSP 2019-2019 IEEE International Conference on Acoustics,
  Speech and Signal Processing (ICASSP)}, pp.\  246--250. IEEE, 2019.

\bibitem[Kilgour et~al.(2019)Kilgour, Zuluaga, Roblek, and
  Sharifi]{kilgour2019frechet}
Kevin Kilgour, Mauricio Zuluaga, Dominik Roblek, and Matthew Sharifi.
\newblock Fr\'echet audio distance: A metric for evaluating music enhancement
  algorithms, 2019.

\bibitem[Kim \& Bello(2019)Kim and Bello]{kim2019adversarial}
Jong~Wook Kim and Juan~Pablo Bello.
\newblock Adversarial learning for improved onsets and frames music
  transcription.
\newblock \emph{International Society forMusic Information Retrieval
  Conference}, pp.\  670--677, 2019.

\bibitem[Kingma \& Ba(2015)Kingma and Ba]{kingma2014adam}
Diederik~P Kingma and Jimmy Ba.
\newblock Adam: A method for stochastic optimization.
\newblock In \emph{Proc. International Conference on Learning Representation
  (ICLR)}, 2015.

\bibitem[Knees et~al.(2015)Knees, Faraldo~P{\'e}rez, Boyer, Vogl, B{\"o}ck,
  H{\"o}rschl{\"a}ger, Le~Goff, et~al.]{knees2015giantsteps}
Peter Knees, {\'A}ngel Faraldo~P{\'e}rez, Herrera Boyer, Richard Vogl,
  Sebastian B{\"o}ck, Florian H{\"o}rschl{\"a}ger, Mickael Le~Goff, et~al.
\newblock Two data sets for tempo estimation and key detection in electronic
  dance music annotated from user corrections.
\newblock In \emph{Proceedings of the 16th International Society for Music
  Information Retrieval Conference (ISMIR); 2015 Oct 26-30; M{\'a}laga,
  Spain.[M{\'a}laga]: International Society for Music Information Retrieval,
  2015. p. 364-70.} International Society for Music Information Retrieval
  (ISMIR), 2015.

\bibitem[Kong et~al.(2021)Kong, Li, Song, Wan, and Wang]{kong2021high}
Qiuqiang Kong, Bochen Li, Xuchen Song, Yuan Wan, and Yuxuan Wang.
\newblock High-resolution piano transcription with pedals by regressing onset
  and offset times.
\newblock \emph{IEEE/ACM Transactions on Audio, Speech, and Language
  Processing}, 29:\penalty0 3707--3717, 2021.

\bibitem[Koo et~al.(2023)Koo, Mart{\'\i}nez-Ram{\'\i}rez, Liao, Uhlich, Lee,
  and Mitsufuji]{koo2022music}
Junghyun Koo, Marco~A Mart{\'\i}nez-Ram{\'\i}rez, Wei-Hsiang Liao, Stefan
  Uhlich, Kyogu Lee, and Yuki Mitsufuji.
\newblock Music mixing style transfer: A contrastive learning approach to
  disentangle audio effects.
\newblock 2023.

\bibitem[Korzeniowski \& Widmer(2017)Korzeniowski and
  Widmer]{widmer2017mtg-giantsteps}
Filip Korzeniowski and Gerhard Widmer.
\newblock End-to-end musical key estimation using a convolutional neural
  network.
\newblock In \emph{2017 25th European Signal Processing Conference (EUSIPCO)},
  pp.\  966--970. IEEE, 2017.

\bibitem[Koutini et~al.(2022)Koutini, Schl{\"u}ter, Eghbal-zadeh, and
  Widmer]{koutini2022patchout}
Khaled Koutini, Jan Schl{\"u}ter, Hamid Eghbal-zadeh, and Gerhard Widmer.
\newblock Efficient training of audio transformers with patchout.
\newblock In \emph{Proc. of Interspeech 2022}, 2022.

\bibitem[Lam et~al.(2023)Lam, Tian, Li, Yin, Feng, Tu, Ji, Xia, Ma, Song, Chen,
  Wang, and Wang]{lam2023efficient}
Max W.~Y. Lam, Qiao Tian, Tang Li, Zongyu Yin, Siyuan Feng, Ming Tu, Yuliang
  Ji, Rui Xia, Mingbo Ma, Xuchen Song, Jitong Chen, Yuping Wang, and Yuxuan
  Wang.
\newblock Efficient neural music generation.
\newblock In \emph{Thirty-seventh Conference on Neural Information Processing
  Systems}, 2023.
\newblock URL \url{https://openreview.net/forum?id=cxazQGSsQa}.

\bibitem[Law et~al.(2009)Law, West, Mandel, Bay, and Downie]{LawWMBD09mtat}
Edith Law, Kris West, Michael~I. Mandel, Mert Bay, and J.~Stephen Downie.
\newblock Evaluation of algorithms using games: The case of music tagging.
\newblock In Keiji Hirata, George Tzanetakis, and Kazuyoshi Yoshii (eds.),
  \emph{Proceedings of the 10th International Society for Music Information
  Retrieval Conference, {ISMIR} 2009, Kobe International Conference Center,
  Kobe, Japan, October 26-30, 2009}, pp.\  387--392. International Society for
  Music Information Retrieval, 2009.
\newblock URL \url{http://ismir2009.ismir.net/proceedings/OS5-5.pdf}.

\bibitem[Lee et~al.(2022)Lee, Kim, Kim, Cho, and Han]{lee2022autoregressive}
Doyup Lee, Chiheon Kim, Saehoon Kim, Minsu Cho, and Wook-Shin Han.
\newblock Autoregressive image generation using residual quantization.
\newblock In \emph{Proc. IEEE Conference on Computer Vision and Pattern
  Recognition (CVPR)}, pp.\  11523--11532, 2022.

\bibitem[Lefford et~al.(2021)Lefford, Bromham, Fazekas, and
  Moffat]{lefford2021context}
M~Nyssim Lefford, Gary Bromham, Gy{\"o}rgy Fazekas, and David Moffat.
\newblock Context aware intelligent mixing systems.
\newblock Journal of the Audio Engineering Society, 2021.

\bibitem[Li et~al.(2019)Li, Liu, Dinesh, Duan, and Sharma]{URMP2019}
Bochen Li, Xinzhao Liu, Karthik Dinesh, Zhiyao Duan, and Gaurav Sharma.
\newblock Creating a multitrack classical music performance dataset for
  multimodal music analysis: Challenges, insights, and applications.
\newblock \emph{IEEE Transactions on Multimedia}, 21\penalty0 (2):\penalty0
  522--535, 2019.
\newblock \doi{10.1109/TMM.2018.2856090}.

\bibitem[Li et~al.(2023{\natexlab{a}})Li, Chen, Yao, Wang, Wang, and
  Wang]{li2023jen1}
Peike Li, Boyu Chen, Yao Yao, Yikai Wang, Allen Wang, and Alex Wang.
\newblock Jen-1: Text-guided universal music generation with omnidirectional
  diffusion models, 2023{\natexlab{a}}.

\bibitem[Li et~al.(2023{\natexlab{b}})Li, Yuan, Zhang, Ma, Chen, Yin, Lin,
  Ragni, Benetos, Gyenge, Dannenberg, Liu, Chen, Xia, Shi, Huang, Guo, and
  Fu]{li2023mert}
Yizhi Li, Ruibin Yuan, Ge~Zhang, Yinghao Ma, Xingran Chen, Hanzhi Yin, Chenghua
  Lin, Anton Ragni, Emmanouil Benetos, Norbert Gyenge, Roger Dannenberg, Ruibo
  Liu, Wenhu Chen, Gus Xia, Yemin Shi, Wenhao Huang, Yike Guo, and Jie Fu.
\newblock Mert: Acoustic music understanding model with large-scale
  self-supervised training, 2023{\natexlab{b}}.

\bibitem[Liu et~al.(2023)Liu, Chen, Yuan, Mei, Liu, Mandic, Wang, and
  Plumbley]{liu2023audioldm}
Haohe Liu, Zehua Chen, Yi~Yuan, Xinhao Mei, Xubo Liu, Danilo Mandic, Wenwu
  Wang, and Mark~D. Plumbley.
\newblock Audioldm: Text-to-audio generation with latent diffusion models.
\newblock In \emph{Proceedings of the 40th International Conference on Machine
  Learning}, ICML'23. JMLR.org, 2023.

\bibitem[Loshchilov \& Hutter(2017)Loshchilov and Hutter]{adamw}
Ilya Loshchilov and Frank Hutter.
\newblock Decoupled weight decay regularization.
\newblock In \emph{International Conference on Learning Representations}, 2017.
\newblock URL \url{https://api.semanticscholar.org/CorpusID:53592270}.

\bibitem[Lu et~al.(2024)Lu, Wang, Kong, and Hung]{lu2024bsroformer}
Wei-Tsung Lu, Ju-Chiang Wang, Qiuqiang Kong, and Yun-Ning Hung.
\newblock Music source separation with band-split rope transformer.
\newblock In \emph{ICASSP 2024 - 2024 IEEE International Conference on
  Acoustics, Speech and Signal Processing (ICASSP)}, pp.\  481--485, 2024.
\newblock \doi{10.1109/ICASSP48485.2024.10446843}.

\bibitem[Ma et~al.(2024)Ma, {\O}land, Ragni, Del~Sette, Saitis, Donahue, Lin,
  Plachouras, Benetos, Shatri, et~al.]{ma2024foundation}
Yinghao Ma, Anders {\O}land, Anton Ragni, Bleiz~MacSen Del~Sette, Charalampos
  Saitis, Chris Donahue, Chenghua Lin, Christos Plachouras, Emmanouil Benetos,
  Elona Shatri, et~al.
\newblock Foundation models for music: A survey.
\newblock \emph{arXiv preprint arXiv:2408.14340}, 2024.

\bibitem[Ma et~al.(2015)Ma, De~Man, Pestana, Black, and
  Reiss]{ma2015intelligent}
Zheng Ma, Brecht De~Man, Pedro~DL Pestana, Dawn~AA Black, and Joshua~D Reiss.
\newblock Intelligent multitrack dynamic range compression.
\newblock \emph{Journal of the Audio Engineering Society}, 63\penalty0
  (6):\penalty0 412--426, 2015.

\bibitem[Mart{\'\i}nez-Ram{\'\i}rez et~al.(2020)Mart{\'\i}nez-Ram{\'\i}rez,
  Benetos, and Reiss]{martinez2020deep}
Marco~A Mart{\'\i}nez-Ram{\'\i}rez, Emmanouil Benetos, and Joshua~D Reiss.
\newblock Deep learning for black-box modeling of audio effects.
\newblock \emph{Applied Sciences}, 10\penalty0 (2):\penalty0 638, 2020.

\bibitem[Mart{\'\i}nez-Ram{\'\i}rez et~al.(2021)Mart{\'\i}nez-Ram{\'\i}rez,
  Stoller, and Moffat]{martinez2021deep}
Marco~A Mart{\'\i}nez-Ram{\'\i}rez, Daniel Stoller, and David Moffat.
\newblock A deep learning approach to intelligent drum mixing with the
  {Wave-U-Net}.
\newblock \emph{Journal of the Audio Engineering Society}, 2021.

\bibitem[Mart{\'\i}nez-Ram{\'\i}rez et~al.(2022)Mart{\'\i}nez-Ram{\'\i}rez,
  Liao, Fabbro, Uhlich, Nagashima, and Mitsufuji]{martinez2022automatic}
Marco~A Mart{\'\i}nez-Ram{\'\i}rez, Wei-Hsiang Liao, Giorgio Fabbro, Stefan
  Uhlich, Chihiro Nagashima, and Yuki Mitsufuji.
\newblock Automatic music mixing with deep learning and out-of-domain data.
\newblock In \emph{ISMIR}, 2022.

\bibitem[McCallum et~al.(2022)McCallum, Korzeniowski, Oramas, Gouyon, and
  Ehmann]{McCallumKOGE22mule}
Matthew~C McCallum, Filip Korzeniowski, Sergio Oramas, Fabien Gouyon, and
  Andreas Ehmann.
\newblock Supervised and unsupervised learning of audio representations for
  music understanding.
\newblock In \emph{ISMIR 2022}, 2022.

\bibitem[Meseguer-Brocal \& Peeters(2019)Meseguer-Brocal and
  Peeters]{meseguer2019conditioned}
Gabriel Meseguer-Brocal and Geoffroy Peeters.
\newblock Conditioned-u-net: Introducing a control mechanism in the u-net for
  multiple source separations.
\newblock 2019.

\bibitem[Mitsufuji et~al.(2022)Mitsufuji, Fabbro, Uhlich, St{\"o}ter,
  D{\'e}fossez, Kim, Choi, Yu, and Cheuk]{mitsufuji2022music}
Yuki Mitsufuji, Giorgio Fabbro, Stefan Uhlich, Fabian-Robert St{\"o}ter,
  Alexandre D{\'e}fossez, Minseok Kim, Woosung Choi, Chin-Yun Yu, and Kin-Wai
  Cheuk.
\newblock {Music Demixing Challenge} 2021.
\newblock \emph{Frontiers in Signal Processing}, 1:\penalty0 808395, 2022.

\bibitem[Modrzejewski et~al.(2023)Modrzejewski, Szachewicz, and
  Rokita]{modrzejewski2023}
Mateusz Modrzejewski, Piotr Szachewicz, and Przemys{\l}aw Rokita.
\newblock Transfer learning with deep neural embeddings for music
  classification tasks.
\newblock In Leszek Rutkowski, Rafa{\l} Scherer, Marcin Korytkowski, Witold
  Pedrycz, Ryszard Tadeusiewicz, and Jacek~M. Zurada (eds.), \emph{Artificial
  Intelligence and Soft Computing}, pp.\  72--81, Cham, 2023. Springer
  International Publishing.
\newblock ISBN 978-3-031-23492-7.

\bibitem[Moffat \& Sandler(2019)Moffat and Sandler]{moffat2019approaches}
David Moffat and Mark~B Sandler.
\newblock Approaches in intelligent music production.
\newblock In \emph{Arts}, volume~8, pp.\  125. MDPI, 2019.

\bibitem[Niizumi et~al.(2022)Niizumi, Takeuchi, Ohishi, Harada, and
  Kashino]{ntt2022mae}
Daisuke Niizumi, Daiki Takeuchi, Yasunori Ohishi, Noboru Harada, and Kunio
  Kashino.
\newblock Masked spectrogram modeling using masked autoencoders for learning
  general-purpose audio representation.
\newblock In \emph{HEAR: Holistic Evaluation of Audio Representations}, pp.\
  1--24. PMLR, 2022.

\bibitem[Nugraha et~al.(2016)Nugraha, Liutkus, and
  Vincent]{nugraha2016multichannel}
Aditya~Arie Nugraha, Antoine Liutkus, and Emmanuel Vincent.
\newblock Multichannel music separation with deep neural networks.
\newblock In \emph{Proc. of 24th European Signal Processing Conference
  (EUSIPCO)}, pp.\  1748--1752, 2016.

\bibitem[Park et~al.(2019)Park, Chan, Zhang, Chiu, Zoph, Cubuk, and
  Le]{specaugment2019}
Daniel~S Park, William Chan, Yu~Zhang, Chung-Cheng Chiu, Barret Zoph, Ekin~D
  Cubuk, and Quoc~V Le.
\newblock Specaugment: A simple data augmentation method for automatic speech
  recognition.
\newblock In \emph{Proc. of Interspeech 2019}, 2019.

\bibitem[Pedregosa et~al.(2011)Pedregosa, Varoquaux, Gramfort, Michel, Thirion,
  Grisel, Blondel, Prettenhofer, Weiss, Dubourg, Vanderplas, Passos,
  Cournapeau, Brucher, Perrot, and Duchesnay]{scikit-learn}
F.~Pedregosa, G.~Varoquaux, A.~Gramfort, V.~Michel, B.~Thirion, O.~Grisel,
  M.~Blondel, P.~Prettenhofer, R.~Weiss, V.~Dubourg, J.~Vanderplas, A.~Passos,
  D.~Cournapeau, M.~Brucher, M.~Perrot, and E.~Duchesnay.
\newblock Scikit-learn: Machine learning in {P}ython.
\newblock \emph{Journal of Machine Learning Research}, 12:\penalty0 2825--2830,
  2011.

\bibitem[Peeters(2004)]{peeters2004large}
Geoffroy Peeters.
\newblock A large set of audio features for sound description (similarity and
  classification) in the {CUIDADO} project.
\newblock \emph{Analysis/Synthesis Team. IRCAM, Paris, France}, 54\penalty0
  (0):\penalty0 1--25, 2004.

\bibitem[Perez et~al.(2018)Perez, Strub, De~Vries, Dumoulin, and
  Courville]{perez2018film}
Ethan Perez, Florian Strub, Harm De~Vries, Vincent Dumoulin, and Aaron
  Courville.
\newblock Film: Visual reasoning with a general conditioning layer.
\newblock In \emph{Proceedings of the AAAI Conference on Artificial
  Intelligence}, volume~32, 2018.

\bibitem[Pons \& Serra(2019)Pons and Serra]{pons2019musicnn}
Jordi Pons and Xavier Serra.
\newblock musicnn: Pre-trained convolutional neural networks for music audio
  tagging.
\newblock \emph{arXiv preprint arXiv:1909.06654}, 2019.

\bibitem[Radford et~al.(2021)Radford, Kim, Hallacy, Ramesh, Goh, Agarwal,
  Sastry, Askell, Mishkin, Clark, et~al.]{radford2021learning}
Alec Radford, Jong~Wook Kim, Chris Hallacy, Aditya Ramesh, Gabriel Goh,
  Sandhini Agarwal, Girish Sastry, Amanda Askell, Pamela Mishkin, Jack Clark,
  et~al.
\newblock Learning transferable visual models from natural language
  supervision.
\newblock In \emph{International Conference on Machine Learning}, pp.\
  8748--8763. PMLR, 2021.

\bibitem[Raffel et~al.(2014)Raffel, McFee, Humphrey, Salamon, Nieto, Liang,
  Ellis, and Raffel]{raffel2014mir_eval}
Colin Raffel, Brian McFee, Eric~J Humphrey, Justin Salamon, Oriol Nieto, Dawen
  Liang, Daniel~PW Ellis, and C~Colin Raffel.
\newblock Mir\_eval: A transparent implementation of common mir metrics.
\newblock In \emph{ISMIR}, volume~10, pp.\  2014, 2014.

\bibitem[Raffel et~al.(2019)Raffel, Shazeer, Roberts, Lee, Narang, Matena,
  Zhou, Li, and Liu]{Raffel2019T5textencoder}
Colin Raffel, Noam~M. Shazeer, Adam Roberts, Katherine Lee, Sharan Narang,
  Michael Matena, Yanqi Zhou, Wei Li, and Peter~J. Liu.
\newblock Exploring the limits of transfer learning with a unified text-to-text
  transformer.
\newblock \emph{J. Mach. Learn. Res.}, 21:\penalty0 140:1--140:67, 2019.
\newblock URL \url{https://api.semanticscholar.org/CorpusID:204838007}.

\bibitem[Rafii et~al.(2017)Rafii, Liutkus, St{\"o}ter, Mimilakis, and
  Bittner]{rafii2017musdb18}
Zafar Rafii, Antoine Liutkus, Fabian-Robert St{\"o}ter, Stylianos~Ioannis
  Mimilakis, and Rachel Bittner.
\newblock Musdb18-a corpus for music separation.
\newblock 2017.

\bibitem[Razavi et~al.(2019)Razavi, van~den Oord, and
  Vinyals]{razavi2019generating}
Ali Razavi, Aaron van~den Oord, and Oriol Vinyals.
\newblock Generating diverse high-fidelity images with {VQ-VAE-2}.
\newblock In \emph{Proc. Advances in Neural Information Processing Systems
  (NeurIPS)}, pp.\  14866--14876, 2019.

\bibitem[Rouard et~al.(2023)Rouard, Massa, and D{\'e}fossez]{rouard2023hybrid}
Simon Rouard, Francisco Massa, and Alexandre D{\'e}fossez.
\newblock Hybrid transformers for music source separation.
\newblock In \emph{ICASSP 2023-2023 IEEE International Conference on Acoustics,
  Speech and Signal Processing (ICASSP)}, pp.\  1--5. IEEE, 2023.

\bibitem[Soleymani et~al.(2013)Soleymani, Caro, Schmidt, Sha, and
  Yang]{soleymani2013emomusic}
Mohammad Soleymani, Micheal~N Caro, Erik~M Schmidt, Cheng-Ya Sha, and Yi-Hsuan
  Yang.
\newblock 1000 songs for emotional analysis of music.
\newblock In \emph{Proceedings of the 2nd ACM international workshop on
  Crowdsourcing for multimedia}, pp.\  1--6, 2013.

\bibitem[Spijkervet \& Burgoyne(2021)Spijkervet and Burgoyne]{janne2021clmr}
Janne Spijkervet and John~Ashley Burgoyne.
\newblock Contrastive learning of musical representations.
\newblock \emph{arXiv preprint arXiv:2103.09410}, 2021.

\bibitem[Stables et~al.(2019)Stables, Reiss, and De~Man]{IMPbook19}
Ryan Stables, Joshua~D. Reiss, and Brecht De~Man.
\newblock \emph{Intelligent Music Production}.
\newblock Focal Press, 2019.

\bibitem[Steinmetz et~al.(2021)Steinmetz, Pons, Pascual, and
  Serr{\`a}]{steinmetz2021automatic}
Christian~J Steinmetz, Jordi Pons, Santiago Pascual, and Joan Serr{\`a}.
\newblock Automatic multitrack mixing with a differentiable mixing console of
  neural audio effects.
\newblock In \emph{ICASSP}. IEEE, 2021.

\bibitem[Steinmetz et~al.(2022)Steinmetz, Vanka, Mart{\'\i}nez-Ram{\'\i}rez,
  and Bromham]{steinmetz2022automix}
Christian~J. Steinmetz, Soumya~Sai Vanka, Marco~A Mart{\'\i}nez-Ram{\'\i}rez,
  and Gary Bromham.
\newblock \emph{Deep Learning for Automatic Mixing}.
\newblock ISMIR, December 2022.
\newblock URL \url{https://dl4am.github.io/tutorial}.

\bibitem[St{\"{o}}ter et~al.(2019)St{\"{o}}ter, Uhlich, Liutkus, and
  Mitsufuji]{stoter19umx}
F.-R. St{\"{o}}ter, S.~Uhlich, A.~Liutkus, and Y.~Mitsufuji.
\newblock Open-unmix - a reference implementation for music source separation.
\newblock \emph{Journal of Open Source Software}, 2019.
\newblock \doi{10.21105/joss.01667}.
\newblock URL \url{https://doi.org/10.21105/joss.01667}.

\bibitem[Su \& Yang(2016)Su and Yang]{su2016escaping}
Li~Su and Yi-Hsuan Yang.
\newblock Escaping from the abyss of manual annotation: New methodology of
  building polyphonic datasets for automatic music transcription.
\newblock In \emph{Music, Mind, and Embodiment: 11th International Symposium,
  CMMR 2015, Plymouth, UK, June 16-19, 2015, Revised Selected Papers 11}, pp.\
  309--321. Springer, 2016.

\bibitem[Takida et~al.(2022)Takida, Shibuya, Liao, Lai, Ohmura, Uesaka, Murata,
  Takahashi, Kumakura, and Mitsufuji]{takida2022sq-vae}
Yuhta Takida, Takashi Shibuya, WeiHsiang Liao, Chieh-Hsin Lai, Junki Ohmura,
  Toshimitsu Uesaka, Naoki Murata, Shusuke Takahashi, Toshiyuki Kumakura, and
  Yuki Mitsufuji.
\newblock {SQ-VAE}: Variational bayes on discrete representation with
  self-annealed stochastic quantization.
\newblock In \emph{International Conference on Machine Learning}, 2022.

\bibitem[Takida et~al.(2024)Takida, Ikemiya, Shibuya, Lai, Shimada, Murata,
  Murata, Uesaka, Uchida, and Mitsufuji]{takida2023hq-vae}
Yuhta Takida, Yukara Ikemiya, Takashi Shibuya, Chieh-Hsin Lai, Kazuki Shimada,
  Naoki Murata, Naoki Murata, Toshimitsu Uesaka, Kengo Uchida, and Yuki
  Mitsufuji.
\newblock Hq-vae: {H}ierarchical discrete representation learning with
  variational bayes.
\newblock \emph{Transactions on Machine Learning Research (TMLR)}, 2024.

\bibitem[Team et~al.(2024)Team, Anil, Borgeaud, Alayrac, Yu, Soricut,
  Schalkwyk, Dai, Hauth, Millican, Silver, Johnson, Antonoglou, Schrittwieser,
  Glaese, Chen, Pitler, Lillicrap, Lazaridou, Firat, Molloy, Isard, Barham,
  Hennigan, Lee, Viola, Reynolds, Xu, Doherty, Collins, Meyer, Rutherford,
  Moreira, Ayoub, Goel, Krawczyk, Du, Chi, Cheng, Ni, Shah, Kane, Chan,
  Faruqui, Severyn, Lin, Li, Cheng, Ittycheriah, Mahdieh, Chen, Sun, Tran,
  Bagri, Lakshminarayanan, Liu, Orban, Güra, Zhou, Song, Boffy, Ganapathy,
  Zheng, Choe, Ágoston Weisz, Zhu, Lu, Gopal, Kahn, Kula, Pitman, Shah,
  Taropa, Merey, Baeuml, Chen, Shafey, Zhang, Sercinoglu, Tucker, Piqueras,
  Krikun, Barr, Savinov, Danihelka, Roelofs, White, Andreassen, von Glehn,
  Yagati, Kazemi, Gonzalez, Khalman, Sygnowski, Frechette, Smith, Culp,
  Proleev, Luan, Chen, Lottes, Schucher, Lebron, Rrustemi, Clay, Crone,
  Kocisky, Zhao, Perz, Yu, Howard, Bloniarz, Rae, Lu, Sifre, Maggioni, Alcober,
  Garrette, Barnes, Thakoor, Austin, Barth-Maron, Wong, Joshi, Chaabouni,
  Fatiha, Ahuja, Tomar, Senter, Chadwick, Kornakov, Attaluri, Iturrate, Liu,
  Li, Cogan, Chen, Jia, Gu, Zhang, Grimstad, Hartman, Garcia, Pillai, Devlin,
  Laskin, de~Las~Casas, Valter, Tao, Blanco, Badia, Reitter, Chen, Brennan,
  Rivera, Brin, Iqbal, Surita, Labanowski, Rao, Winkler, Parisotto, Gu,
  Olszewska, Addanki, Miech, Louis, Teplyashin, Brown, Catt, Balaguer, Xiang,
  Wang, Ashwood, Briukhov, Webson, Ganapathy, Sanghavi, Kannan, Chang,
  Stjerngren, Djolonga, Sun, Bapna, Aitchison, Pejman, Michalewski, Yu, Wang,
  Love, Ahn, Bloxwich, Han, Humphreys, Sellam, Bradbury, Godbole, Samangooei,
  Damoc, Kaskasoli, Arnold, Vasudevan, Agrawal, Riesa, Lepikhin, Tanburn,
  Srinivasan, Lim, Hodkinson, Shyam, Ferret, Hand, Garg, Paine, Li, Li, Giang,
  Neitz, Abbas, York, Reid, Cole, Chowdhery, Das, Rogozińska, Nikolaev,
  Sprechmann, Nado, Zilka, Prost, He, Monteiro, Mishra, Welty, Newlan, Jia,
  Allamanis, Hu, de~Liedekerke, Gilmer, Saroufim, Rijhwani, Hou, Shrivastava,
  Baddepudi, Goldin, Ozturel, Cassirer, Xu, Sohn, Sachan, Amplayo, Swanson,
  Petrova, Narayan, Guez, Brahma, Landon, Patel, Zhao, Villela, Wang, Jia,
  Rahtz, Giménez, Yeung, Keeling, Georgiev, Mincu, Wu, Haykal, Saputro,
  Vodrahalli, Qin, Cankara, Sharma, Fernando, Hawkins, Neyshabur, Kim, Hutter,
  Agrawal, Castro-Ros, van~den Driessche, Wang, Yang, yiin Chang, Komarek,
  McIlroy, Lučić, Zhang, Farhan, Sharman, Natsev, Michel, Bansal, Qiao, Cao,
  Shakeri, Butterfield, Chung, Rubenstein, Agrawal, Mensch, Soparkar, Lenc,
  Chung, Pope, Maggiore, Kay, Jhakra, Wang, Maynez, Phuong, Tobin, Tacchetti,
  Trebacz, Robinson, Katariya, Riedel, Bailey, Xiao, Ghelani, Aroyo, Slone,
  Houlsby, Xiong, Yang, Gribovskaya, Adler, Wirth, Lee, Li, Kagohara,
  Pavagadhi, Bridgers, Bortsova, Ghemawat, Ahmed, Liu, Powell, Bolina, Iinuma,
  Zablotskaia, Besley, Chung, Dozat, Comanescu, Si, Greer, Su, Polacek,
  Kaufman, Tokumine, Hu, Buchatskaya, Miao, Elhawaty, Siddhant, Tomasev, Xing,
  Greer, Miller, Ashraf, Roy, Zhang, Ma, Filos, Besta, Blevins, Klimenko, Yeh,
  Changpinyo, Mu, Chang, Pajarskas, Muir, Cohen, Lan, Haridasan, Marathe,
  Hansen, Douglas, Samuel, Wang, Austin, Lan, Jiang, Chiu, Lorenzo, Sjösund,
  Cevey, Gleicher, Avrahami, Boral, Srinivasan, Selo, May, Aisopos, Hussenot,
  Soares, Baumli, Chang, Recasens, Caine, Pritzel, Pavetic, Pardo, Gergely,
  Frye, Ramasesh, Horgan, Badola, Kassner, Roy, Dyer, Campos, Tomala, Tang,
  Badawy, White, Mustafa, Lang, Jindal, Vikram, Gong, Caelles, Hemsley,
  Thornton, Feng, Stokowiec, Zheng, Thacker, Çağlar Ünlü, Zhang, Saleh,
  Svensson, Bileschi, Patil, Anand, Ring, Tsihlas, Vezer, Selvi, Shevlane,
  Rodriguez, Kwiatkowski, Daruki, Rong, Dafoe, FitzGerald, Gu-Lemberg, Khan,
  Hendricks, Pellat, Feinberg, Cobon-Kerr, Sainath, Rauh, Hashemi, Ives,
  Hasson, Noland, Cao, Byrd, Hou, Wang, Sottiaux, Paganini, Lespiau, Moufarek,
  Hassan, Shivakumar, van Amersfoort, Mandhane, Joshi, Goyal, Tung, Brock,
  Sheahan, Misra, Li, Rakićević, Dehghani, Liu, Mittal, Oh, Noury, Sezener,
  Huot, Lamm, Cao, Chen, Mudgal, Stella, Brooks, Vasudevan, Liu, Chain,
  Melinkeri, Cohen, Wang, Seymore, Zubkov, Goel, Yue, Krishnakumaran, Albert,
  Hurley, Sano, Mohananey, Joughin, Filonov, Kępa, Eldawy, Lim, Rishi,
  Badiezadegan, Bos, Chang, Jain, Padmanabhan, Puttagunta, Krishna, Baker,
  Kalb, Bedapudi, Kurzrok, Lei, Yu, Litvin, Zhou, Wu, Sobell, Siciliano, Papir,
  Neale, Bragagnolo, Toor, Chen, Anklin, Wang, Feng, Gholami, Ling, Liu,
  Walter, Moghaddam, Kishore, Adamek, Mercado, Mallinson, Wandekar, Cagle,
  Ofek, Garrido, Lombriser, Mukha, Sun, Mohammad, Matak, Qian, Peswani, Janus,
  Yuan, Schelin, David, Garg, He, Duzhyi, Älgmyr, Lottaz, Li, Yadav, Xu,
  Chinien, Shivanna, Chuklin, Li, Spadine, Wolfe, Mohamed, Das, Dai, He, von
  Dincklage, Upadhyay, Maurya, Chi, Krause, Salama, Rabinovitch, M, Selvan,
  Dektiarev, Ghiasi, Guven, Gupta, Liu, Sharma, Shtacher, Paul, Akerlund,
  Aubet, Huang, Zhu, Zhu, Teixeira, Fritze, Bertolini, Marinescu, Bölle,
  Paulus, Gupta, Latkar, Chang, Sanders, Wilson, Wu, Tan, Thiet, Doshi, Lall,
  Mishra, Chen, Luong, Benjamin, Lee, Andrejczuk, Rabiej, Ranjan, Styrc, Yin,
  Simon, Harriott, Bansal, Robsky, Bacon, Greene, Mirylenka, Zhou, Sarvana,
  Goyal, Andermatt, Siegler, Horn, Israel, Pongetti, Chen, Selvatici, Silva,
  Wang, Tolins, Guu, Yogev, Cai, Agostini, Shah, Nguyen, Donnaile, Pereira,
  Friso, Stambler, Kurzrok, Kuang, Romanikhin, Geller, Yan, Jang, Lee, Fica,
  Malmi, Tan, Banica, Balle, Pham, Huang, Avram, Shi, Singh, Hidey, Ahuja,
  Saxena, Dooley, Potharaju, O'Neill, Gokulchandran, Foley, Zhao, Dusenberry,
  Liu, Mehta, Kotikalapudi, Safranek-Shrader, Goodman, Kessinger, Globen,
  Kolhar, Gorgolewski, Ibrahim, Song, Eichenbaum, Brovelli, Potluri, Lahoti,
  Baetu, Ghorbani, Chen, Crawford, Pal, Sridhar, Gurita, Mujika, Petrovski,
  Cedoz, Li, Chen, Santo, Goyal, Punjabi, Kappaganthu, Kwak, LV, Velury,
  Choudhury, Hall, Shah, Figueira, Thomas, Lu, Zhou, Kumar, Jurdi, Chikkerur,
  Ma, Yu, Kwak, Ähdel, Rajayogam, Choma, Liu, Barua, Ji, Park, Hellendoorn,
  Bailey, Bilal, Zhou, Khatir, Sutton, Rzadkowski, Macintosh, Shagin, Medina,
  Liang, Zhou, Shah, Bi, Dankovics, Banga, Lehmann, Bredesen, Lin, Hoffmann,
  Lai, Chung, Yang, Balani, Bražinskas, Sozanschi, Hayes, Alcalde, Makarov,
  Chen, Stella, Snijders, Mandl, Kärrman, Nowak, Wu, Dyck, Vaidyanathan, R,
  Mallet, Rudominer, Johnston, Mittal, Udathu, Christensen, Verma, Irving,
  Santucci, Elsayed, Davoodi, Georgiev, Tenney, Hua, Cideron, Leurent,
  Alnahlawi, Georgescu, Wei, Zheng, Scandinaro, Jiang, Snoek, Sundararajan,
  Wang, Ontiveros, Karo, Cole, Rajashekhar, Tumeh, Ben-David, Jain, Uesato,
  Datta, Bunyan, Wu, Zhang, Stanczyk, Zhang, Steiner, Naskar, Azzam, Johnson,
  Paszke, Chiu, Elias, Mohiuddin, Muhammad, Miao, Lee, Vieillard, Park, Zhang,
  Stanway, Garmon, Karmarkar, Dong, Lee, Kumar, Zhou, Evens, Isaac, Irving,
  Loper, Fink, Arkatkar, Chen, Shafran, Petrychenko, Chen, Jia, Levskaya, Zhu,
  Grabowski, Mao, Magni, Yao, Snaider, Casagrande, Palmer, Suganthan, Castaño,
  Giannoumis, Kim, Rybiński, Sreevatsa, Prendki, Soergel, Goedeckemeyer,
  Gierke, Jafari, Gaba, Wiesner, Wright, Wei, Vashisht, Kulizhskaya, Hoover,
  Le, Li, Iwuanyanwu, Liu, Ramirez, Khorlin, Cui, LIN, Wu, Aguilar, Pallo,
  Chakladar, Perng, Abellan, Zhang, Dasgupta, Kushman, Penchev, Repina, Wu,
  van~der Weide, Ponnapalli, Kaplan, Simsa, Li, Dousse, Yang, Piper, Ie,
  Pasumarthi, Lintz, Vijayakumar, Andor, Valenzuela, Lui, Paduraru, Peng, Lee,
  Zhang, Greene, Nguyen, Kurylowicz, Hardin, Dixon, Janzer, Choo, Feng, Zhang,
  Singhal, Du, McKinnon, Antropova, Bolukbasi, Keller, Reid, Finchelstein,
  Raad, Crocker, Hawkins, Dadashi, Gaffney, Franko, Bulanova, Leblond, Chung,
  Askham, Cobo, Xu, Fischer, Xu, Sorokin, Alberti, Lin, Evans, Dimitriev,
  Forbes, Banarse, Tung, Omernick, Bishop, Sterneck, Jain, Xia, Amid, Piccinno,
  Wang, Banzal, Mankowitz, Polozov, Krakovna, Brown, Bateni, Duan, Firoiu,
  Thotakuri, Natan, Geist, tan Girgin, Li, Ye, Roval, Tojo, Kwong, Lee-Thorp,
  Yew, Sinopalnikov, Ramos, Mellor, Sharma, Wu, Miller, Sonnerat, Vnukov,
  Greig, Beattie, Caveness, Bai, Eisenschlos, Korchemniy, Tsai, Jasarevic,
  Kong, Dao, Zheng, Liu, Yang, Zhu, Teh, Sanmiya, Gladchenko, Trdin, Toyama,
  Rosen, Tavakkol, Xue, Elkind, Woodman, Carpenter, Papamakarios, Kemp, Kafle,
  Grunina, Sinha, Talbert, Wu, Owusu-Afriyie, Du, Thornton, Pont-Tuset,
  Narayana, Li, Fatehi, Wieting, Ajmeri, Uria, Ko, Knight, Héliou, Niu, Gu,
  Pang, Li, Levine, Stolovich, Santamaria-Fernandez, Goenka, Yustalim, Strudel,
  Elqursh, Deck, Lee, Li, Levin, Hoffmann, Holtmann-Rice, Bachem, Arora, Koh,
  Yeganeh, Põder, Tariq, Sun, Ionita, Seyedhosseini, Tafti, Liu, Gulati, Liu,
  Ye, Chrzaszcz, Wang, Sethi, Li, Brown, Singh, Fan, Parisi, Stanton,
  Koverkathu, Choquette-Choo, Li, Lu, Ittycheriah, Shroff, Varadarajan,
  Bahargam, Willoughby, Gaddy, Desjardins, Cornero, Robenek, Mittal, Albrecht,
  Shenoy, Moiseev, Jacobsson, Ghaffarkhah, Rivière, Walton, Crepy, Parrish,
  Zhou, Farabet, Radebaugh, Srinivasan, van~der Salm, Fidjeland, Scellato,
  Latorre-Chimoto, Klimczak-Plucińska, Bridson, de~Cesare, Hudson,
  Mendolicchio, Walker, Morris, Mauger, Guseynov, Reid, Odoom, Loher, Cotruta,
  Yenugula, Grewe, Petrushkina, Duerig, Sanchez, Yadlowsky, Shen, Globerson,
  Webb, Dua, Li, Bhupatiraju, Hurt, Qureshi, Agarwal, Shani, Eyal, Khare,
  Belle, Wang, Tekur, Kale, Wei, Sang, Saeta, Liechty, Sun, Zhao, Lee, Nayak,
  Fritz, Vuyyuru, Aslanides, Vyas, Wicke, Ma, Eltyshev, Martin, Cate, Manyika,
  Amiri, Kim, Xiong, Kang, Luisier, Tripuraneni, Madras, Guo, Waters, Wang,
  Ainslie, Baldridge, Zhang, Pruthi, Bauer, Yang, Mansour, Gelman, Xu,
  Polovets, Liu, Cai, Chen, Sheng, Xue, Ozair, Angermueller, Li, Sinha, Wang,
  Wiesinger, Koukoumidis, Tian, Iyer, Gurumurthy, Goldenson, Shah, Blake, Yu,
  Urbanowicz, Palomaki, Fernando, Durden, Mehta, Momchev, Rahimtoroghi,
  Georgaki, Raul, Ruder, Redshaw, Lee, Zhou, Jalan, Li, Hechtman, Schuh, Nasr,
  Milan, Mikulik, Franco, Green, Nguyen, Kelley, Mahendru, Hu, Howland, Vargas,
  Hui, Bansal, Rao, Ghiya, Wang, Ye, Sarr, Preston, Elish, Li, Kaku, Gupta,
  Pasupat, Juan, Someswar, M., Chen, Amini, Fabrikant, Chu, Dong, Muthal,
  Buthpitiya, Jauhari, Hua, Khandelwal, Hitron, Ren, Rinaldi, Drath, Dabush,
  Jiang, Godhia, Sachs, Chen, Fan, Taitelbaum, Noga, Dai, Wang, Liang, Hamer,
  Ferng, Elkind, Atias, Lee, Listík, Carlen, van~de Kerkhof, Pikus, Zaher,
  Müller, Zykova, Stefanec, Gatsko, Hirnschall, Sethi, Xu, Ahuja, Tsai,
  Stefanoiu, Feng, Dhandhania, Katyal, Gupta, Parulekar, Pitta, Zhao, Bhatia,
  Bhavnani, Alhadlaq, Li, Danenberg, Tu, Pine, Filippova, Ghosh, Limonchik,
  Urala, Lanka, Clive, Sun, Li, Wu, Hongtongsak, Li, Thakkar, Omarov,
  Majmundar, Alverson, Kucharski, Patel, Jain, Zabelin, Pelagatti, Kohli,
  Kumar, Kim, Sankar, Shah, Ramachandruni, Zeng, Bariach, Weidinger, Vu,
  Andreev, He, Hui, Kashem, Subramanya, Hsiao, Hassabis, Kavukcuoglu, Sadovsky,
  Le, Strohman, Wu, Petrov, Dean, and Vinyals]{Geminiteam2024gemini}
Gemini Team, Rohan Anil, Sebastian Borgeaud, Jean-Baptiste Alayrac, Jiahui Yu,
  Radu Soricut, Johan Schalkwyk, Andrew~M. Dai, Anja Hauth, Katie Millican,
  David Silver, Melvin Johnson, Ioannis Antonoglou, Julian Schrittwieser,
  Amelia Glaese, Jilin Chen, Emily Pitler, Timothy Lillicrap, Angeliki
  Lazaridou, Orhan Firat, James Molloy, Michael Isard, Paul~R. Barham, Tom
  Hennigan, Benjamin Lee, Fabio Viola, Malcolm Reynolds, Yuanzhong Xu, Ryan
  Doherty, Eli Collins, Clemens Meyer, Eliza Rutherford, Erica Moreira, Kareem
  Ayoub, Megha Goel, Jack Krawczyk, Cosmo Du, Ed~Chi, Heng-Tze Cheng, Eric Ni,
  Purvi Shah, Patrick Kane, Betty Chan, Manaal Faruqui, Aliaksei Severyn,
  Hanzhao Lin, YaGuang Li, Yong Cheng, Abe Ittycheriah, Mahdis Mahdieh, Mia
  Chen, Pei Sun, Dustin Tran, Sumit Bagri, Balaji Lakshminarayanan, Jeremiah
  Liu, Andras Orban, Fabian Güra, Hao Zhou, Xinying Song, Aurelien Boffy,
  Harish Ganapathy, Steven Zheng, HyunJeong Choe, Ágoston Weisz, Tao Zhu,
  Yifeng Lu, Siddharth Gopal, Jarrod Kahn, Maciej Kula, Jeff Pitman, Rushin
  Shah, Emanuel Taropa, Majd~Al Merey, Martin Baeuml, Zhifeng Chen, Laurent~El
  Shafey, Yujing Zhang, Olcan Sercinoglu, George Tucker, Enrique Piqueras,
  Maxim Krikun, Iain Barr, Nikolay Savinov, Ivo Danihelka, Becca Roelofs,
  Anaïs White, Anders Andreassen, Tamara von Glehn, Lakshman Yagati, Mehran
  Kazemi, Lucas Gonzalez, Misha Khalman, Jakub Sygnowski, Alexandre Frechette,
  Charlotte Smith, Laura Culp, Lev Proleev, Yi~Luan, Xi~Chen, James Lottes,
  Nathan Schucher, Federico Lebron, Alban Rrustemi, Natalie Clay, Phil Crone,
  Tomas Kocisky, Jeffrey Zhao, Bartek Perz, Dian Yu, Heidi Howard, Adam
  Bloniarz, Jack~W. Rae, Han Lu, Laurent Sifre, Marcello Maggioni, Fred
  Alcober, Dan Garrette, Megan Barnes, Shantanu Thakoor, Jacob Austin, Gabriel
  Barth-Maron, William Wong, Rishabh Joshi, Rahma Chaabouni, Deeni Fatiha, Arun
  Ahuja, Gaurav~Singh Tomar, Evan Senter, Martin Chadwick, Ilya Kornakov,
  Nithya Attaluri, Iñaki Iturrate, Ruibo Liu, Yunxuan Li, Sarah Cogan, Jeremy
  Chen, Chao Jia, Chenjie Gu, Qiao Zhang, Jordan Grimstad, Ale~Jakse Hartman,
  Xavier Garcia, Thanumalayan~Sankaranarayana Pillai, Jacob Devlin, Michael
  Laskin, Diego de~Las~Casas, Dasha Valter, Connie Tao, Lorenzo Blanco,
  Adrià~Puigdomènech Badia, David Reitter, Mianna Chen, Jenny Brennan, Clara
  Rivera, Sergey Brin, Shariq Iqbal, Gabriela Surita, Jane Labanowski, Abhi
  Rao, Stephanie Winkler, Emilio Parisotto, Yiming Gu, Kate Olszewska, Ravi
  Addanki, Antoine Miech, Annie Louis, Denis Teplyashin, Geoff Brown, Elliot
  Catt, Jan Balaguer, Jackie Xiang, Pidong Wang, Zoe Ashwood, Anton Briukhov,
  Albert Webson, Sanjay Ganapathy, Smit Sanghavi, Ajay Kannan, Ming-Wei Chang,
  Axel Stjerngren, Josip Djolonga, Yuting Sun, Ankur Bapna, Matthew Aitchison,
  Pedram Pejman, Henryk Michalewski, Tianhe Yu, Cindy Wang, Juliette Love,
  Junwhan Ahn, Dawn Bloxwich, Kehang Han, Peter Humphreys, Thibault Sellam,
  James Bradbury, Varun Godbole, Sina Samangooei, Bogdan Damoc, Alex Kaskasoli,
  Sébastien M.~R. Arnold, Vijay Vasudevan, Shubham Agrawal, Jason Riesa,
  Dmitry Lepikhin, Richard Tanburn, Srivatsan Srinivasan, Hyeontaek Lim, Sarah
  Hodkinson, Pranav Shyam, Johan Ferret, Steven Hand, Ankush Garg, Tom~Le
  Paine, Jian Li, Yujia Li, Minh Giang, Alexander Neitz, Zaheer Abbas, Sarah
  York, Machel Reid, Elizabeth Cole, Aakanksha Chowdhery, Dipanjan Das,
  Dominika Rogozińska, Vitaliy Nikolaev, Pablo Sprechmann, Zachary Nado, Lukas
  Zilka, Flavien Prost, Luheng He, Marianne Monteiro, Gaurav Mishra, Chris
  Welty, Josh Newlan, Dawei Jia, Miltiadis Allamanis, Clara~Huiyi Hu, Raoul
  de~Liedekerke, Justin Gilmer, Carl Saroufim, Shruti Rijhwani, Shaobo Hou,
  Disha Shrivastava, Anirudh Baddepudi, Alex Goldin, Adnan Ozturel, Albin
  Cassirer, Yunhan Xu, Daniel Sohn, Devendra Sachan, Reinald~Kim Amplayo, Craig
  Swanson, Dessie Petrova, Shashi Narayan, Arthur Guez, Siddhartha Brahma,
  Jessica Landon, Miteyan Patel, Ruizhe Zhao, Kevin Villela, Luyu Wang, Wenhao
  Jia, Matthew Rahtz, Mai Giménez, Legg Yeung, James Keeling, Petko Georgiev,
  Diana Mincu, Boxi Wu, Salem Haykal, Rachel Saputro, Kiran Vodrahalli, James
  Qin, Zeynep Cankara, Abhanshu Sharma, Nick Fernando, Will Hawkins, Behnam
  Neyshabur, Solomon Kim, Adrian Hutter, Priyanka Agrawal, Alex Castro-Ros,
  George van~den Driessche, Tao Wang, Fan Yang, Shuo yiin Chang, Paul Komarek,
  Ross McIlroy, Mario Lučić, Guodong Zhang, Wael Farhan, Michael Sharman,
  Paul Natsev, Paul Michel, Yamini Bansal, Siyuan Qiao, Kris Cao, Siamak
  Shakeri, Christina Butterfield, Justin Chung, Paul~Kishan Rubenstein, Shivani
  Agrawal, Arthur Mensch, Kedar Soparkar, Karel Lenc, Timothy Chung, Aedan
  Pope, Loren Maggiore, Jackie Kay, Priya Jhakra, Shibo Wang, Joshua Maynez,
  Mary Phuong, Taylor Tobin, Andrea Tacchetti, Maja Trebacz, Kevin Robinson,
  Yash Katariya, Sebastian Riedel, Paige Bailey, Kefan Xiao, Nimesh Ghelani,
  Lora Aroyo, Ambrose Slone, Neil Houlsby, Xuehan Xiong, Zhen Yang, Elena
  Gribovskaya, Jonas Adler, Mateo Wirth, Lisa Lee, Music Li, Thais Kagohara,
  Jay Pavagadhi, Sophie Bridgers, Anna Bortsova, Sanjay Ghemawat, Zafarali
  Ahmed, Tianqi Liu, Richard Powell, Vijay Bolina, Mariko Iinuma, Polina
  Zablotskaia, James Besley, Da-Woon Chung, Timothy Dozat, Ramona Comanescu,
  Xiance Si, Jeremy Greer, Guolong Su, Martin Polacek, Raphaël~Lopez Kaufman,
  Simon Tokumine, Hexiang Hu, Elena Buchatskaya, Yingjie Miao, Mohamed
  Elhawaty, Aditya Siddhant, Nenad Tomasev, Jinwei Xing, Christina Greer, Helen
  Miller, Shereen Ashraf, Aurko Roy, Zizhao Zhang, Ada Ma, Angelos Filos, Milos
  Besta, Rory Blevins, Ted Klimenko, Chih-Kuan Yeh, Soravit Changpinyo, Jiaqi
  Mu, Oscar Chang, Mantas Pajarskas, Carrie Muir, Vered Cohen, Charline~Le Lan,
  Krishna Haridasan, Amit Marathe, Steven Hansen, Sholto Douglas, Rajkumar
  Samuel, Mingqiu Wang, Sophia Austin, Chang Lan, Jiepu Jiang, Justin Chiu,
  Jaime~Alonso Lorenzo, Lars~Lowe Sjösund, Sébastien Cevey, Zach Gleicher,
  Thi Avrahami, Anudhyan Boral, Hansa Srinivasan, Vittorio Selo, Rhys May,
  Konstantinos Aisopos, Léonard Hussenot, Livio~Baldini Soares, Kate Baumli,
  Michael~B. Chang, Adrià Recasens, Ben Caine, Alexander Pritzel, Filip
  Pavetic, Fabio Pardo, Anita Gergely, Justin Frye, Vinay Ramasesh, Dan Horgan,
  Kartikeya Badola, Nora Kassner, Subhrajit Roy, Ethan Dyer, Víctor~Campos
  Campos, Alex Tomala, Yunhao Tang, Dalia~El Badawy, Elspeth White, Basil
  Mustafa, Oran Lang, Abhishek Jindal, Sharad Vikram, Zhitao Gong, Sergi
  Caelles, Ross Hemsley, Gregory Thornton, Fangxiaoyu Feng, Wojciech Stokowiec,
  Ce~Zheng, Phoebe Thacker, Çağlar Ünlü, Zhishuai Zhang, Mohammad Saleh,
  James Svensson, Max Bileschi, Piyush Patil, Ankesh Anand, Roman Ring,
  Katerina Tsihlas, Arpi Vezer, Marco Selvi, Toby Shevlane, Mikel Rodriguez,
  Tom Kwiatkowski, Samira Daruki, Keran Rong, Allan Dafoe, Nicholas FitzGerald,
  Keren Gu-Lemberg, Mina Khan, Lisa~Anne Hendricks, Marie Pellat, Vladimir
  Feinberg, James Cobon-Kerr, Tara Sainath, Maribeth Rauh, Sayed~Hadi Hashemi,
  Richard Ives, Yana Hasson, Eric Noland, Yuan Cao, Nathan Byrd, Le~Hou, Qingze
  Wang, Thibault Sottiaux, Michela Paganini, Jean-Baptiste Lespiau, Alexandre
  Moufarek, Samer Hassan, Kaushik Shivakumar, Joost van Amersfoort, Amol
  Mandhane, Pratik Joshi, Anirudh Goyal, Matthew Tung, Andrew Brock, Hannah
  Sheahan, Vedant Misra, Cheng Li, Nemanja Rakićević, Mostafa Dehghani,
  Fangyu Liu, Sid Mittal, Junhyuk Oh, Seb Noury, Eren Sezener, Fantine Huot,
  Matthew Lamm, Nicola~De Cao, Charlie Chen, Sidharth Mudgal, Romina Stella,
  Kevin Brooks, Gautam Vasudevan, Chenxi Liu, Mainak Chain, Nivedita Melinkeri,
  Aaron Cohen, Venus Wang, Kristie Seymore, Sergey Zubkov, Rahul Goel, Summer
  Yue, Sai Krishnakumaran, Brian Albert, Nate Hurley, Motoki Sano, Anhad
  Mohananey, Jonah Joughin, Egor Filonov, Tomasz Kępa, Yomna Eldawy, Jiawern
  Lim, Rahul Rishi, Shirin Badiezadegan, Taylor Bos, Jerry Chang, Sanil Jain,
  Sri Gayatri~Sundara Padmanabhan, Subha Puttagunta, Kalpesh Krishna, Leslie
  Baker, Norbert Kalb, Vamsi Bedapudi, Adam Kurzrok, Shuntong Lei, Anthony Yu,
  Oren Litvin, Xiang Zhou, Zhichun Wu, Sam Sobell, Andrea Siciliano, Alan
  Papir, Robby Neale, Jonas Bragagnolo, Tej Toor, Tina Chen, Valentin Anklin,
  Feiran Wang, Richie Feng, Milad Gholami, Kevin Ling, Lijuan Liu, Jules
  Walter, Hamid Moghaddam, Arun Kishore, Jakub Adamek, Tyler Mercado, Jonathan
  Mallinson, Siddhinita Wandekar, Stephen Cagle, Eran Ofek, Guillermo Garrido,
  Clemens Lombriser, Maksim Mukha, Botu Sun, Hafeezul~Rahman Mohammad, Josip
  Matak, Yadi Qian, Vikas Peswani, Pawel Janus, Quan Yuan, Leif Schelin, Oana
  David, Ankur Garg, Yifan He, Oleksii Duzhyi, Anton Älgmyr, Timothée Lottaz,
  Qi~Li, Vikas Yadav, Luyao Xu, Alex Chinien, Rakesh Shivanna, Aleksandr
  Chuklin, Josie Li, Carrie Spadine, Travis Wolfe, Kareem Mohamed, Subhabrata
  Das, Zihang Dai, Kyle He, Daniel von Dincklage, Shyam Upadhyay, Akanksha
  Maurya, Luyan Chi, Sebastian Krause, Khalid Salama, Pam~G Rabinovitch, Pavan
  Kumar~Reddy M, Aarush Selvan, Mikhail Dektiarev, Golnaz Ghiasi, Erdem Guven,
  Himanshu Gupta, Boyi Liu, Deepak Sharma, Idan~Heimlich Shtacher, Shachi Paul,
  Oscar Akerlund, François-Xavier Aubet, Terry Huang, Chen Zhu, Eric Zhu,
  Elico Teixeira, Matthew Fritze, Francesco Bertolini, Liana-Eleonora
  Marinescu, Martin Bölle, Dominik Paulus, Khyatti Gupta, Tejasi Latkar, Max
  Chang, Jason Sanders, Roopa Wilson, Xuewei Wu, Yi-Xuan Tan, Lam~Nguyen Thiet,
  Tulsee Doshi, Sid Lall, Swaroop Mishra, Wanming Chen, Thang Luong, Seth
  Benjamin, Jasmine Lee, Ewa Andrejczuk, Dominik Rabiej, Vipul Ranjan,
  Krzysztof Styrc, Pengcheng Yin, Jon Simon, Malcolm~Rose Harriott, Mudit
  Bansal, Alexei Robsky, Geoff Bacon, David Greene, Daniil Mirylenka, Chen
  Zhou, Obaid Sarvana, Abhimanyu Goyal, Samuel Andermatt, Patrick Siegler, Ben
  Horn, Assaf Israel, Francesco Pongetti, Chih-Wei~"Louis" Chen, Marco
  Selvatici, Pedro Silva, Kathie Wang, Jackson Tolins, Kelvin Guu, Roey Yogev,
  Xiaochen Cai, Alessandro Agostini, Maulik Shah, Hung Nguyen, Noah~Ó
  Donnaile, Sébastien Pereira, Linda Friso, Adam Stambler, Adam Kurzrok,
  Chenkai Kuang, Yan Romanikhin, Mark Geller, ZJ~Yan, Kane Jang, Cheng-Chun
  Lee, Wojciech Fica, Eric Malmi, Qijun Tan, Dan Banica, Daniel Balle, Ryan
  Pham, Yanping Huang, Diana Avram, Hongzhi Shi, Jasjot Singh, Chris Hidey,
  Niharika Ahuja, Pranab Saxena, Dan Dooley, Srividya~Pranavi Potharaju, Eileen
  O'Neill, Anand Gokulchandran, Ryan Foley, Kai Zhao, Mike Dusenberry, Yuan
  Liu, Pulkit Mehta, Ragha Kotikalapudi, Chalence Safranek-Shrader, Andrew
  Goodman, Joshua Kessinger, Eran Globen, Prateek Kolhar, Chris Gorgolewski,
  Ali Ibrahim, Yang Song, Ali Eichenbaum, Thomas Brovelli, Sahitya Potluri,
  Preethi Lahoti, Cip Baetu, Ali Ghorbani, Charles Chen, Andy Crawford, Shalini
  Pal, Mukund Sridhar, Petru Gurita, Asier Mujika, Igor Petrovski, Pierre-Louis
  Cedoz, Chenmei Li, Shiyuan Chen, Niccolò~Dal Santo, Siddharth Goyal, Jitesh
  Punjabi, Karthik Kappaganthu, Chester Kwak, Pallavi LV, Sarmishta Velury,
  Himadri Choudhury, Jamie Hall, Premal Shah, Ricardo Figueira, Matt Thomas,
  Minjie Lu, Ting Zhou, Chintu Kumar, Thomas Jurdi, Sharat Chikkerur, Yenai Ma,
  Adams Yu, Soo Kwak, Victor Ähdel, Sujeevan Rajayogam, Travis Choma, Fei Liu,
  Aditya Barua, Colin Ji, Ji~Ho Park, Vincent Hellendoorn, Alex Bailey, Taylan
  Bilal, Huanjie Zhou, Mehrdad Khatir, Charles Sutton, Wojciech Rzadkowski,
  Fiona Macintosh, Konstantin Shagin, Paul Medina, Chen Liang, Jinjing Zhou,
  Pararth Shah, Yingying Bi, Attila Dankovics, Shipra Banga, Sabine Lehmann,
  Marissa Bredesen, Zifan Lin, John~Eric Hoffmann, Jonathan Lai, Raynald Chung,
  Kai Yang, Nihal Balani, Arthur Bražinskas, Andrei Sozanschi, Matthew Hayes,
  Héctor~Fernández Alcalde, Peter Makarov, Will Chen, Antonio Stella,
  Liselotte Snijders, Michael Mandl, Ante Kärrman, Paweł Nowak, Xinyi Wu,
  Alex Dyck, Krishnan Vaidyanathan, Raghavender R, Jessica Mallet, Mitch
  Rudominer, Eric Johnston, Sushil Mittal, Akhil Udathu, Janara Christensen,
  Vishal Verma, Zach Irving, Andreas Santucci, Gamaleldin Elsayed, Elnaz
  Davoodi, Marin Georgiev, Ian Tenney, Nan Hua, Geoffrey Cideron, Edouard
  Leurent, Mahmoud Alnahlawi, Ionut Georgescu, Nan Wei, Ivy Zheng, Dylan
  Scandinaro, Heinrich Jiang, Jasper Snoek, Mukund Sundararajan, Xuezhi Wang,
  Zack Ontiveros, Itay Karo, Jeremy Cole, Vinu Rajashekhar, Lara Tumeh, Eyal
  Ben-David, Rishub Jain, Jonathan Uesato, Romina Datta, Oskar Bunyan, Shimu
  Wu, John Zhang, Piotr Stanczyk, Ye~Zhang, David Steiner, Subhajit Naskar,
  Michael Azzam, Matthew Johnson, Adam Paszke, Chung-Cheng Chiu, Jaume~Sanchez
  Elias, Afroz Mohiuddin, Faizan Muhammad, Jin Miao, Andrew Lee, Nino
  Vieillard, Jane Park, Jiageng Zhang, Jeff Stanway, Drew Garmon, Abhijit
  Karmarkar, Zhe Dong, Jong Lee, Aviral Kumar, Luowei Zhou, Jonathan Evens,
  William Isaac, Geoffrey Irving, Edward Loper, Michael Fink, Isha Arkatkar,
  Nanxin Chen, Izhak Shafran, Ivan Petrychenko, Zhe Chen, Johnson Jia, Anselm
  Levskaya, Zhenkai Zhu, Peter Grabowski, Yu~Mao, Alberto Magni, Kaisheng Yao,
  Javier Snaider, Norman Casagrande, Evan Palmer, Paul Suganthan, Alfonso
  Castaño, Irene Giannoumis, Wooyeol Kim, Mikołaj Rybiński, Ashwin
  Sreevatsa, Jennifer Prendki, David Soergel, Adrian Goedeckemeyer, Willi
  Gierke, Mohsen Jafari, Meenu Gaba, Jeremy Wiesner, Diana~Gage Wright, Yawen
  Wei, Harsha Vashisht, Yana Kulizhskaya, Jay Hoover, Maigo Le, Lu~Li, Chimezie
  Iwuanyanwu, Lu~Liu, Kevin Ramirez, Andrey Khorlin, Albert Cui, Tian LIN,
  Marcus Wu, Ricardo Aguilar, Keith Pallo, Abhishek Chakladar, Ginger Perng,
  Elena~Allica Abellan, Mingyang Zhang, Ishita Dasgupta, Nate Kushman, Ivo
  Penchev, Alena Repina, Xihui Wu, Tom van~der Weide, Priya Ponnapalli,
  Caroline Kaplan, Jiri Simsa, Shuangfeng Li, Olivier Dousse, Fan Yang, Jeff
  Piper, Nathan Ie, Rama Pasumarthi, Nathan Lintz, Anitha Vijayakumar, Daniel
  Andor, Pedro Valenzuela, Minnie Lui, Cosmin Paduraru, Daiyi Peng, Katherine
  Lee, Shuyuan Zhang, Somer Greene, Duc~Dung Nguyen, Paula Kurylowicz, Cassidy
  Hardin, Lucas Dixon, Lili Janzer, Kiam Choo, Ziqiang Feng, Biao Zhang,
  Achintya Singhal, Dayou Du, Dan McKinnon, Natasha Antropova, Tolga Bolukbasi,
  Orgad Keller, David Reid, Daniel Finchelstein, Maria~Abi Raad, Remi Crocker,
  Peter Hawkins, Robert Dadashi, Colin Gaffney, Ken Franko, Anna Bulanova,
  Rémi Leblond, Shirley Chung, Harry Askham, Luis~C. Cobo, Kelvin Xu, Felix
  Fischer, Jun Xu, Christina Sorokin, Chris Alberti, Chu-Cheng Lin, Colin
  Evans, Alek Dimitriev, Hannah Forbes, Dylan Banarse, Zora Tung, Mark
  Omernick, Colton Bishop, Rachel Sterneck, Rohan Jain, Jiawei Xia, Ehsan Amid,
  Francesco Piccinno, Xingyu Wang, Praseem Banzal, Daniel~J. Mankowitz, Alex
  Polozov, Victoria Krakovna, Sasha Brown, MohammadHossein Bateni, Dennis Duan,
  Vlad Firoiu, Meghana Thotakuri, Tom Natan, Matthieu Geist, Ser tan Girgin,
  Hui Li, Jiayu Ye, Ofir Roval, Reiko Tojo, Michael Kwong, James Lee-Thorp,
  Christopher Yew, Danila Sinopalnikov, Sabela Ramos, John Mellor, Abhishek
  Sharma, Kathy Wu, David Miller, Nicolas Sonnerat, Denis Vnukov, Rory Greig,
  Jennifer Beattie, Emily Caveness, Libin Bai, Julian Eisenschlos, Alex
  Korchemniy, Tomy Tsai, Mimi Jasarevic, Weize Kong, Phuong Dao, Zeyu Zheng,
  Frederick Liu, Fan Yang, Rui Zhu, Tian~Huey Teh, Jason Sanmiya, Evgeny
  Gladchenko, Nejc Trdin, Daniel Toyama, Evan Rosen, Sasan Tavakkol, Linting
  Xue, Chen Elkind, Oliver Woodman, John Carpenter, George Papamakarios, Rupert
  Kemp, Sushant Kafle, Tanya Grunina, Rishika Sinha, Alice Talbert, Diane Wu,
  Denese Owusu-Afriyie, Cosmo Du, Chloe Thornton, Jordi Pont-Tuset, Pradyumna
  Narayana, Jing Li, Saaber Fatehi, John Wieting, Omar Ajmeri, Benigno Uria,
  Yeongil Ko, Laura Knight, Amélie Héliou, Ning Niu, Shane Gu, Chenxi Pang,
  Yeqing Li, Nir Levine, Ariel Stolovich, Rebeca Santamaria-Fernandez, Sonam
  Goenka, Wenny Yustalim, Robin Strudel, Ali Elqursh, Charlie Deck, Hyo Lee,
  Zonglin Li, Kyle Levin, Raphael Hoffmann, Dan Holtmann-Rice, Olivier Bachem,
  Sho Arora, Christy Koh, Soheil~Hassas Yeganeh, Siim Põder, Mukarram Tariq,
  Yanhua Sun, Lucian Ionita, Mojtaba Seyedhosseini, Pouya Tafti, Zhiyu Liu,
  Anmol Gulati, Jasmine Liu, Xinyu Ye, Bart Chrzaszcz, Lily Wang, Nikhil Sethi,
  Tianrun Li, Ben Brown, Shreya Singh, Wei Fan, Aaron Parisi, Joe Stanton,
  Vinod Koverkathu, Christopher~A. Choquette-Choo, Yunjie Li, TJ~Lu, Abe
  Ittycheriah, Prakash Shroff, Mani Varadarajan, Sanaz Bahargam, Rob
  Willoughby, David Gaddy, Guillaume Desjardins, Marco Cornero, Brona Robenek,
  Bhavishya Mittal, Ben Albrecht, Ashish Shenoy, Fedor Moiseev, Henrik
  Jacobsson, Alireza Ghaffarkhah, Morgane Rivière, Alanna Walton, Clément
  Crepy, Alicia Parrish, Zongwei Zhou, Clement Farabet, Carey Radebaugh,
  Praveen Srinivasan, Claudia van~der Salm, Andreas Fidjeland, Salvatore
  Scellato, Eri Latorre-Chimoto, Hanna Klimczak-Plucińska, David Bridson,
  Dario de~Cesare, Tom Hudson, Piermaria Mendolicchio, Lexi Walker, Alex
  Morris, Matthew Mauger, Alexey Guseynov, Alison Reid, Seth Odoom, Lucia
  Loher, Victor Cotruta, Madhavi Yenugula, Dominik Grewe, Anastasia
  Petrushkina, Tom Duerig, Antonio Sanchez, Steve Yadlowsky, Amy Shen, Amir
  Globerson, Lynette Webb, Sahil Dua, Dong Li, Surya Bhupatiraju, Dan Hurt,
  Haroon Qureshi, Ananth Agarwal, Tomer Shani, Matan Eyal, Anuj Khare,
  Shreyas~Rammohan Belle, Lei Wang, Chetan Tekur, Mihir~Sanjay Kale, Jinliang
  Wei, Ruoxin Sang, Brennan Saeta, Tyler Liechty, Yi~Sun, Yao Zhao, Stephan
  Lee, Pandu Nayak, Doug Fritz, Manish~Reddy Vuyyuru, John Aslanides, Nidhi
  Vyas, Martin Wicke, Xiao Ma, Evgenii Eltyshev, Nina Martin, Hardie Cate,
  James Manyika, Keyvan Amiri, Yelin Kim, Xi~Xiong, Kai Kang, Florian Luisier,
  Nilesh Tripuraneni, David Madras, Mandy Guo, Austin Waters, Oliver Wang,
  Joshua Ainslie, Jason Baldridge, Han Zhang, Garima Pruthi, Jakob Bauer, Feng
  Yang, Riham Mansour, Jason Gelman, Yang Xu, George Polovets, Ji~Liu, Honglong
  Cai, Warren Chen, XiangHai Sheng, Emily Xue, Sherjil Ozair, Christof
  Angermueller, Xiaowei Li, Anoop Sinha, Weiren Wang, Julia Wiesinger,
  Emmanouil Koukoumidis, Yuan Tian, Anand Iyer, Madhu Gurumurthy, Mark
  Goldenson, Parashar Shah, MK~Blake, Hongkun Yu, Anthony Urbanowicz,
  Jennimaria Palomaki, Chrisantha Fernando, Ken Durden, Harsh Mehta, Nikola
  Momchev, Elahe Rahimtoroghi, Maria Georgaki, Amit Raul, Sebastian Ruder,
  Morgan Redshaw, Jinhyuk Lee, Denny Zhou, Komal Jalan, Dinghua Li, Blake
  Hechtman, Parker Schuh, Milad Nasr, Kieran Milan, Vladimir Mikulik, Juliana
  Franco, Tim Green, Nam Nguyen, Joe Kelley, Aroma Mahendru, Andrea Hu, Joshua
  Howland, Ben Vargas, Jeffrey Hui, Kshitij Bansal, Vikram Rao, Rakesh Ghiya,
  Emma Wang, Ke~Ye, Jean~Michel Sarr, Melanie~Moranski Preston, Madeleine
  Elish, Steve Li, Aakash Kaku, Jigar Gupta, Ice Pasupat, Da-Cheng Juan, Milan
  Someswar, Tejvi M., Xinyun Chen, Aida Amini, Alex Fabrikant, Eric Chu, Xuanyi
  Dong, Amruta Muthal, Senaka Buthpitiya, Sarthak Jauhari, Nan Hua, Urvashi
  Khandelwal, Ayal Hitron, Jie Ren, Larissa Rinaldi, Shahar Drath, Avigail
  Dabush, Nan-Jiang Jiang, Harshal Godhia, Uli Sachs, Anthony Chen, Yicheng
  Fan, Hagai Taitelbaum, Hila Noga, Zhuyun Dai, James Wang, Chen Liang, Jenny
  Hamer, Chun-Sung Ferng, Chenel Elkind, Aviel Atias, Paulina Lee, Vít
  Listík, Mathias Carlen, Jan van~de Kerkhof, Marcin Pikus, Krunoslav Zaher,
  Paul Müller, Sasha Zykova, Richard Stefanec, Vitaly Gatsko, Christoph
  Hirnschall, Ashwin Sethi, Xingyu~Federico Xu, Chetan Ahuja, Beth Tsai, Anca
  Stefanoiu, Bo~Feng, Keshav Dhandhania, Manish Katyal, Akshay Gupta, Atharva
  Parulekar, Divya Pitta, Jing Zhao, Vivaan Bhatia, Yashodha Bhavnani, Omar
  Alhadlaq, Xiaolin Li, Peter Danenberg, Dennis Tu, Alex Pine, Vera Filippova,
  Abhipso Ghosh, Ben Limonchik, Bhargava Urala, Chaitanya~Krishna Lanka, Derik
  Clive, Yi~Sun, Edward Li, Hao Wu, Kevin Hongtongsak, Ianna Li, Kalind
  Thakkar, Kuanysh Omarov, Kushal Majmundar, Michael Alverson, Michael
  Kucharski, Mohak Patel, Mudit Jain, Maksim Zabelin, Paolo Pelagatti, Rohan
  Kohli, Saurabh Kumar, Joseph Kim, Swetha Sankar, Vineet Shah, Lakshmi
  Ramachandruni, Xiangkai Zeng, Ben Bariach, Laura Weidinger, Tu~Vu, Alek
  Andreev, Antoine He, Kevin Hui, Sheleem Kashem, Amar Subramanya, Sissie
  Hsiao, Demis Hassabis, Koray Kavukcuoglu, Adam Sadovsky, Quoc Le, Trevor
  Strohman, Yonghui Wu, Slav Petrov, Jeffrey Dean, and Oriol Vinyals.
\newblock Gemini: A family of highly capable multimodal models, 2024.
\newblock URL \url{https://arxiv.org/abs/2312.11805}.

\bibitem[Thickstun et~al.(2016)Thickstun, Harchaoui, and
  Kakade]{Thickstun2016LearningFO}
John Thickstun, Za{\"i}d Harchaoui, and Sham~M. Kakade.
\newblock Learning features of music from scratch.
\newblock In \emph{ICLR}, volume abs/1611.09827, 2016.

\bibitem[Toyama et~al.(2023)Toyama, Akama, Ikemiya, Takida, Liao, and
  Mitsufuji]{toyama2023}
Keisuke Toyama, Taketo Akama, Yukara Ikemiya, Yuhta Takida, Wei-Hsiang Liao,
  and Yuki Mitsufuji.
\newblock Automatic music transcription with hierarchical frequency-time
  transformer.
\newblock In \emph{Proceedings of the 24th International Society for Music
  Information Retrieval Conference}, pp.\  215--222, 2023.

\bibitem[Tzanetakis \& Cook(2002)Tzanetakis and Cook]{tzanetakis2002gtzan}
George Tzanetakis and Perry Cook.
\newblock Musical genre classification of audio signals.
\newblock \emph{IEEE Transactions on speech and audio processing}, 10\penalty0
  (5):\penalty0 293--302, 2002.

\bibitem[Tzanetakis et~al.(2007)Tzanetakis, Jones, and
  McNally]{tzanetakis2007stereo}
George Tzanetakis, Randy Jones, and Kirk McNally.
\newblock Stereo panning features for classifying recording production style.
\newblock In \emph{ISMIR}, pp.\  441--444, 2007.

\bibitem[Uhlich et~al.(2017)Uhlich, Porcu, Giron, Enenkl, Kemp, Takahashi, and
  Mitsufuji]{stefan2017improving}
Stefan Uhlich, Marcello Porcu, Franck Giron, Michael Enenkl, Thomas Kemp, Naoya
  Takahashi, and Yuki Mitsufuji.
\newblock Improving music source separation based on deep neural networks
  through data augmentation and network blending.
\newblock In \emph{Proc. of IEEE ICASSP}, pp.\  261--265, 2017.
\newblock \doi{10.1109/ICASSP.2017.7952158}.

\bibitem[van~den Oord et~al.(2017)van~den Oord, Vinyals, and
  Kavukcuoglu]{van2017neural}
A{\"a}ron van~den Oord, Oriol Vinyals, and Koray Kavukcuoglu.
\newblock Neural discrete representation learning.
\newblock In \emph{Proc. Advances in Neural Information Processing Systems
  (NeurIPS)}, pp.\  6306--6315, 2017.

\bibitem[Vanka et~al.(2023)Vanka, Safi, Rolland, and
  Fazekas]{vanka2023adoption}
Soumya~Sai Vanka, Maryam Safi, Jean-Baptiste Rolland, and George Fazekas.
\newblock Adoption of ai technology in the music mixing workflow: An
  investigation.
\newblock 2023.

\bibitem[Vanka et~al.(2024)Vanka, Steinmetz, Rolland, Reiss, and
  Fazekas]{vanka2024diff}
Soumya~Sai Vanka, Christian Steinmetz, Jean-Baptiste Rolland, Joshua Reiss, and
  George Fazekas.
\newblock Diff-{MST}: Differentiable mixing style transfer.
\newblock In \emph{ISMIR}, 2024.

\bibitem[Vaswani et~al.(2017)Vaswani, Shazeer, Parmar, Uszkoreit, Jones, Gomez,
  Kaiser, and Polosukhin]{vaswani2017attention}
Ashish Vaswani, Noam Shazeer, Niki Parmar, Jakob Uszkoreit, Llion Jones,
  Aidan~N Gomez, \L~ukasz Kaiser, and Illia Polosukhin.
\newblock Attention is all you need.
\newblock In I.~Guyon, U.~Von Luxburg, S.~Bengio, H.~Wallach, R.~Fergus,
  S.~Vishwanathan, and R.~Garnett (eds.), \emph{Advances in Neural Information
  Processing Systems}, volume~30. Curran Associates, Inc., 2017.
\newblock URL
  \url{https://proceedings.neurips.cc/paper_files/paper/2017/file/3f5ee243547dee91fbd053c1c4a845aa-Paper.pdf}.

\bibitem[Wang et~al.(2022)Wang, Luc, Wu, Recasens, Smaira, Brock, Jaegle,
  Alayrac, Dieleman, Carreira, et~al.]{wang2022slowfast}
Luyu Wang, Pauline Luc, Yan Wu, Adria Recasens, Lucas Smaira, Andrew Brock,
  Andrew Jaegle, Jean-Baptiste Alayrac, Sander Dieleman, Joao Carreira, et~al.
\newblock Towards learning universal audio representations.
\newblock In \emph{ICASSP 2022-2022 IEEE International Conference on Acoustics,
  Speech and Signal Processing (ICASSP)}, pp.\  4593--4597. IEEE, 2022.

\bibitem[Wilkins et~al.(2018)Wilkins, Seetharaman, Wahl, and
  Pardo]{wilkins2018vocalset}
Julia Wilkins, Prem Seetharaman, Alison Wahl, and Bryan Pardo.
\newblock Vocalset: A singing voice dataset.
\newblock In \emph{ISMIR}, pp.\  468--474, 2018.

\bibitem[Wu* et~al.(2023)Wu*, Chen*, Zhang*, Hui*, Berg-Kirkpatrick, and
  Dubnov]{laionclap2023}
Yusong Wu*, Ke~Chen*, Tianyu Zhang*, Yuchen Hui*, Taylor Berg-Kirkpatrick, and
  Shlomo Dubnov.
\newblock Large-scale contrastive language-audio pretraining with feature
  fusion and keyword-to-caption augmentation.
\newblock In \emph{IEEE International Conference on Acoustics, Speech and
  Signal Processing, ICASSP}, 2023.

\bibitem[Xi et~al.(2018)Xi, Bittner, Pauwels, Ye, and Bello]{GuitarSet2018}
Qingyang Xi, Rachel~M Bittner, Johan Pauwels, Xuzhou Ye, and Juan~Pablo Bello.
\newblock Guitarset: A dataset for guitar transcription.
\newblock In \emph{ISMIR}, pp.\  453--460, 2018.

\bibitem[Yamamoto et~al.(2022)Yamamoto, Nam, and Terasawa]{YamamotoNT22}
Yuya Yamamoto, Juhan Nam, and Hiroko Terasawa.
\newblock Deformable {CNN} and imbalance-aware feature learning for singing
  technique classification.
\newblock In Hanseok Ko and John H.~L. Hansen (eds.), \emph{Interspeech 2022,
  23rd Annual Conference of the International Speech Communication Association,
  Incheon, Korea, 18-22 September 2022}, pp.\  2778--2782. {ISCA}, 2022.
\newblock \doi{10.21437/INTERSPEECH.2022-11137}.
\newblock URL \url{https://doi.org/10.21437/Interspeech.2022-11137}.

\bibitem[Yang et~al.(2023)Yang, Tian, Tan, Huang, Liu, Chang, Shi, Zhao, Bian,
  Wu, Zhao, Watanabe, and Meng]{yang2023uniaudio}
Dongchao Yang, Jinchuan Tian, Xu~Tan, Rongjie Huang, Songxiang Liu, Xuankai
  Chang, Jiatong Shi, Sheng Zhao, Jiang Bian, Xixin Wu, Zhou Zhao, Shinji
  Watanabe, and Helen Meng.
\newblock Uniaudio: An audio foundation model toward universal audio
  generation, 2023.

\bibitem[Yuan et~al.(2023)Yuan, Ma, Li, Zhang, Chen, Yin, Zhuo, Liu, Huang,
  Tian, et~al.]{yuan2023marble}
Ruibin Yuan, Yinghao Ma, Yizhi Li, Ge~Zhang, Xingran Chen, Hanzhi Yin, Le~Zhuo,
  Yiqi Liu, Jiawen Huang, Zeyue Tian, et~al.
\newblock Marble: Music audio representation benchmark for universal
  evaluation.
\newblock \emph{arXiv preprint arXiv:2306.10548}, 2023.

\bibitem[Zeghidour et~al.(2022)Zeghidour, Luebs, Omran, Skoglund, and
  Tagliasacchi]{zeghidour2022soundstream}
Neil Zeghidour, Alejandro Luebs, Ahmed Omran, Jan Skoglund, and Marco
  Tagliasacchi.
\newblock Soundstream: An end-to-end neural audio codec.
\newblock \emph{{IEEE/ACM} Trans. Audio, Speech, Lang. Process.}, 30:\penalty0
  495--507, 2022.

\bibitem[Zhong et~al.(2023{\natexlab{a}})Zhong, Hirano, Shimada, Tateishi,
  Takahashi, and Mitsufuji]{zhong2023tagging}
Zhi Zhong, Masato Hirano, Kazuki Shimada, Kazuya Tateishi, Shusuke Takahashi,
  and Yuki Mitsufuji.
\newblock An attention-based approach to hierarchical multi-label music
  instrument classification.
\newblock In \emph{ICASSP 2023-2023 IEEE International Conference on Acoustics,
  Speech and Signal Processing (ICASSP)}, pp.\  1--5. IEEE, 2023{\natexlab{a}}.

\bibitem[Zhong et~al.(2023{\natexlab{b}})Zhong, Shi, Hirano, Shimada, Tateishi,
  Shibuya, Takahashi, and Mitsufuji]{zhong2023extending}
Zhi Zhong, Hao Shi, Masato Hirano, Kazuki Shimada, Kazuya Tateishi, Takashi
  Shibuya, Shusuke Takahashi, and Yuki Mitsufuji.
\newblock Extending audio masked autoencoders toward audio restoration.
\newblock In \emph{2023 IEEE Workshop on Applications of Signal Processing to
  Audio and Acoustics (WASPAA)}, pp.\  1--5, 2023{\natexlab{b}}.
\newblock \doi{10.1109/WASPAA58266.2023.10248171}.

\end{thebibliography}
\bibliographystyle{tmlr}

\appendix
\section{Related Work}
\label{sec:app_related_works}

\subsection{Understanding models}
Understanding models based on supervised learning (SL) has shown good performance on music tagging tasks~\citep{zhong2023tagging, koutini2022patchout, McCallumKOGE22mule}, but they have difficulties into addressing tasks that involve unseen annotations during inference~\citep{McCallumKOGE22mule, ntt2022mae}.

Self-supervised learning (SSL), however, learns features without human annotation. For example, contrastive learning reflects data similarity and dissimilarity in the learned representation~\citep{janne2021clmr, McCallumKOGE22mule}. Masked reconstruction asks the model to predict the missing data and has been widely applied to language~\citep{Delvin2018bert}, speech~\citep{hsu2021hubert}, sound~\citep{ntt2022mae}, and music tasks~\citep{chen2023pac}.
MERT~\citep{li2023mert} further extended masked reconstruction to both acoustic and music features.

SSL methods achieved high performance across a wider range of tasks than SL methods. While a comprehensive set of downstream tasks, including music tagging~\citep{li2023mert, McCallumKOGE22mule, ntt2022mae}, music source separation~\citep{chen2023pac, li2023mert} and music bandwidth extension~\citep{zhong2023extending}, has been examined, some tasks related to music production, such as music transcription and music mixing, have not been well-investigated. 

\subsection{Generative models}
Most understanding models are encoder-only models, in which generation is not covered. In contrast, generative models have the ability of generation as well as learning representation~\citep{openai2020imagegpt,dhariwal2020jukebox}. 
Some generative models are based on auto-regressive modeling, thus can execute not only generation but tasks such as music continuation~\citep{dhariwal2020jukebox, agostinelli2023musiclm, copet2023musicgen}. Recent diffusion-based models can even execute music inpainting~\citep{Forsgren_Martiros_2022riffusion, li2023jen1, huang2023noise2music} and music style transfer~\citep{liu2023audioldm}. Multi-tasking can also be achieved by task-augmentation~\citep{li2023jen1} or explicit design~\citep{yang2023uniaudio}. Generation can be conditioned by text, lyrics, and melody~\citep{copet2023musicgen, agostinelli2023musiclm, dhariwal2020jukebox}. The use of generative models greatly extends the coverage of applicable music downstream tasks.

\subsection{Neural tokenizer in generative models}
Vector quantization (VQ)~\citep{van2017neural} has been a promising step for music generative modeling. Tokens obtained by VQ can be better characterized with generative models than raw signals. Building music foundation models on top of tokenizers has become a mainstream approach~\citep{agostinelli2023musiclm,copet2023musicgen}.

VQ-VAE-2~\citep{razavi2019generating} first extended VQ learning to hierarchical discrete representation learning in computer vision, which prompted the emergence of the pioneering music generation model~\citep{dhariwal2020jukebox}.
It was shown to learn global and local information in top and bottom levels, respectively.
Another variant of VQ aimed at learning structured discrete representation is residual VQ, which assigns multiple codes to encoded vectors in a residual manner~\citep{zeghidour2022soundstream,lee2022autoregressive}. Residual VQ was initially proposed for neural audio codec~\citep{zeghidour2022soundstream,defossez2023encodec} and shown to result in coarse-to-fine representation~\citep{lee2022autoregressive}. 

HQ-VAE was proposed to encompass two advanced VQ schemes including RVQ within the variational Bayes framework~\citep{takida2023hq-vae}. The unified scheme enhances the codebook utilization of the VQ-based models due to the effect of self-annealing~\citep{takida2022sq-vae}. The authors constructed a hierarchical latent space with three levels and jointly trained the levels on an audio dataset. The model achieved local-to-global representation similar to VQ-VAE-2, thus being freed from layer collapse.

\section{Architecture of {\mfm}}
\label{sec:app_details_of_mfm}
\mfm is a generative model that generates music samples conditioned on given text prompts. We train the model using an \textcolor{c_revise}{internal dataset contains around 115k studio-quality library music tracks sampled at 44.1 kHz. Their lengths vary from 30s to 150s, their tempos vary between 50bpm to 200bpm, and the total length is around 4,000h. 90\% of the dataset is non-vocal. Although there are more than 50 genres included in the dataset, it is biased toward orchestral and western music.} To generate music samples, we use the three priors in \mfm consecutively from the top level to the bottom level. It is important to note that during music generation, we use the token sequences sampled from top and middle priors for conditioning. For training and feature extraction, however, we rely on ground-truth tokens. 

Given a text prompt, we first obtain the CLAP embedding $\bm{y}_{text} = \text{CLAP.text\_embedding}(x)$. Conditioned on $\bm{y}_{text}$, the top prior transformer generates a token sequence $\bm{Z}_1$ in an auto-regressive manner. Subsequently, $\bm{Z}_1$ is fed into the middle conditional transformer along with $\bm{y}_{text}$ to generate $\bm{Z}_2$. The length of $\bm{Z}_2$ is four times that of $\bm{Z}_1$.
Similarly, we generate $\bm{Z}_3$ using the bottom conditional prior by conditioning it with $\bm{Z}_1$, $\bm{Z}_2$ and $\bm{y}_{text}$. The length of $\bm{Z}_3$ is once again four times that of $\bm{Z}_2$. 
Finally, all the three token sequences $\bm{Z}_{1:3}$ are fed into the decoder of the stage-1 model for audio reconstruction.

\subsection{Stage-1 model}
\label{sec:app_details_of_mfm_stage1}

The stage-1 model for \mfm is based on the aforementioned SQ-VAE-2 structure, as illustrated in Figure \ref{fig:mfm_overall}(a).
It is a three-level SQ-VAE-2 (i.e., $L=3$) autoencoder.
It comprises three encoding blocks, denoted as $encoder_{bottom}$, $encoder_{middle}$, and $encoder_{top}$, where the \textit{bottom} layer processes the input audio sampled at 44.1 kHz.
Each encoding block consists of 1-D convolutional layers with a strided convolution for down-sampling. The down-sampling ratios are set to 8, 4, and 4 for $encoder_{bottom}$, $encoder_{middle}$, and $encoder_{top}$, respectively. The stage-1 model has three decoding blocks, $decoder_{bottom}$, $decoder_{middle}$, and $decoder_{top}$. Each is a mirrored version of the encoding block with the same resolution.
We train the stage-1 model on top of the HQ-VAE framework on $4,000$h of $44.1$-kHz studio-quality library music.
The token sequence generated by the stage-1 model is used to train the prior $P(\bm{Z}_{1:3})$, as presented in the subsequent section.

\subsection{Stage-2 model}
\label{sec:app_details_of_mfm_stage2}

For the stage-2 model, we followed Jukebox \citep{dhariwal2020jukebox}, which generates hierarchical tokens with the same down-sampling rates as the stage-1 model (i.e, a token in the top/middle/bottom level compresses 128/32/8 audio samples).
Specifically, we train three transformers in an auto-regressive manner to model $P(\bm{Z}_1|\bm{y}_{\text{audio}})$, $P(\bm{Z}_2|\bm{Z}_1, \bm{y}_{\text{audio}})$, and $P(\bm{Z}_3|\bm{Z}_{1:2}, \bm{y}_{\text{audio}})$, which we call the top prior, middle conditional prior, and bottom conditional prior, respectively.

Initially, we train a sparse transformer to learn the probability distribution of top-level token sequence $\bm{Z}_1$ in an auto-regressive manner, conditioned on $\bm{y}_{\text{audio}}$ (i.e.,  $P_{\bm{\pi}_1}(\bm{Z}_1 | \bm{y}_{\text{audio}})$). 
Subsequently, we train other sparse transformers to model $P_{\bm{\pi}_i}(\bm{Z}_l | \bm{Z}_{<l}, \bm{y}_{\text{audio}})$. These transformers are conditioned by token sequences from upper levels.
We used similar configurations of Jukebox's transformers, but modified some hyperparameters, as shown in Table \ref{tab:prior_comparison_vs_jukebox}. The primary distinction between them lies in the conditioning mechanism for each prior. We modified conditioning modules to use $\bm{y}_{\text{audio}}$, instead of the original conditioning inputs employed in Jukebox such as artists, genres, and lyrics. 
In contrast to Jukebox, the bottom conditional transformer in our model is conditioned on all the upper token sequences $\bm{Z}_{<l}$, diverging from Jukebox's approach, where the $l^{th}$ transformer is conditioned on the adjacent upper token sequences $\bm{Z}_{l-1}$.
This difference is a result of the hierarchical structure obtained with SQ-VAE-2, which has a tight interrelation between different levels, unlike Jukebox's independently trained multi-level token sequences.

\begin{table}[t!]
\centering
\caption{Hyper-parameter comparison on Jukebox's and \mfm's top-level prior. Hyper-parameter marked with * indicates that we used different setting from Jukebox. Otherwise we used same configuration.}
\footnotesize
\begin{tabular}{lccccccc}
    \toprule
    & \multicolumn{3}{c}{Jukebox's} &\phantom{a}& \multicolumn{3}{c}{\mfm's} \\ \cmidrule{2-4} \cmidrule{6-8}
    & Top & Middle & Bottom && Top & Middle & Bottom \\
    \midrule
    Sample length 
    & 1048576 & 262144 & 65536 
    && 1048576 & 262144 & 65536 \\
    Context length 
    & 8192 & 8192 & 8192 
    && 8192 & 8192 & 8192\\
    Transformer width 
    & 4800 & 1920 & 1920 
    && 4800 & 3200 & 2880\\
    Transformer self-attention layers 
    & 72 & 72 & 72 
    && 72 & 72 & 72 \\
    Attention heads
    & 8 & 1 & 1
    && 8 & 4 & 4\\ 
    Factorized attention shape & 
    (128, 64) & (128, 64) & (128, 64) &&
    (128, 64) & (128, 64) & (128, 64) \\
    Encoder-decoder attention layers 
    & 7 & - & - 
    && 7 & - & - \\
    Up-sampling modules 
    & - & 1 & 1
    && - & 1 & 2\\
    Up-sampling-module residual block width 
    & - & 1024 & 1024 
    && - & 1024 & 1024 \\
    Up-sampling modules residual blocks 
    & - & 16 & 16 
    && - & 16 & 16 \\
    Up-sampling-module conv filter size 
    & - & 3 & 3 
    && - & 3 & 3 \\ 
    Up-sampling-module conv channels 
    & - & 1024 & 1024 
    && - & 1024 & 1024 \\
    Up-sampling-module dilation growth rate 
    & - & 3 & 3 
    && - & 3 & 3 \\
    Up-sampling-module dilation cycle 
    & - & 8 & 8 
    && - & 8 & 8 \\
    Initialization scale 
    & 0.002 & 0.004 & 0.008
    && 0.002 & 0.004 & 0.004\\
    Encoder initialization scale 
    & 0.014 & - & -  
    && 0.014 & - & - \\
    Batch size$^*$ 
    & 512 & 192 & 184
    && 240 & 240 & 240 \\
    Training step$^*$ 
    & 310500 & 265000 & 279000
    && 88000 & 152000 & 272000 \\
    Learning rate$^*$ 
    & 0.00015 & 0.0003 & 0.0003 
    && 0.0001 & 0.0002 & 0.0002 \\
    Adam $\beta_2$ 
    & 0.925 & 0.95 & 0.95 
    && 0.925 & 0.95 & 0.95\\
    Weight decay 
    & 0.002 & 0.01 & 0.01
    && 0.002 & 0.01 & 0.01\\
    \bottomrule
\end{tabular}
\label{tab:prior_comparison_vs_jukebox}
\end{table}

We explain the entire conditioning mechanism using the bottom conditional prior as an example, which allows us to address all the detail (see Figure \ref{fig:mfm_overall}).
We first obtain the hierarchical discrete representations $\bm{Z}_1$, $\bm{Z}_2$, and $\bm{Z}_3$ by applying the pre-trained stage-1 model to an input musical audio $x$. The goal of training is to make the model execute next-token prediction on $\bm{Z}_3$, conditioned on $\bm{Z}_1$, $\bm{Z}_2$ and the CLAP embedding $\bm{y}_{audio}=CLAP.audio\_embedding(x)$.
We condition the transformer in a frame-by-frame manner. Since each level has a different time resolution, each token sequence from upper level is first converted into embedding sequence and then up-sampled to the next time resolution by a transposed convolution. 
We use two up-sampling modules: one for $\bm{Z}_1 \rightarrow \bm{Z}_2$ and the other for $\bm{Z}_1, \bm{Z}_2 \rightarrow \bm{Z}_3$.
The conditioning module $\bm{Z}_1 \rightarrow \bm{Z}_2$ up-samples the embedding sequence $\bm{Z}_1$ to match the resolution of $\bm{Z}_2$. Similarly, up-sampling module $\bm{Z}_1, \bm{Z}_2 \rightarrow \bm{Z}_3$ takes $\bm{Z}_2$ as input, along with the up-sampled embeddings of $\bm{Z}_1$, generating further up-sampled embeddings tailored to $\bm{Z}_3$'s resolution. The resulting frame-level embedding is used to condition the transformer for $\bm{Z}_3$.

For the actual next-token prediction task, we shift the token sequence $\bm{Z}_3$ by one position to the right and embed each token to a continuous vector using an embedding layer.
The first token, which is empty, is optionally filled with $\bm{y}_{audio}$ to make the system conditioned on the CLAP embedding.
With the frame-level aggregated conditioning vectors, namely, up-sampled embeddings from the upper layers, $\bm{y}_{audio}$, and the time positional embeddings, the transformer is trained to estimate $\bm{Z}_3$ in an auto-regressive manner.

\subsection{Evaluation on Music Generation}
\label{sec:app_eval_music_generation}

For evaluation, we used the MusicCaps dataset \citep{agostinelli2023musiclm}, which includes 5,521 \footnote{We used a subset of 5,514 samples due to the presence of expired or invalid URLs in MusicCaps items. Efforts were made to retrieve the missing files; however, three samples are absent despite our efforts.} pairs of audio clips and their text captions written by musicians.
Subsequently, we used \mfm to generate $10$-s audio samples, conditioned upon textual captions of MusicCaps.
Following the objective evaluation methodology\citep{copet2023musicgen}, we evaluated \mfm's performance on MusicCaps using three objective metrics: the Fréchet audio distance (FAD) \citep{kilgour2019frechet}, Kullback-Leibler Divergence (KL), and CLAP \citep{laionclap2023} score. 

Table \ref{tab:ttm} compares \mfm in terms of the three metrics with other SOTA auto-regressive methods, namely MusicLM \citep{agostinelli2023musiclm}, MeLoDy \citep{lam2023efficient}, and MusicGen \citep{copet2023musicgen}. 

We used FAD to measure the overall quality of the generated samples. It is a reference-free metric that compares embedding statistics computed on an evaluation set with embedding statistics computed on the clean dataset. A lower FAD indicates a higher degree of similarity between the embedding statistics from generated audio samples and the embedding statistics from the ground-truth (i.e., MusicCaps).
We used the official VGGish based-FAD computation module\footnote{\url{https://github.com/google-research/google-research/tree/master/frechet_audio_distance}} to extract embeddings from audio.
Since VGGish was trained using $16$-kHz audio, all the audio data were resampled to $16$ kHz.

We also computed KL-divergence, which compares the two probability distributions of labels. The labels are estimated by applying a pre-trained audio classifier to the generated and original audio clips. Following \citep{copet2023musicgen}, we used a SOTA audio classifier called PaSST. A lower KL-divergence indicates that audio content generated from a model aligns more closely with the characteristics of the reference. We report the mean of KL-divergence. 

Finally, we calculated the CLAP consistency loss to provide a consistent benchmark aligned with SOTA methodologies, even though it might not be entirely fair since \mfm was trained using CLAP embeddings. We computed the cosine similarity between the audio CLAP embedding from the generated clip and text CLAP embedding from the original text caption from MusicCaps. A higher CLAP score indicates a stronger alignment between the generated clip and original text description.

\mfm achieved an FAD (denoted as \textsc{Fad}$_{vgg}$) of 4.98, a performance comparable to that of the other models.
We only report the KL-divergence computed using PaSST as the audio classifier in Table \ref{tab:ttm}. \mfm attained a KL Divergence (denoted as \textsc{KL}) of 1.45, which is slightly higher than MusicGen-large. 
Finally, we report the CLAP consistency score (denoted as \textsc{Clap}$_{scr}$). \mfm's CLAP score was the highest, but it should be noted that \mfm was trained using CLAP embeddings, as we mentioned earlier. 

While the evaluation result of \mfm is slightly inferior to the SOTA, it is important to note that \mfm generates audio with a higher sampling rate. Table \ref{tab:ttm} shows the target sampling rate of each model denoted as \textsc{Target SR}. While \mfm generated $44.1$ kHz directly, the other models generated audio with lower sampling rates. Furthermore, the evaluation protocol requires \mfm to be evaluated in a much lower sampling rate. 
Despite this challenge, we designed \mfm to operate at a studio quality level (i.e., $44.1$ kHz), enabling its potential use in most downstream tasks, which we describe in the following sections.

\begin{table}[t!]
  \caption{Objective evaluation of \mfm and other auto-regressive generative models on MusicCaps. Model marked with * indicates that we generated audio samples and computed metrics using same protocol. \label{tab:ttm}}
  \centering
  \footnotesize
  \begin{tabular}{lrrrr}
    \toprule
    \multicolumn{1}{c}{} & \multicolumn{4}{c}{\footnotesize\textsc{MusicCaps} Test Set}\\ \cmidrule{2-5}
    \text{Model} & \text{Target SR} & \textsc{Fad}$_{vgg} \downarrow$ & \textsc{Kl} $\downarrow$ & \textsc{Clap}$_{scr} \uparrow$\\
    \midrule
    MusicLM & 24kHz & 4.0 & - & - \\
    MeLoDy & 24kHz & 5.41 & - & - \\
    MusicGen-large & 32kHz & 3.8 & 1.22 & 0.31 \\
    MusicGen-large (public \tablefootnote{Publicly released MusicGen-large trained on non-vocal dataset: \url{https://github.com/facebookresearch/audiocraft/blob/main/model_cards/MUSICGEN_MODEL_CARD.md}, })$^*$ & 32kHz & 5.88 & 1.41 & 0.27 \\
    \midrule
    \mfm$^*$ & 44.1kHz & 4.98 & 1.45 & 0.39 \\
    \bottomrule
  \end{tabular}

\end{table}

\section{Music Tagging}
\label{sec:app_mir}
We implemented the attention-based aggregator described in Section~\ref{ssec:dt_pp_1} as a 1-layer standard transformer. 
There is only single attention head in the attention layer. Features from different priors are processed by an input normalization layer then converted to $768$ dimensions with a linear layer.
Similar to \citet{CastellonDL21calm}, we conducted a grid search for the method of input normalization (BatchNorm or LayerNorm), and the dropout ratio in MLP (0.10, 0.25, 0.50, 0.75). For all single-label tasks, by default we set the label smoothing to $0.1$. The batch size was set to $256$ for MTAT and Nsynth for their large amounts of data, and 64 for others. Unless specifically mentioned, the learning rate was 5e$^{-5}$.
\begin{table}
    \caption{Datasets for music tagging. Except MTAT (multi-label) and EmoMusic (2-axis regression), all other tasks are single-label. Segment length of VocalSet is after pre-processing.}
    \label{tab:mir_datasets}
    \centering
    \footnotesize
    {
    \begin{tabular}{ccccccc}
    \toprule
    Dataset & Task & Num. of Classes & Num. of Segments & Segment Length & Sampling Rate\\
    \midrule
    MTAT & Auto tagging & 50 &  $\sim$ 25~k & 29~s & 16~kHz\\
    \multirow{2}*{Nsynth} & Pitch & 128 & \multirow{2}*{$\sim$306~k} & \multirow{2}*{4~s} & \multirow{2}*{16~kHz}\\
    ~ & Instrument & 11 & ~ & ~ & ~\\
    EmoMusic & Emotion regression & 
n/a & $\sim$740 & 45~s & 44.1~kHz\\
    GiantSteps & Key estimation & 24 & $\sim$2.1~k & 120~s & 44.1~kHz\\
    GTZAN & Genre & 10 & $\sim$900 & 30~s & 22.05~kHz\\
    \multirow{2}*{VocalSet} & Singer & 20 & \multirow{2}*{$\sim$7.5~k} & \multirow{2}*{3~s} & \multirow{2}*{44.1~kHz} \\
    ~ & Vocal technique & 10 & ~ & ~ & ~ \\
    \bottomrule
    \end{tabular}}
\end{table}
\subsection{Comparison of Features}
Since \mfm is trained with an upstream auto-regressive modeling task, where both CLAP-conditional and unconditional cases are introduced, it is still unclear which setting should be used for downstream tasks. We thus conducted a preliminary study with MTAT, a dataset designed for coarse-grained tagging tasks, and Nsynth-pitch, a dataset designed for fine-grained classification tasks. In this experiment, features from the top prior were aggregated by average pooling, and the learning rate was set to 1e$^{-4}$. As mentioned above, the CLAP audio encoder receives the same clip of music as \mfm and encodes it into a single feature vector. The results in table \ref{tab:mir_exp_preliminary} indicates that CLAP performs well for coarse concepts, while unconditional feature extraction results in better accuracy for pitch estimation. On top of them, the CLAP-conditional feature extraction achieved better scores in both tasks. We will thus use CLAP-conditional feature extraction as the default feature extraction for time-invariant retrieval tasks.

\begin{table}[tb]
    \caption{Preliminary study on impact of CLAP conditioning in music tagging. Acc.: accuracy. mAP: mean average precision (\textbf{bold}: top score).}
    \label{tab:mir_exp_preliminary}
    \centering
     \footnotesize
     {
    \begin{tabular}{ccc}
    \toprule
         Features & MTAT mAP & Nsynth-pitch Acc.\\
    \midrule
        CLAP & 39.6 & 48.8\\
         Unconditional \mfm & 39.0 & 90.6\\
         CLAP-conditional \mfm & \textbf{39.9} & \textbf{91.5}\\
    \bottomrule
    \end{tabular}}
\end{table}

\subsection{Investigation on Feature Aggregation}
Next, we conducted ablation studies to verify the effectiveness of the feature aggregation pipeline. The learning rate for attention-based and average pooling is 5e$^{-5}$ and 1e$^{-4}$, respectively. As shown in Table \ref{tab:mir_exp_ablation}, we have the following observations: (1) the feature aggregation pipeline outperformed the average pooing baselines; (2) applying the same pipeline to non-hierarchical features did not bring as much performance gain as the hierarchical case, showing the hierarchical features contain complementary information for music tagging. The bottom prior did not contribute to a better performance. We assume this is because the bottom prior processes high-frequency components which are less important for the tasks we tested. It is also worth mentioning that many SOTA classification models have been working with $F_s$ = 16 kHz, \textit{e.g.}, \cite{ntt2022mae, McCallumKOGE22mule, wang2022slowfast}, neglecting high frequency components.

\begin{table}[tb]
\caption{Ablation study on feature aggregation pipeline in music tagging (\textbf{bold}: top score)}
\label{tab:mir_exp_ablation}
    \centering
    \footnotesize
    \begin{tabular}{cccc}
    \toprule
        Features & Aggregation & MTAT ROC-AUC & mAP\\
    \midrule
        Top & Average pooling & 91.1 & 39.9\\
        Top & Attention & 91.1 & 40.4\\
        Mid. & Average pooling & 91.1 & 39.4\\
        Mid. & Attention & 91.3 & 40.9 \\
        Top + mid. & Average pooling & 91.5 & 40.6\\
        Top + mid. & Attention & \textbf{91.7} & \textbf{41.5}\\
    \bottomrule
    \end{tabular}
\end{table}
\section{Music Transcription}\label{sec:app_amt}

Music transcription requires accurate predictions of both pitch and temporal information for optimal performance. Current models~\citep{toyama2023, gardner2022} rely exclusively on spectrograms as input. 
There have been attempts in applying extra feature to music transcription. For example, \cite{donahue2022} used Jukebox for melody transcription. However, the HookTheory dataset used by \cite{donahue2022} does not include octave information, which makes it differ from the typical music transcription. 
There are two ways of modeling the typical music transcription task: piano-roll-based music transcription~\citep{hawthorne2018, kim2019adversarial, gardner2022, Cheuk_IJCNN2021, Thickstun2016LearningFO, kelz2019deep, toyama2023} and token-based music transcription~\citep{gardner2022}. The former transforms spectrograms into piano rolls in which both have the same time resolution. The latter transforms spectrograms into a series of tokens indicating the note pitch and note on and off locations. 
We selected the relatively well-established piano-roll-based music transcription to test the effectiveness of intermediate representation extracted from a music foundation model.

\subsection{Piano Transcription with Linear Layers}\label{sec: app_amt_linear_piano}
\begin{figure}[tb]
    \centering
    \includegraphics[width=0.95\textwidth,keepaspectratio,clip]{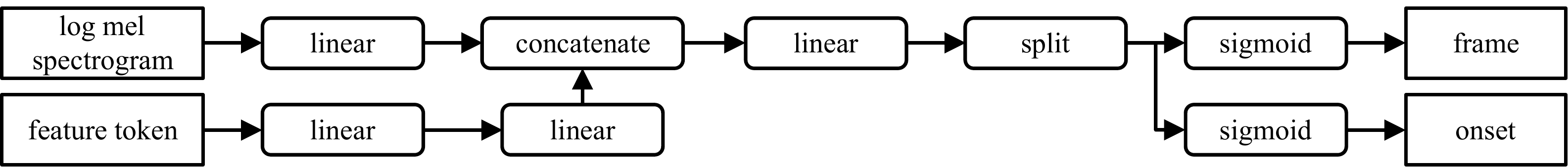}
    \caption{Model architectures of linear music transcription for piano}
    \label{fig:amt_architecture_simple}
\end{figure}

\begin{figure}[tb]
    \centering
    \subfloat[Linear transcription model using \mfm \textcolor{c_revise}{or Jukebox}]{
    \includegraphics[width=0.45\textwidth,keepaspectratio,clip]{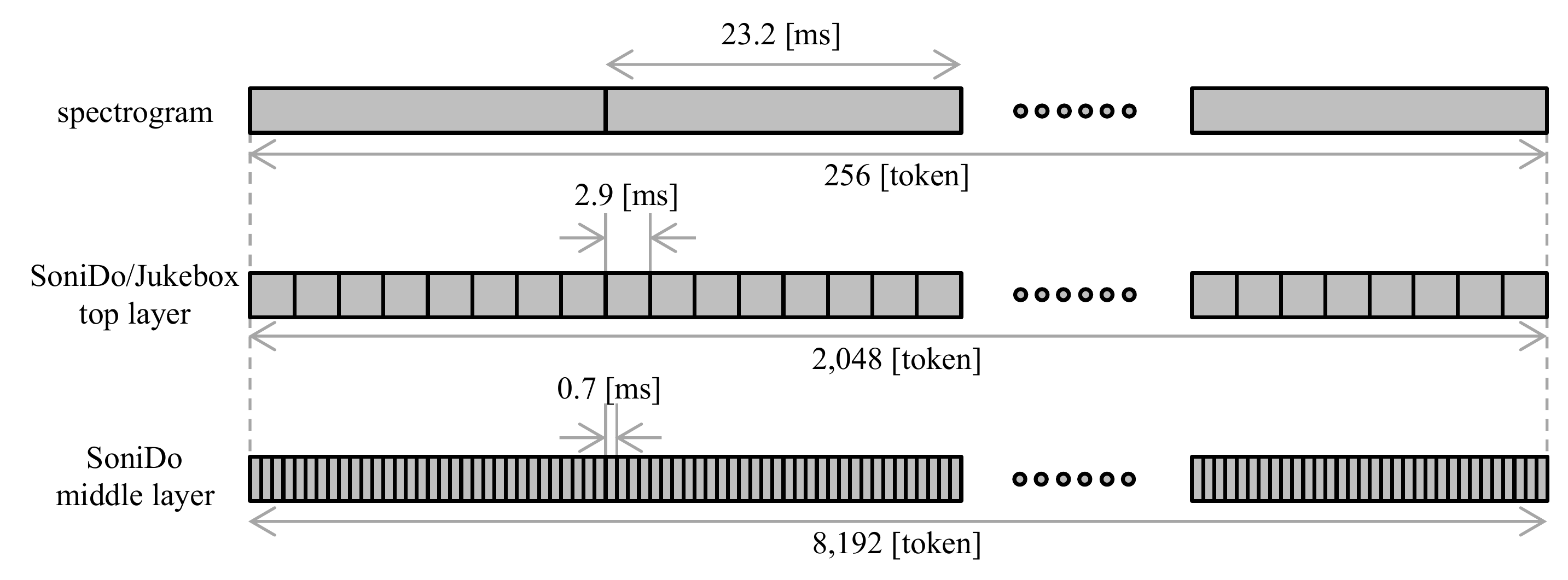}}
    \subfloat[hFT-Transformer using \mfm \textcolor{c_revise}{or Jukebox}]{
    \includegraphics[width=0.45\textwidth,keepaspectratio,clip]{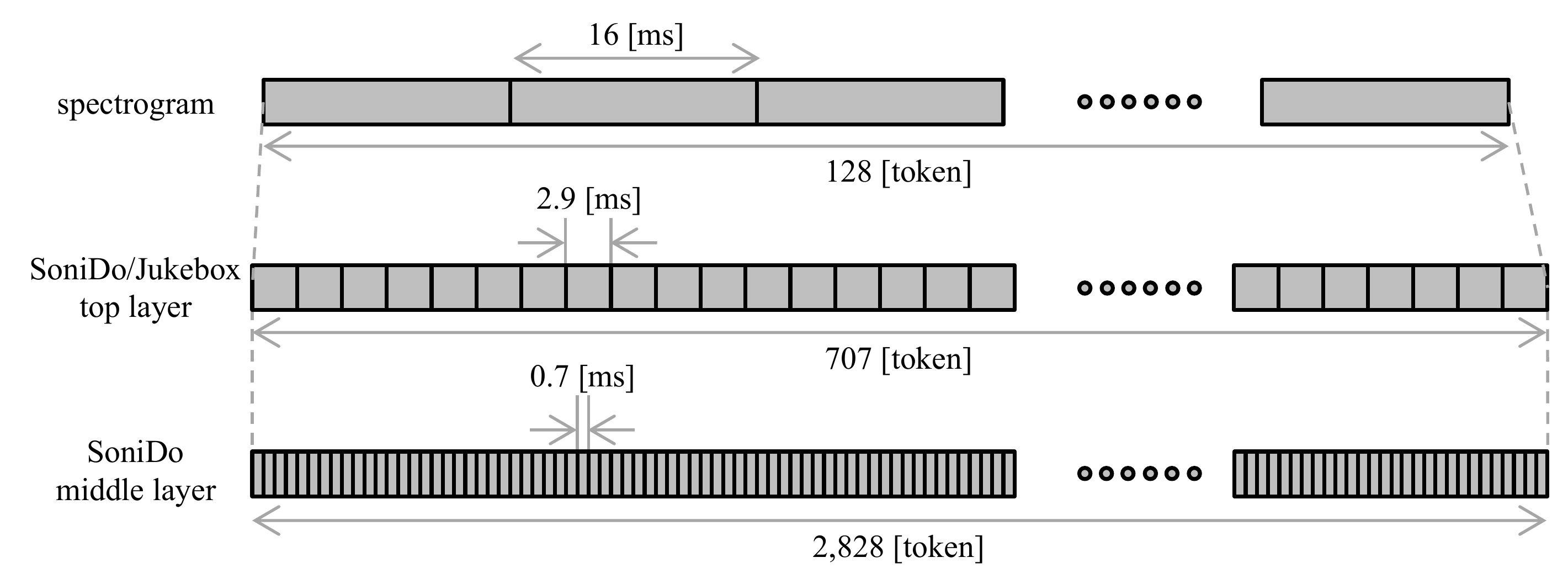}}
    \\
    \subfloat[Linear transcription model using \musicgen]{
    \includegraphics[width=0.45\textwidth,keepaspectratio,clip]{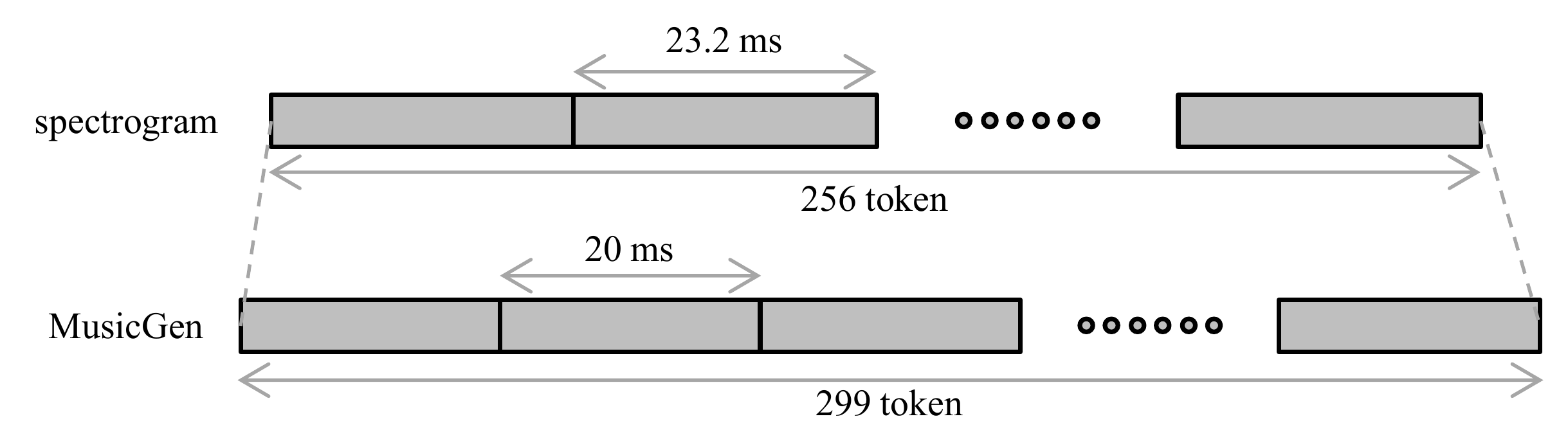}}
    \subfloat[hFT-Transformer using \musicgen]{
    \includegraphics[width=0.45\textwidth,keepaspectratio,clip]{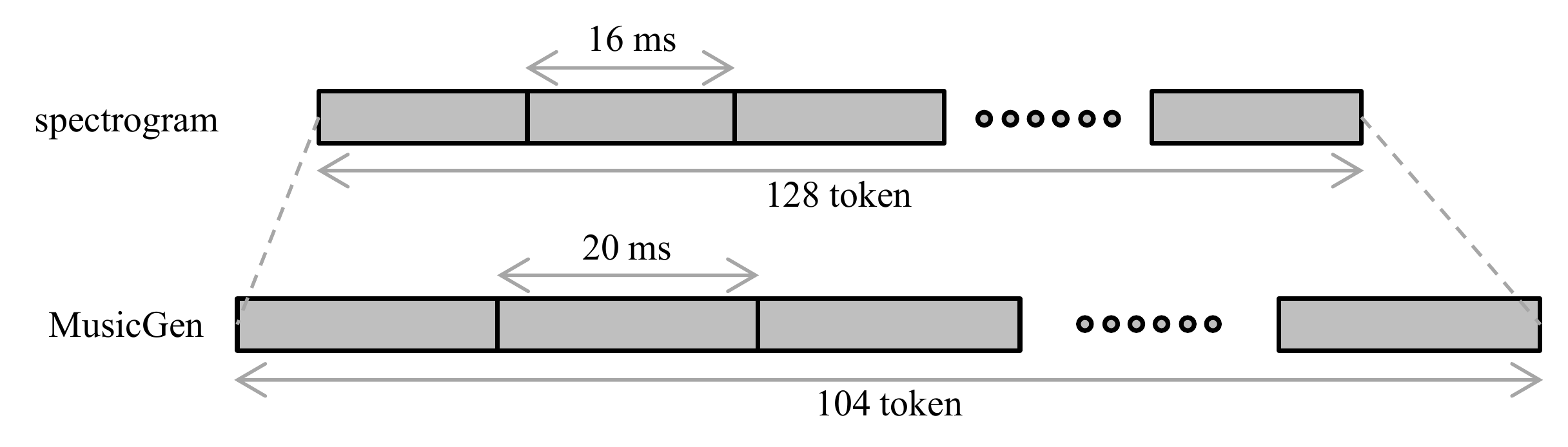}}
    \caption{Feature alignment schemes for different models in music transcription}
    \label{fig:amt_alignment}
\end{figure}

To understand the usefulness of the extracted intermediate features, we started with piano transcription by probing those features with linear layers. The architecture used for this experiment is shown in Figure~\ref{fig:amt_architecture_simple}.
The inputs are log mel spectrogram and the \mfm features.
Input audio signals are down-sampled to $22.05$ kHz then converted to a $256$-bin mel spectrogram using a $2048$-point Hann window with a $512$ hop size.
The alignment between the mel spectrogram and extracted features is illustrated in Figures \ref{fig:amt_alignment}(a) and (c), respectively.

The mel spectrogram is a 3D tensor ($B$, $N$, $F$), where $B=8$ is the batch size, $N=256$ is the number of frames, and $F=256$ is the number of frequency bins. A linear layer converts the tensor into ($B$, $N$, $Z$) along the last axis, where $Z=512$ is a common size for latent embeddings.
The \mfm, \musicgen\textcolor{c_revise}{, and Jukebox} features are formed as a 3D tensor ($B$, $V$, $G$), where $V$ is the number of tokens per $N$ frames (2,048 for the top layer of \mfm \textcolor{c_revise}{and for Jukebox}, 8,192 for the middle layer of \mfm, and 299 for \musicgen), and $G$ is the embedding size of features (4,800 for the top layer of \mfm \textcolor{c_revise}{and for Jukebox}, 3,200 for the middle layer of \mfm, 2,048 for \musicgenl, and 1,024 for \musicgens).
The first linear layer converts $G$ to $Z$, then the second linear layer converts $V$ to $N$ to make the tensor into the shape of ($B$, $N$, $Z$).
The tensors are then concatenated along the last axis. If the top layer of \mfm, middle layer of \mfm, or \musicgen is used, the dimension of the last axis is $2Z$, whereas if both layers of \mfm are used, the dimension is $3Z$.
The following linear layer is used to convert $2Z$ or $3Z$ to $2P$, where $P=88$ is the number of pitches.
Finally, the tensor is split into two ($B$, $N$, $P$) tensors followed by a sigmoid activation to indicate the estimated \textit{frame} and \textit{onset} information.
We followed~\citet{kong2021high} to extract the precise timing of onsets from datasets, in which we set the hyper-parameter to control the target sharpness $J$ to $3$.
The loss function for \textit{frame} and \textit{onset} is binary cross entropy.

We trained multiple models with the following input: (1) spectrogram only, (2) spectrogram and the top layer of the \mfm features, (3) spectrogram and the middle layer of the \mfm features, (4) spectrogram and both the top and middle layers of the \mfm features, (5) spectrogram and the \musicgenl features, (6) spectrogram and the \musicgens features\textcolor{c_revise}{, and (7) spectrogram and the Jukebox features}.
For each model, we trained for 50 epochs on one A100 graphics processing unit (GPU), using Adam~\citep{kingma2014adam} optimizer with a learning rate of 1e$^{-4}$.
\texttt{PyTorch} \texttt{ReduceLROnPlateu} was used for learning-rate scheduling with default parameters.
We chose to use the checkpoint with the highest F1 score in the validation split for each model.

As listed in Table \ref{tab:amt_linear_piano}, the model using the \mfm features, the \musicgen features\textcolor{c_revise}{, or the Jukebox features} outperformed the spectrogram-only baseline.
In particular, the contribution of \mfm's middle layer was higher than that of the top layer. These results suggest that the \mfm features are promising for music transcription tasks.

\subsection{Linear Instrument-agnostic Music Transcription}\label{app: amt_linear_IAAMT}
Following the setup described previously in \ref{sec: app_amt_linear_piano}, we also investigated the effectiveness of the extracted features with instrument-agnostic music transcription. We wanted to study if the features of music foundation model are also applicable to a multi-instrument transcription scenario.
In this experiment, we reused the model illustrated in Figure~\ref{fig:amt_architecture_simple} and trained it on the URMP dataset, which contains strings, woodwinds and brass instruments~\citep{URMP2019}.
Unlike piano transcription, we do not need to split the final output into frame and onset prediction for instrument-agnostic transcription. This is due to the fact that some musical instruments, such as violins and flutes, sometimes produce ambiguous onset attacks. Applying onset prediction to these instruments can be detrimental to the F1 score~\citep{cheuk2021reconvat}.

We validated the model performance on the Bach10 dataset~\citep{Bach10} and selected the best checkpoint based on the validation performance. We then evaluated the best checkpoint on GuitarSet~\citep{GuitarSet2018}, Su~\citep{su2016escaping}, and TRIOS~\citep{TRIOS2014} datasets.
Table~\ref{tab:iaamt} shows the F1 scores obtained using the best checkpoint.
The results indicate that features of music foundation model, regardless of \mfm or \musicgen, boost instrument-agnostic transcription performance compared with the baseline model, which uses only the spectrogram as the input features. This indicates that the model trained using these features has a stronger generalizability across different datasets. \mfm generally outperformed \musicgen on Bach10, GuitarSet, and TRIOS, while \musicgen performed better on Su. This difference may be due to the different datasets on which \mfm and \musicgen were trained. We will investigate this further in the future.

\begin{table}[tbp]
  \centering
  \footnotesize
  \caption{F1 scores on different datasets when best validation checkpoint was obtained using Bach10 as validation set.}
  \begin{tabular}{ccccccc}
    \toprule
    &\multicolumn{6}{c}{\text{Best validation checkpoint }} \\ 
    \cmidrule{2-7}
    \text{Dataset} & \text{Base} & \text{\musicgen Small} & \text{\musicgen Large} & \text{\mfm Top} & \text{\mfm Middle} & \text{\mfm Top+Middle} \\
    \hline
    Bach10 & 43.3 & 68.4 & 60.7 & 69.4 & \textbf{74.4} & 72.5 \\
    GuitarSet & 34.8 & 49.7 & 38.6 & 43.0 & \textbf{50.1} & 44.9 \\
    Su & 13.5 & \textbf{32.7} & 31.7 & 21.1 & 27.3 & 24.4 \\
    TRIOS & 20.8 & 38.2 & 31.4 & 34.9 & \textbf{40.7} & 39.4 \\
    \bottomrule
  \end{tabular}
  \label{tab:iaamt}
\end{table}

\subsection{hFT-Transformer}
\label{sec:app_amt_hft_transformer}
\begin{figure}[tb]
    \centering
    \includegraphics[width=0.95\textwidth,keepaspectratio,clip]{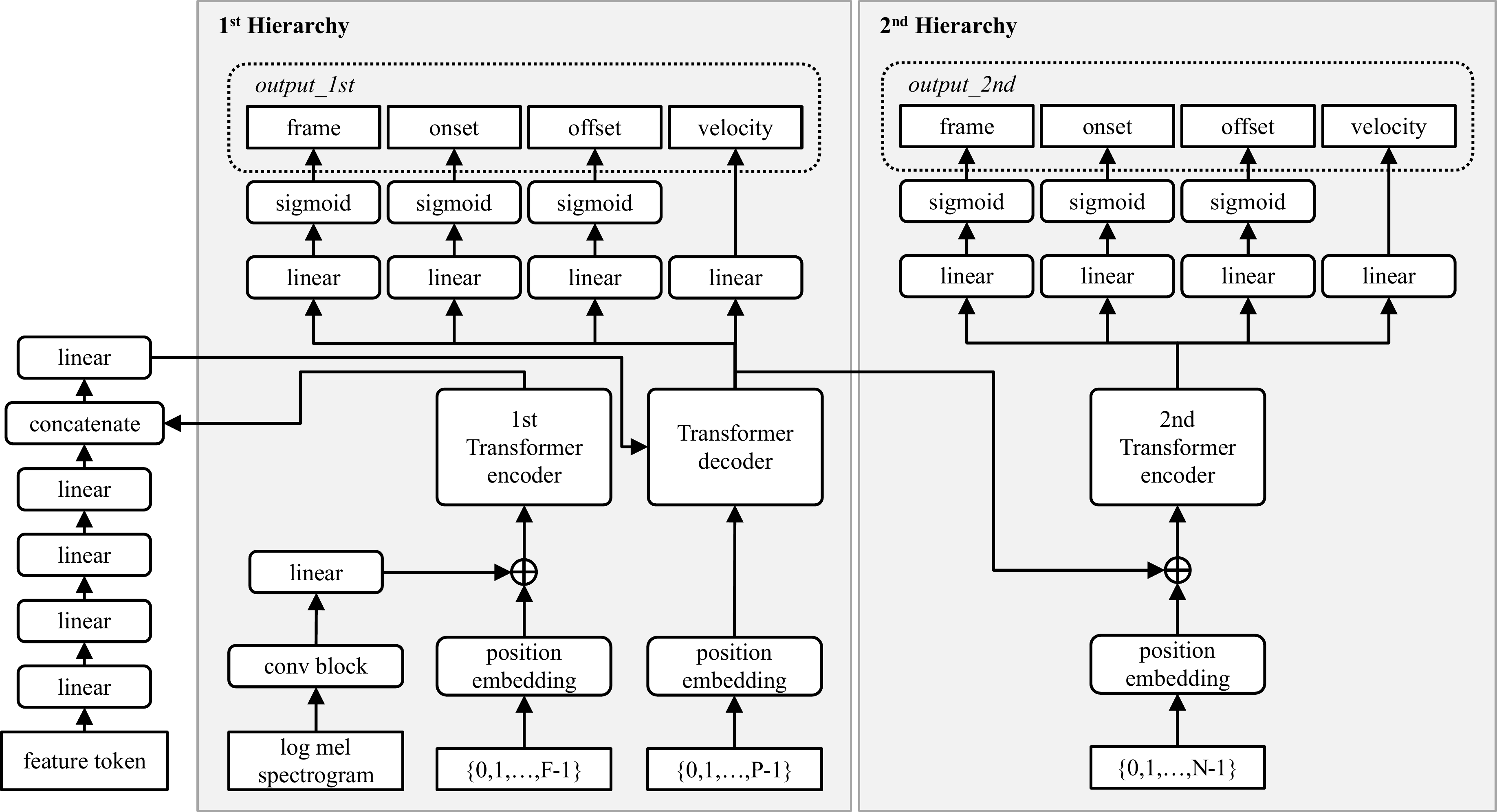}
    \caption{Modified version of hFT-Transformer for \mfm, \musicgen\textcolor{c_revise}{, and Jukebox} feature tokens}
    \label{fig:amt_architecture_hft_transformer}   
\end{figure}

The detailed model architecture of hFT-Transformer has been described in~\citep{toyama2023}.
Figures~\ref{fig:amt_alignment}(b) and (d) show the feature-spectrogram alignment scheme for hFT-Transformer.
The feature tokens and spectrogram cannot be perfectly aligned due to the different sampling frequencies in hFT-Transformer ($16$ kHz), \mfm ($44.1$ kHz), and \musicgen ($32$ kHz).
The $N$ frame in the spectrogram corresponds to roughly $707$ tokens of the top layer of the \mfm features \textcolor{c_revise}{and of the Jukebox features}, $2,828$ tokens of the middle layer of the \mfm features, and $104$ tokens of the \musicgen features.
Figure~\ref{fig:amt_architecture_hft_transformer} shows the modified hFT-Transformer that accepts the \mfm or \musicgen features as the additional input.
Following the setup in~\ref{sec: app_amt_linear_piano}, the \mfm and \musicgen feature tokens have three dimensions ($B$, $V$, $G$), where $B=8$ is the batch size, $V$ is the number of tokens per number of frames $N=128$,
 and $G$ is the embedding size of the feature tokens.
The first linear layer for the feature tokens reduces $G$ to $Z'=256$; the second linear layer reduces $V$ to $N$ then reshapes the tensor to ($B$, $N$, $F''=16$, $Z''=16$); the third linear layer changes $Z''$ to $Z=256$, then the last linear layer changes $F''$ to $F'=128$.
Thus, a tensor with shape ($B$, $N$, $F'$, $Z$) is obtained.
When using both the top and middle layers of the \mfm features, we form such tensors for each layer following the pipeline above.
The tensor(s) and output of the first encoder are then concatenated on the third axis.
Finally, the size of the concatenated tensor is reduced to $256$ ($F$ in~\citep{toyama2023}) by a linear layer.
These $F'$, $F''$, $Z'$, and $Z''$ were determined from preliminary experiments.

We trained the following models that have different inputs: (1) spectrogram only, (2) spectrogram and the top layer of the \mfm features, (3) spectrogram and the middle layer of the \mfm features, (4) spectrogram and both the top and middle layer of tthe \mfm features, (5) spectrogram and the \musicgenl features, (6) spectrogram and the \musicgens features, \textcolor{c_revise}{and (7) spectrogram and the Jukebox features}, the same as the experiment described in \ref{sec: app_amt_linear_piano}. We trained the models for 50 epochs on one A100 GPU.
For the other conditions, we followed~\citep{toyama2023}.
To confirm if \textcolor{c_revise}{these} features are useful when there are less training data, we train the models using 100, 50, 25, and 10\% of training data.
We chose the checkpoint with the highest F1 score in the validation split for further evaluation.
\begin{figure}
\centering
\subfloat[\centering Training Loss (100\%)]{{\includegraphics[width=0.49\textwidth]{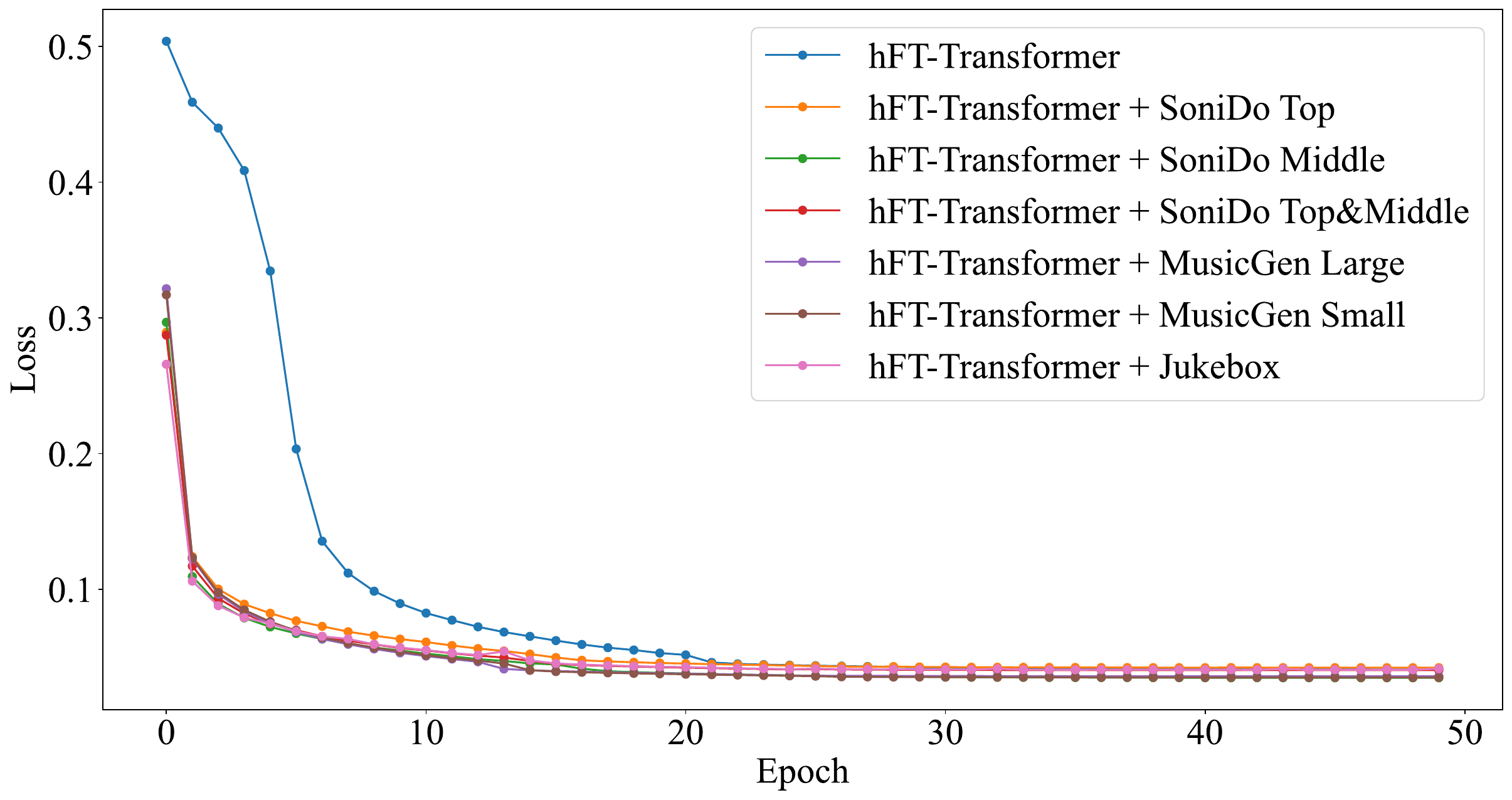}}}
\subfloat[\centering Validation Loss (100\%)]{{\includegraphics[width=0.49\textwidth]{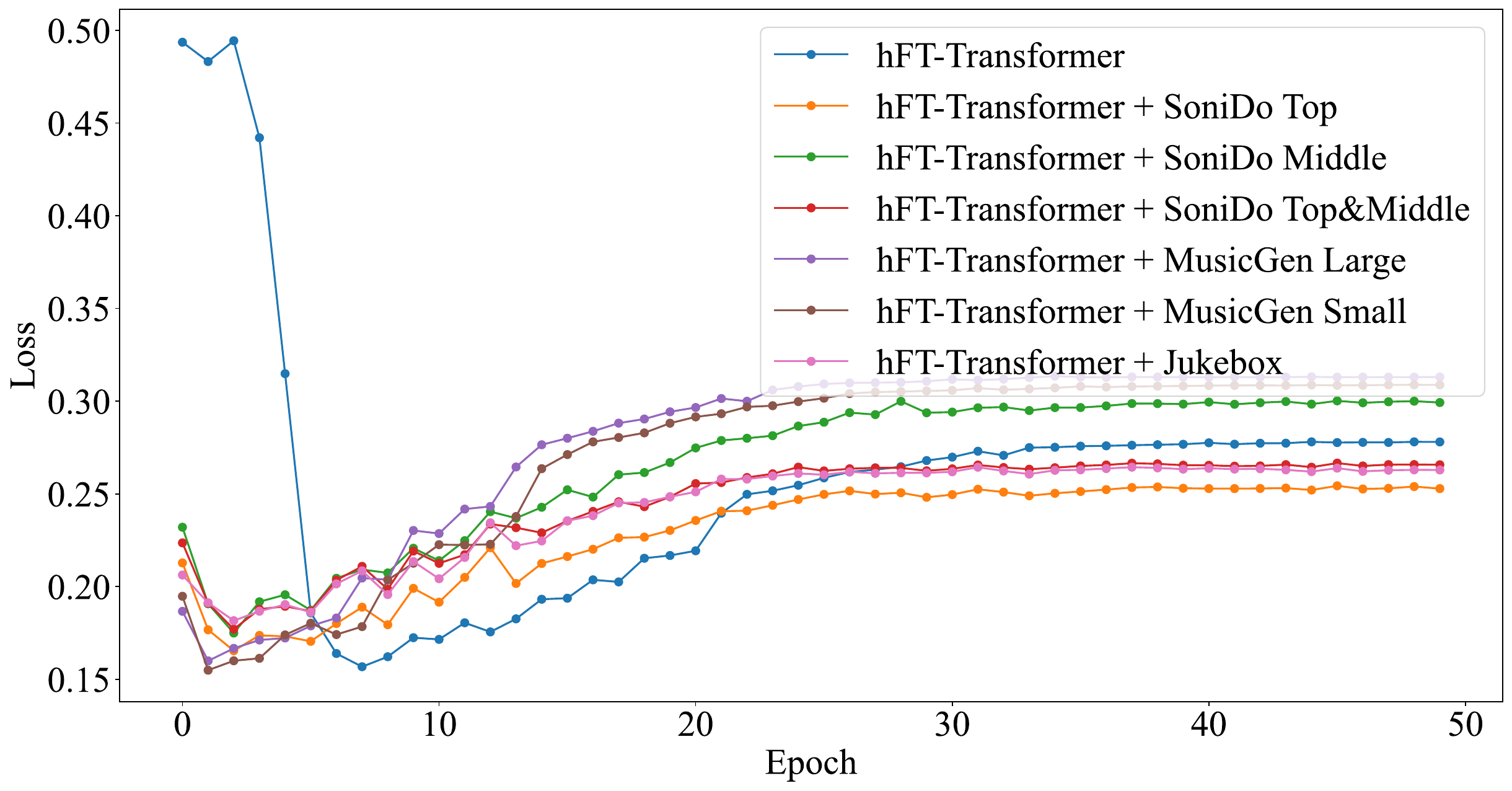}}}\\
\subfloat[\centering Training Loss (50\%)]{{\includegraphics[width=0.49\textwidth]{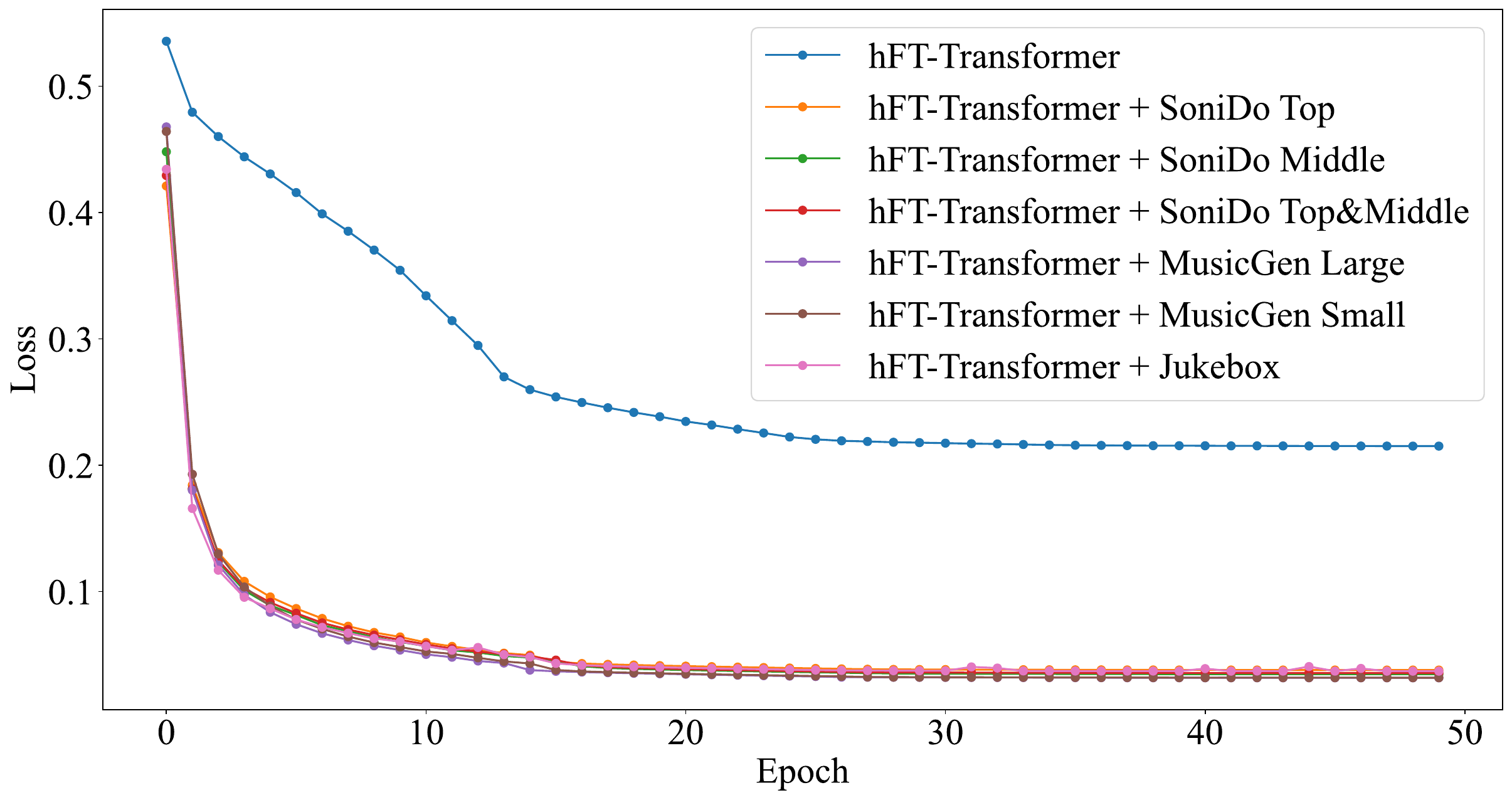}}}
\subfloat[\centering Validation Loss (50\%)]{{\includegraphics[width=0.49\textwidth]{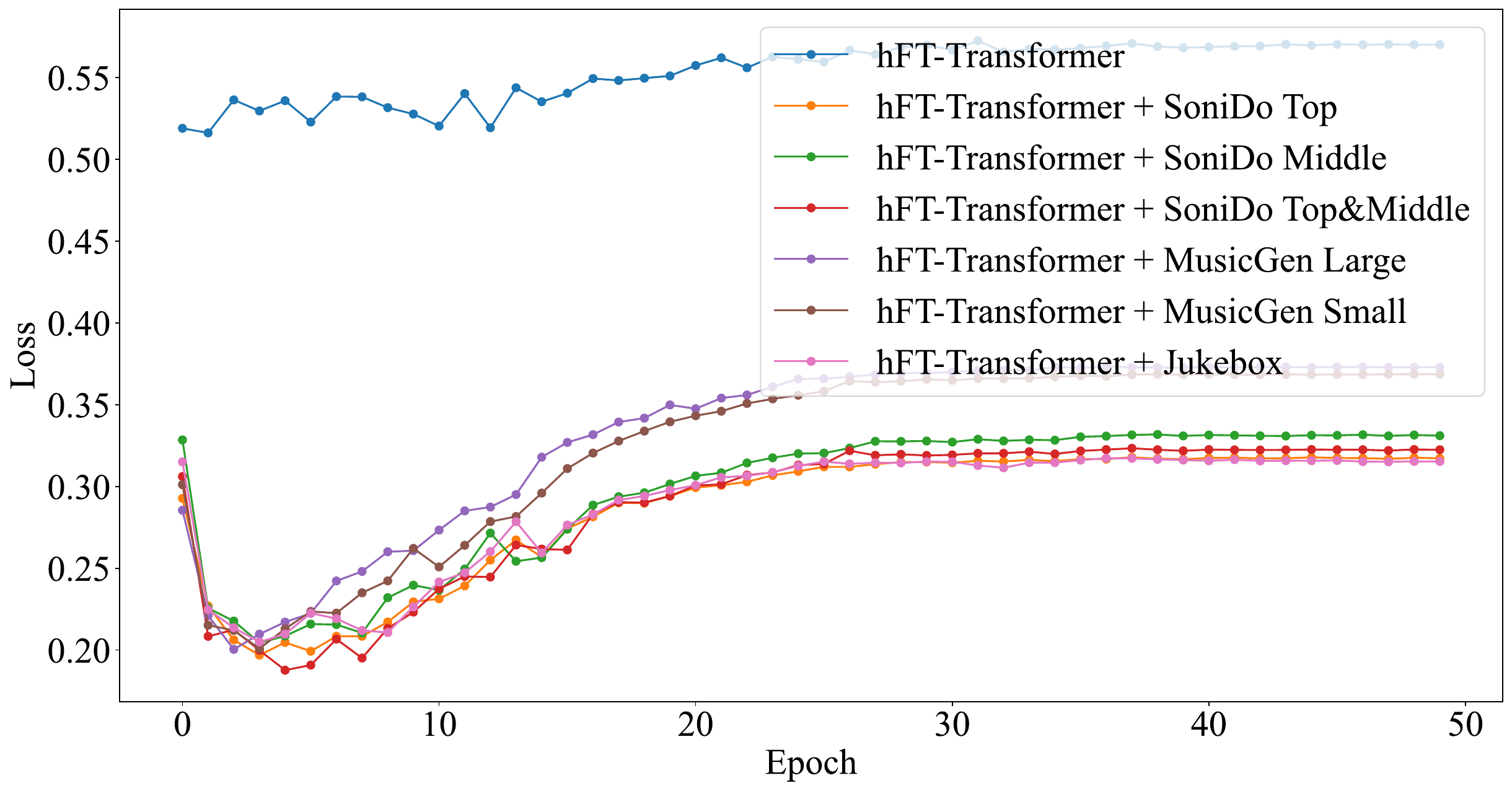}}}\\
\subfloat[\centering Training Loss (25\%)]{{\includegraphics[width=0.49\textwidth]{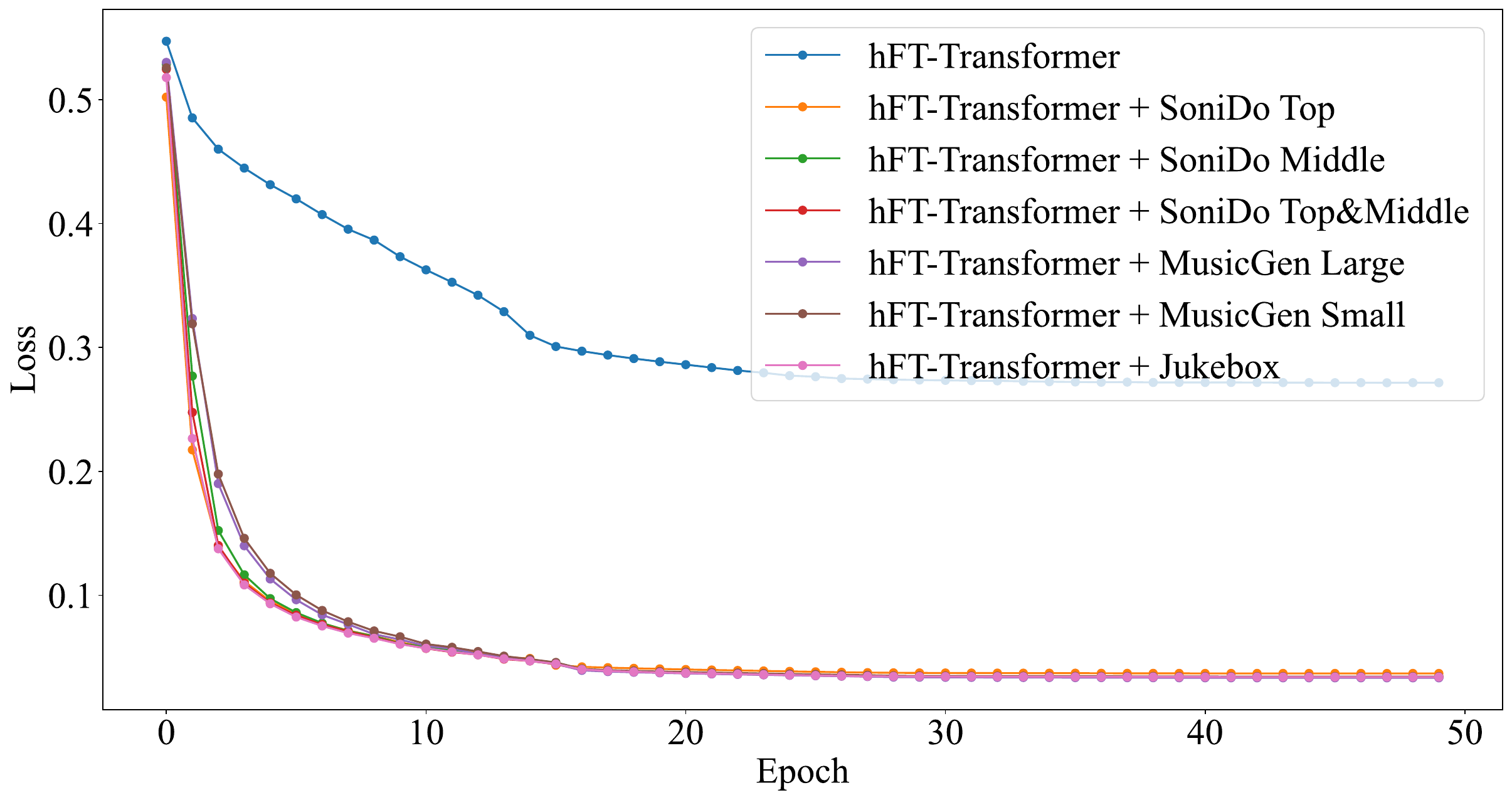}}}
\subfloat[\centering Validation Loss (25\%)]{{\includegraphics[width=0.49\textwidth]{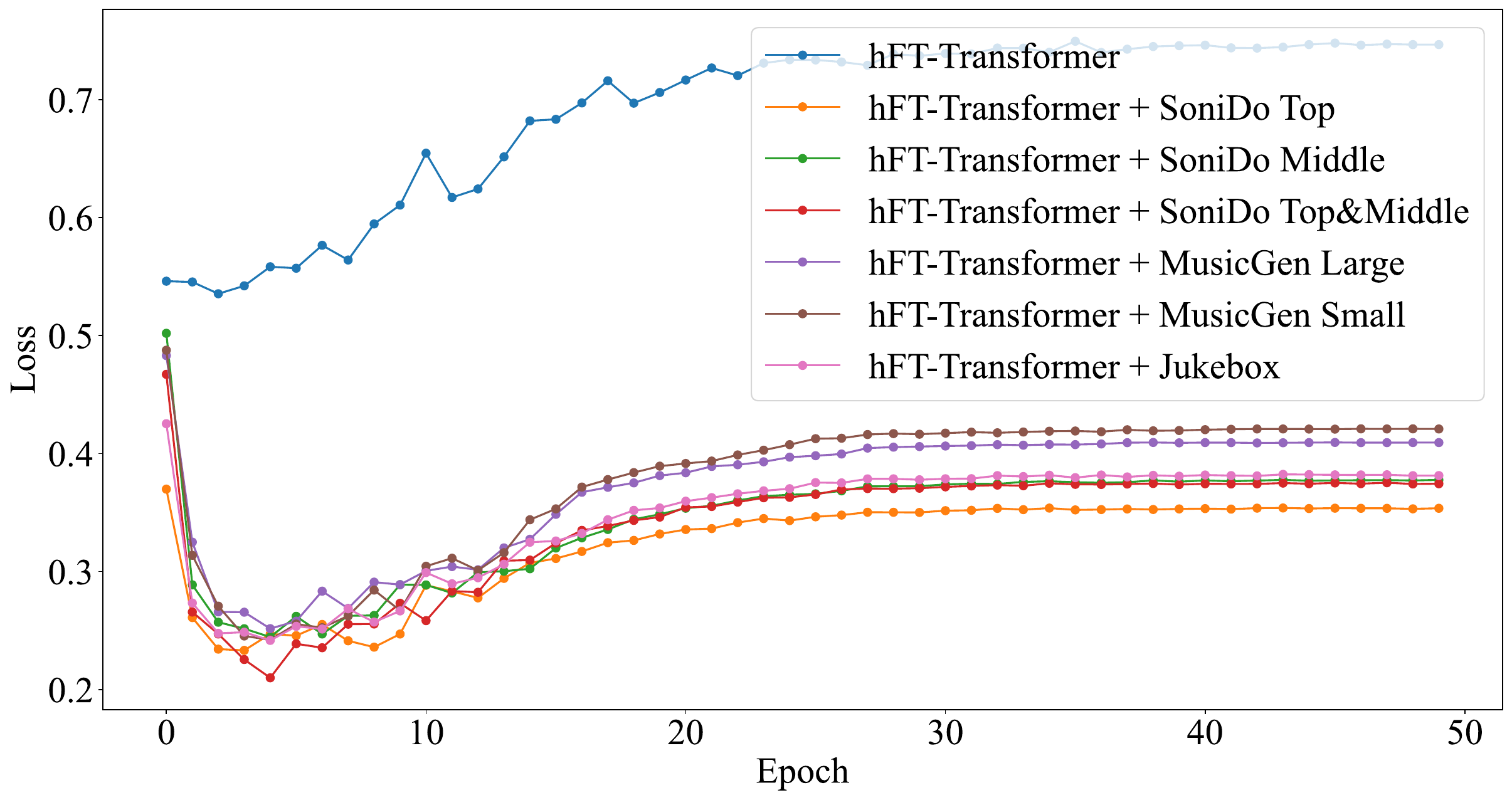}}}\\
\subfloat[\centering Training Loss (10\%)]{{\includegraphics[width=0.49\textwidth]{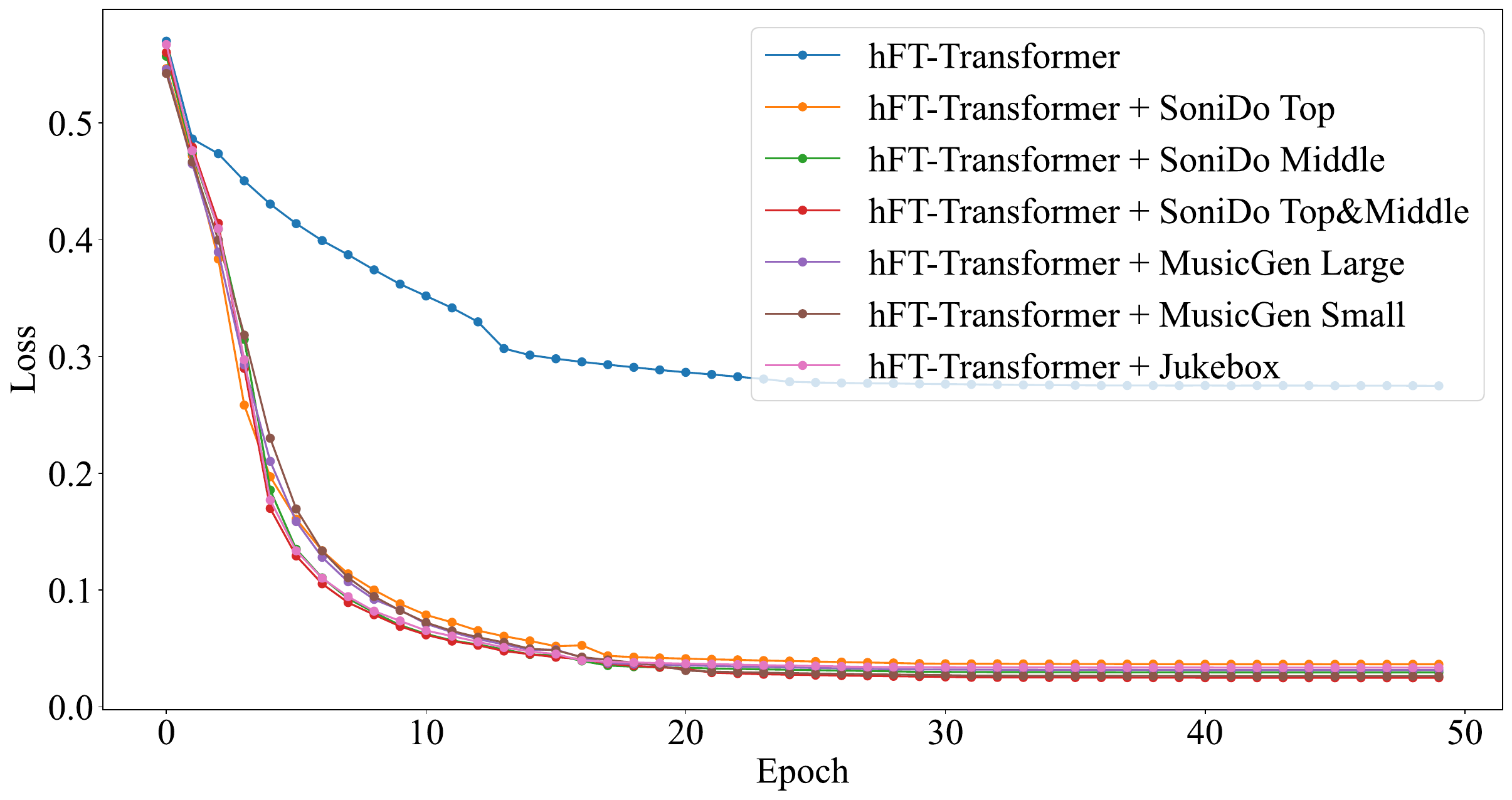}}}
\subfloat[\centering Validation Loss (10\%)]{{\includegraphics[width=0.49\textwidth]{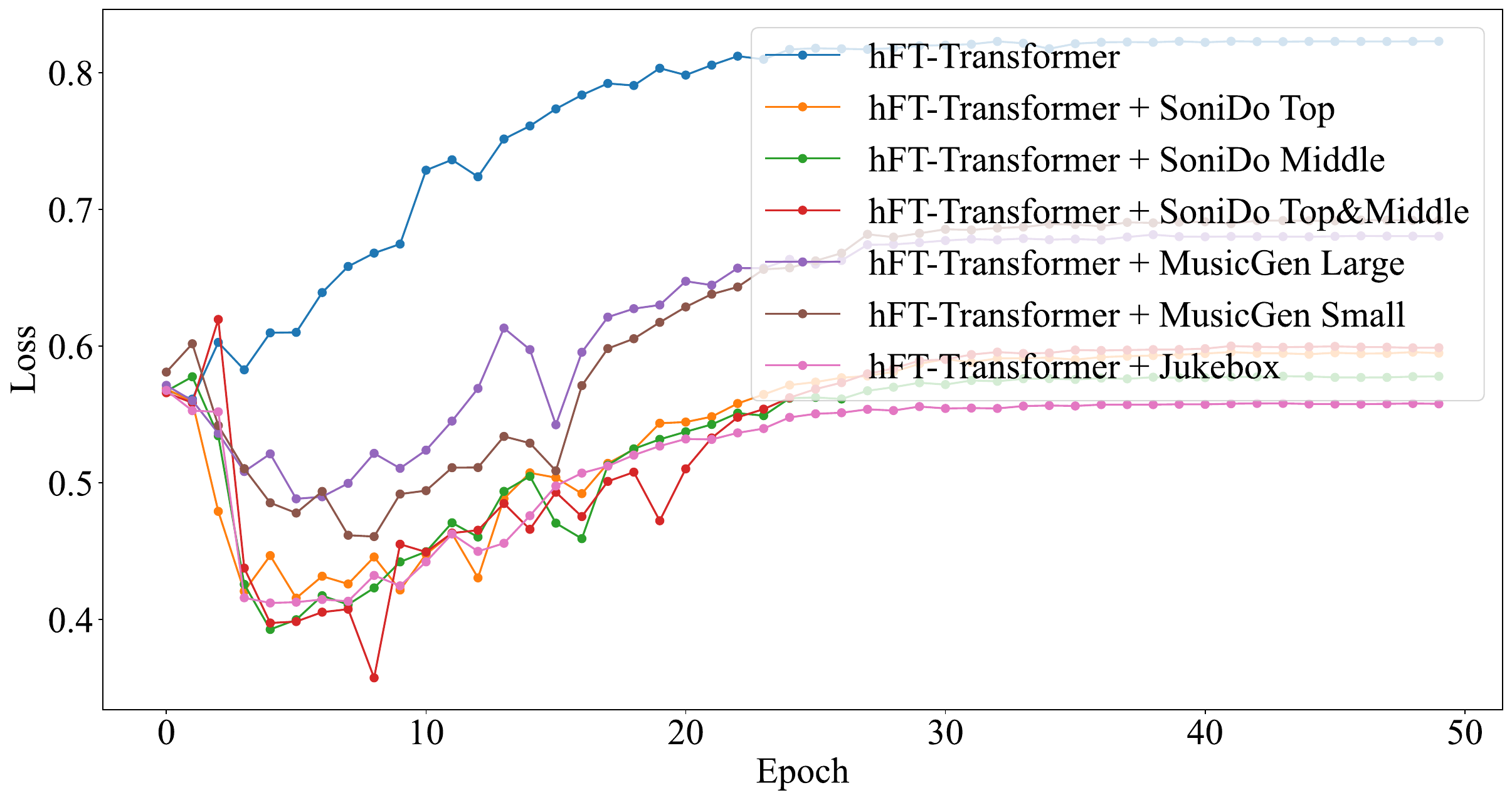}}}\\
\caption{\textcolor{c_revise}{Loss curves for hFT-Transformer in music transcription task}}
\label{fig:amt_loss_curve_hft}
\end{figure}

Figure~\ref{fig:amt_loss_curve_hft} shows the loss curves of training and validation.
The models using the \mfm, \musicgen \textcolor{c_revise}{or Jukebox} features reached a lower loss value at an earlier epoch compared with the baseline model using the spectrogram only as input.
Table~\ref{tab:amt_result_hft_full} lists the scores on the test set of MAPS.
When all the training data were available, the models using the \mfm, \musicgen\textcolor{c_revise}{, and Jukebox} features, except the model using the \musicgenl features outperformed the baseline.
When the models were trained with scarce data, the performance of the models using the \mfm, \musicgen\textcolor{c_revise}{, and Jukebox} features was superior to that of the model using the spectrogram only.
When the training data size was 50 and 25\%, the performance of the models using the \mfm features was still comparable to the baseline model trained with 100\% data.

\begin{table}[htb]
    \centering
    \caption{Evaluation results on MAPS in music transcription (\textbf{bold}: best score, \underline{underline}: second-best score). ``Note'' refers to note-wise estimation. First row is of hFT-Transformer~\cite{toyama2023}.}
    \small
    \resizebox{0.95\linewidth}{!}{
    \begin{tabular}{clcccc}
    \toprule
    Training data&Input&Frame&Note&Note w/ Offset&Note w/ Offset\&Velocity\\
    \midrule
    \multirow{7}{*}{100\%}&Spectrogram&82.89&85.14&66.34&48.20\\
    &Spectrogram + \mfm Top&83.92&\underline{86.45}&\underline{68.27}&\textbf{51.34}\\
    &Spectrogram + \mfm Middle&83.47&86.13&67.93&\underline{51.24}\\
    &Spectrogram + \mfm Top + \mfm Middle&\underline{84.16}&85.96&67.37&50.98\\
    &Spectrogram + \musicgenl&81.53&85.14&66.28&48.69\\
    &Spectrogram + \musicgens&82.94&85.97&\textbf{68.27}&50.42\\
    &\textcolor{c_revise}{Spectrogram + Jukebox}&\textcolor{c_revise}{\textbf{84.23}}&\textcolor{c_revise}{\textbf{86.54}}&\textcolor{c_revise}{68.26}&\textcolor{c_revise}{50.46}\\
    \midrule
    \multirow{7}{*}{50\%}&Spectrogram&39.12&23.34&13.22&9.12\\
    &Spectrogram + \mfm Top&\underline{83.35}&\underline{85.51}&\underline{65.84}&\underline{47.40}\\
    &Spectrogram + \mfm Middle&82.52&85.46&65.40&47.25\\
    &Spectrogram + \mfm Top + \mfm Middle&\textbf{83.37}&85.33&\textbf{67.04}&\textbf{49.32}\\
    &Spectrogram + \musicgenl&82.26&84.58&65.50&47.07\\
    &Spectrogram + \musicgens&82.24&85.21&65.32&46.72\\
    &\textcolor{c_revise}{Spectrogram + Jukebox}&\textcolor{c_revise}{83.33}&\textcolor{c_revise}{\textbf{85.58}}&\textcolor{c_revise}{64.62}&\textcolor{c_revise}{46.20}\\
    \midrule
    \multirow{7}{*}{25\%}&Spectrogram&12.88&1.61&0.66&1.01\\
    &Spectrogram + \mfm Top&\underline{81.71}&\textbf{84.70}&\underline{63.00}&\textbf{45.50}\\
    &Spectrogram + \mfm Middle&81.65&\underline{84.59}&62.19&43.91\\
    &Spectrogram + \mfm Top + \mfm Middle&81.44&83.79&62.61&\underline{44.71}\\
    &Spectrogram + \musicgenl&78.98&82.36&58.64&39.62\\
    &Spectrogram + \musicgens&79.39&82.23&60.36&41.02\\
    &\textcolor{c_revise}{Spectrogram + Jukebox}&\textcolor{c_revise}{\textbf{81.96}}&\textcolor{c_revise}{84.47}&\textcolor{c_revise}{\textbf{63.63}}&\textcolor{c_revise}{44.63}\\
    \midrule
    \multirow{7}{*}{10\%}&Spectrogram&9.83&0.59&0.17&0.46\\
    &Spectrogram + \mfm Top&65.91&66.64&39.88&25.87\\
    &Spectrogram + \mfm Middle&\underline{70.77}&\underline{74.02}&45.75&29.46\\
    &Spectrogram + \mfm Top + \mfm Middle&\textbf{71.57}&\textbf{75.00}&\textbf{46.18}&\textbf{30.63}\\
    &Spectrogram + \musicgenl&61.81&63.27&37.03&24.01\\
    &Spectrogram + \musicgens&63.73&65.90&39.00&24.94\\
    &\textcolor{c_revise}{Spectrogram + Jukebox}&\textcolor{c_revise}{70.43}&\textcolor{c_revise}{73.76}&\textcolor{c_revise}{\underline{45.80}}&\textcolor{c_revise}{\underline{30.42}}\\
    \bottomrule
    \end{tabular}
    }
    \label{tab:amt_result_hft_full}
\end{table}

\section{Music source separation}
\label{sec:app_mss_ds}

\subsection{Details of UMX with \mfm}
\label{sec:app_mss_umx_detail}

\citet{huang2022investigating} investigated various speech enhancement (SE) systems that use self-supervised learning (SSL) features and discussed the challenge that the SSL features may have lost some local signal information necessary for estimating lower-level features (e.g., spectrograms, waveform).
Following the above observation, \citet{hung22_interspeech} proposed to combine spectrograms with SSL features in their SE system to avoid such problem.
Therefore, we hypothesize that the features extracted with a large-scale foundation model could serve as auxiliary information for a neural network on music source separation.
However, it remains unclear how to integrate the extracted features into the network. Therefore, we investigated several integration strategies.

\begin{figure*}[tb]
  \centering
  \includegraphics[width=15.0cm]{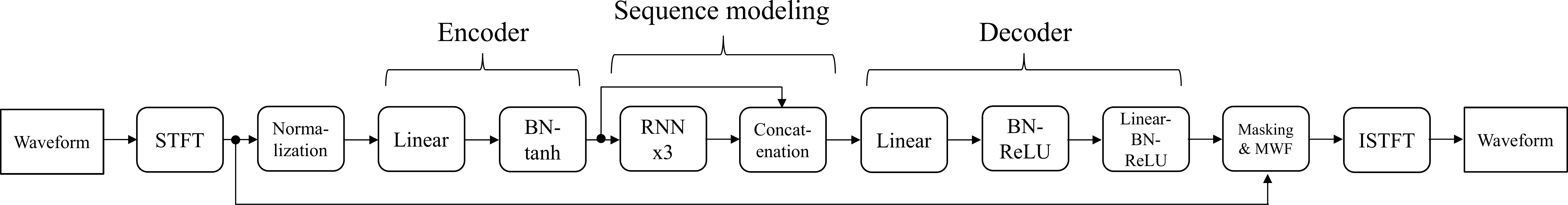}
  \vspace{-0mm}
  \caption{Original architecture of UMX}
\label{fig:umx}
\vspace{-0mm}
\end{figure*}

\begin{figure*}[tb]
  \centering
  \includegraphics[width=15.0cm]{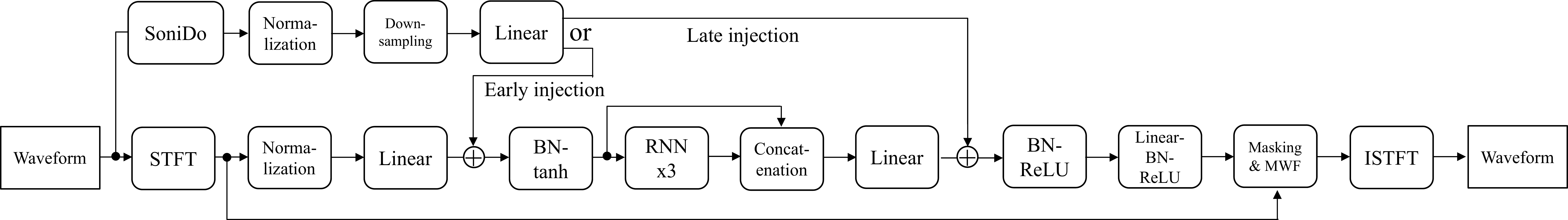}
  \vspace{-0mm}
  \caption{Architecture with \mfm using UMX}
\label{fig:umx_mfm}
\vspace{-0mm}
\end{figure*}

Figures \ref{fig:umx} and \ref{fig:umx_mfm} show the architectures of the original UMX and UMX with \mfm, respectively. UMX starts with an input audio waveform and converts it into an STFT spectrogram. The magnitude part of the spectrogram is normalized to a mean of $0$ and standard deviation of $1$, based on statistics collected from the entire training dataset before training.
The normalized magnitude spectrogram is then passed through an ``encoder'' block, which includes a linear layer, batch normalization (BN) layer, and hyperbolic tangent function. A ``sequence modeling'' block, consisting of three RNN layers along with a skip connection structure is then used. It processes the encoded features and combines their input and output features.
The resulting combined features are then transformed into time-frequency masks in the ``decoder'' block, which is made up of several linear layers, BN layers, and rectified linear unit (ReLU) activations. These time-frequency masks are used to modify the magnitude spectrogram, and a multi-channel Wiener filter (MWF)~\cite{nugraha2016multichannel,stefan2017improving} is optionally applied to the spectrogram.
Finally, an inverse STFT yields the separated audio waveform.

All learnable parameters are optimized with the mean-squared error of magnitude spectrograms.
We propose using the \mfm features while keeping the UMX architecture unchanged.
As shown in Figure~\ref{fig:umx_mfm}, \mfm is introduced to extract features from the input audio waveform.
These features are then adjusted to have a standard statistical distribution computed across the entire training dataset similarly to the normalization used for the magnitude spectrograms in UMX. Subsequently, they are processed through a down-sampling block, followed by a linear layer. The down-sampling block is introduced because of the time resolution difference between \mfm and UMX; \mfm generates a feature for every $128$ waveform samples. In contrast, the original branch of the UMX model processes every $1024$ waveform samples, as determined by the STFT hop size. We explain the design of the down-sampling block in the next paragraph.
The down-sampled features are then converted from the dimension of $4800$ to $512$ by the linear layer and summed up with the original features from the UMX branch.
\begin{figure}
\centering
    \subfloat[\centering Original branch]{{\includegraphics[width=.30\textwidth]{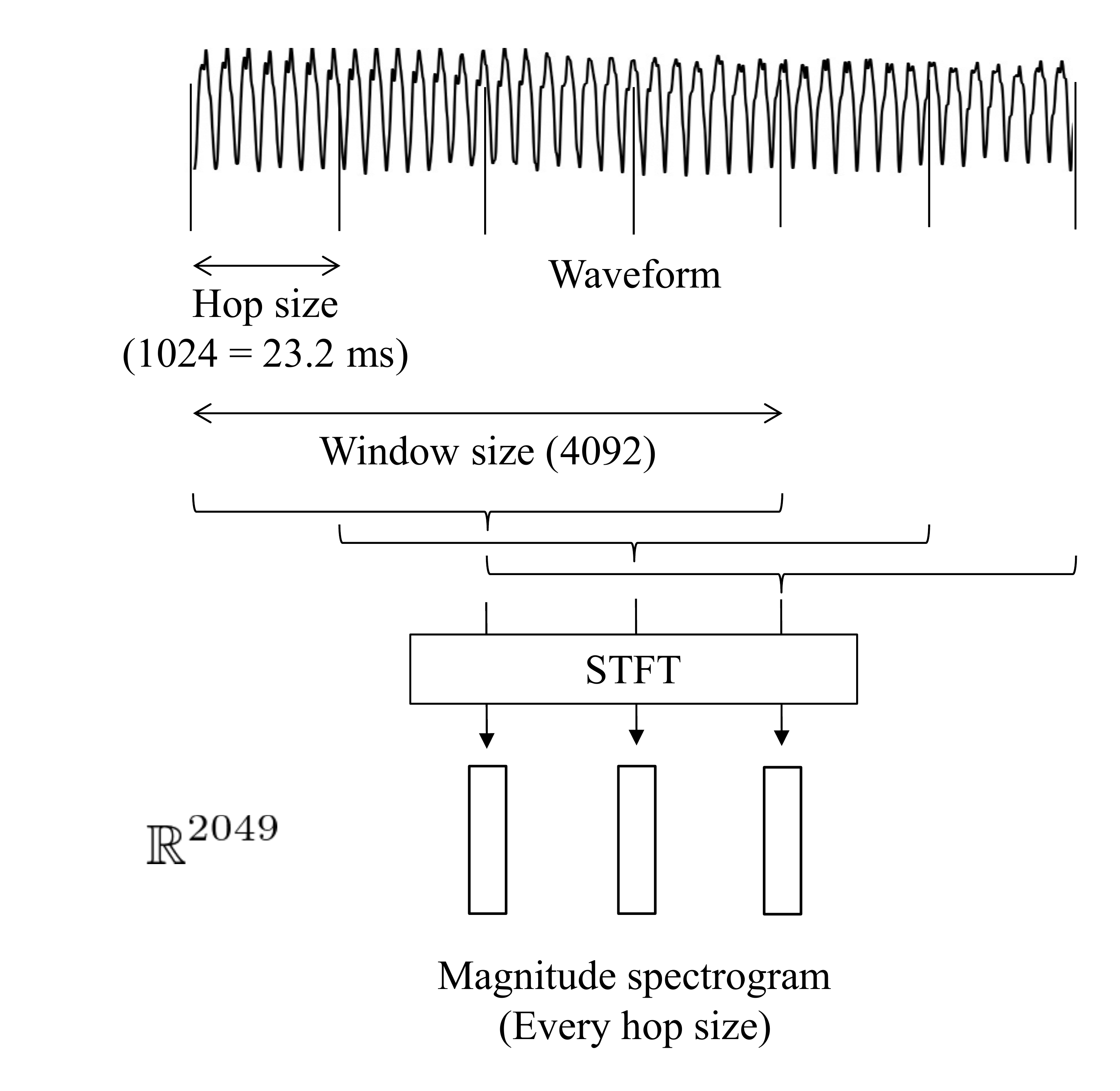} }}
    \subfloat[\centering Average/Max pooling]{{\includegraphics[width=.30\textwidth]{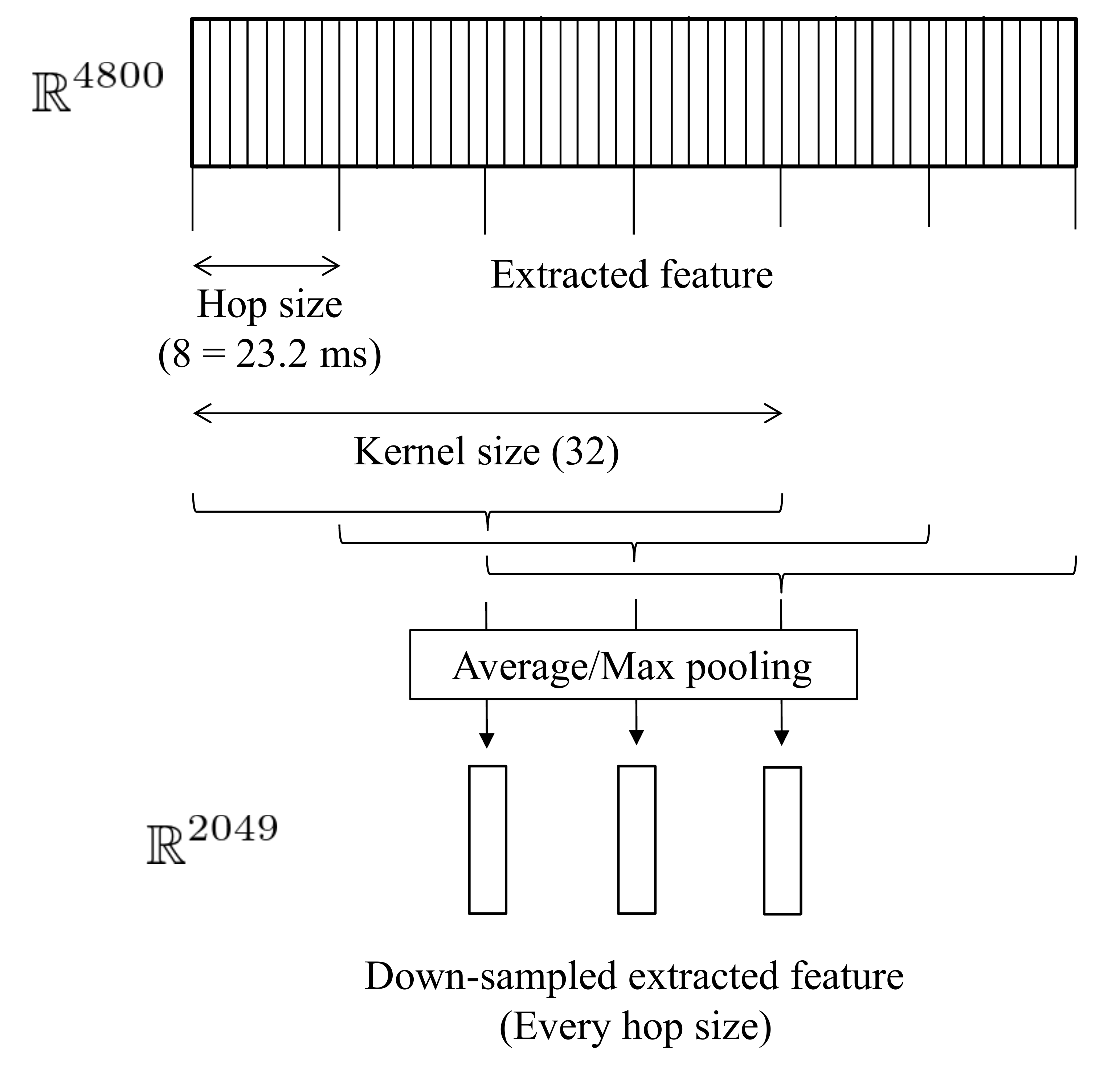} }}    
    \subfloat[\centering Unfolding]{{\includegraphics[width=.30\textwidth]{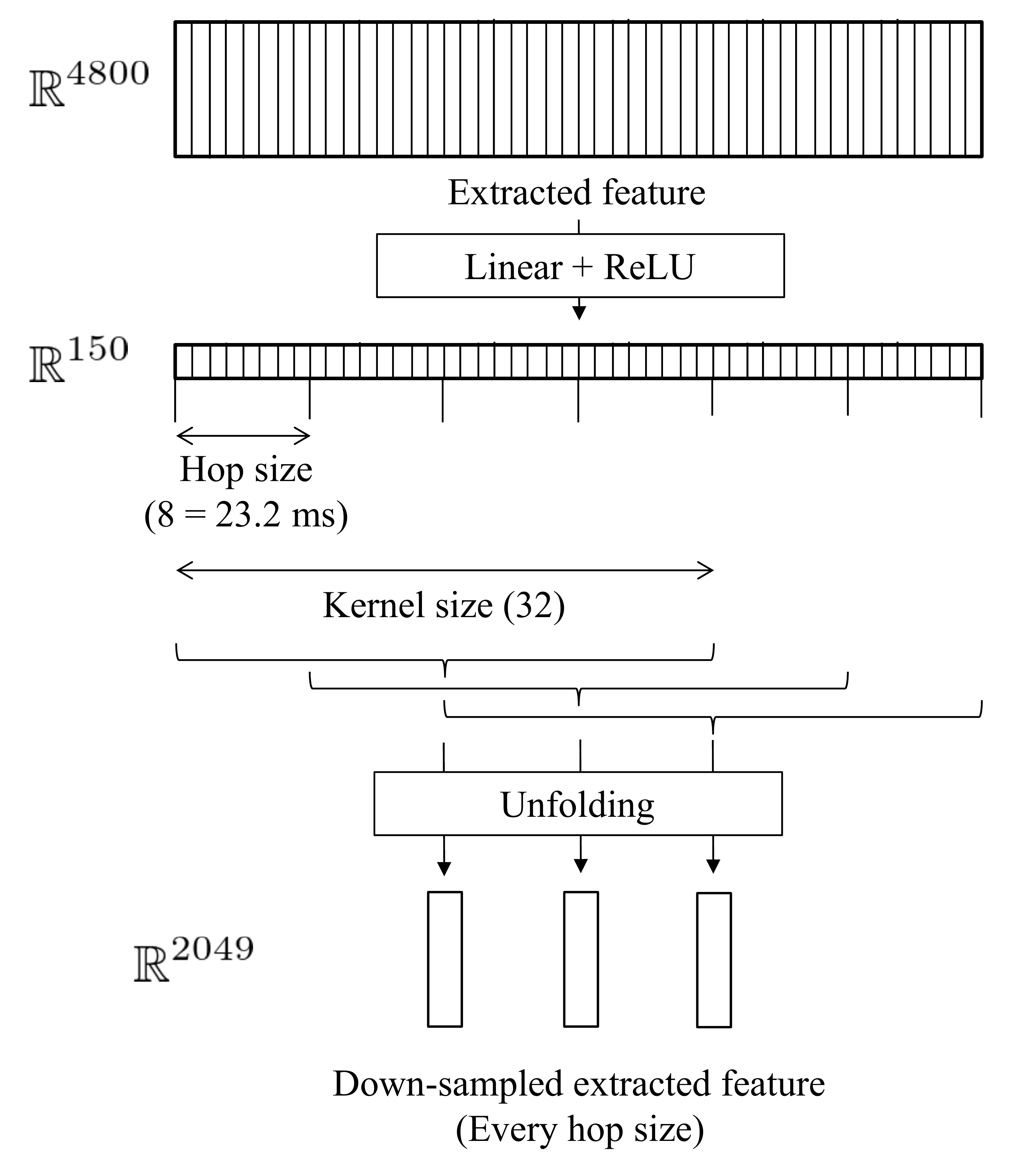} }}    
\caption{Down-sampling blocks to align time resolution.}
\label{fig:down_sampling}
\end{figure}

In the original UMX architecture, the input is the spectrogram of STFT. Since all modules in UMX work with the time resolution of STFT, the down-sampling block is required to align the time resolution between the \mfm features and STFT. 
In the ablation study, we explored three different down-sampling operations: max pooling (MP), average pooling (AP), and unfolding (UF).
Figure \ref{fig:down_sampling} illustrates the three different operations for the down-sampling block.

AP averages the multiple \mfm features with the hop size and kernel size corresponding to the hop size and the window size of STFT, respectively.
MP works similarly to AP but replaces the average pooling with a max pooling operation.
In UF, the dimension of the \mfm feature sequence is first converted from $4800$ to $150$ with a linear layer and ReLU function.
Next, the feature sequence corresponding to the window size of STFT are concatenated every STFT hop size.
After the concatenation, the dimension of the feature returns to $4800$ again.

As well as the design of the down-sampling block, we further evaluated two methods for injecting the down-sampled features into UMX, as shown in Figure ~\ref{fig:umx_mfm}. With the ``early injection'' (EI) method, the \mfm features are injected into UMX's encoder block. With the ``late injection'' (LI) method, the extracted features are injected into the decoder block.

We explored the optimal down-sampling and injection method for UMX on a vocal extraction task using MUSDB18.
We followed the original training configuration of UMX for all experiments except that we disabled data augmentation for simplicity.
To purely evaluate the trained components, MWF was skipped in all experiments.

\subsection{Results: Ablation study for UMX with \mfm}
\label{sec:app_mss_umx_ablation}
The results of the aforementioned ablation study are summarized in Table~\ref{tab:result_vocal}.
\begin{table}[tb]
  \centering
  \caption{Evaluation results of music source separation on vocal extraction task}
 \footnotesize
 { 
 \begin{tabular}{l|cc}
    \toprule

        \multirow{2}{*}{Method} &
        \multicolumn{2}{c}{SDR [dB]}  
        
        \\ 
        & Vocals & Accompaniment
        \\ 
        \midrule
        Open-Unmix (UMX) & 
         5.71 & 
         11.57 \\
        \midrule
        UMX with \mfm (MP, EI) &        
         5.76 &
         11.86 \\
        UMX with \mfm (AP, EI) &        
         5.55 &
         11.71 \\
        UMX with \mfm (UF, EI) &        
         \textbf{5.92} &
         \textbf{11.92} \\
        UMX with \mfm (UF, LI) &        
         5.83 &
         11.68 \\
        \bottomrule
    \end{tabular}
}
\label{tab:result_vocal}
\end{table}

For the down-sampling block, MP and AP did not improve the SDR score, whereas UF achieved a $0.34$~dB improvement for vocals and $0.36$~dB improvement for accompaniment.
The results indicate that UF is the proper choice for the down-sampling block.
We assume that the lower-level information is lost during the pooling operation over the temporal axis in MP and AP, while UF preserves such information by stacking multiple tokens into one frame in the unfolding manner.

Table~\ref{tab:result_vocal} shows that EI is superior to LI.
This suggests that it is the sequence modeling block in UMX that effectively used the \mfm features to improve separation performance. The observation also inspired us to inject \mfm features into the transformer block in HTDemucs.

\begin{figure}
\centering
    \subfloat[\centering Training loss]{{\includegraphics[width=.45\textwidth]{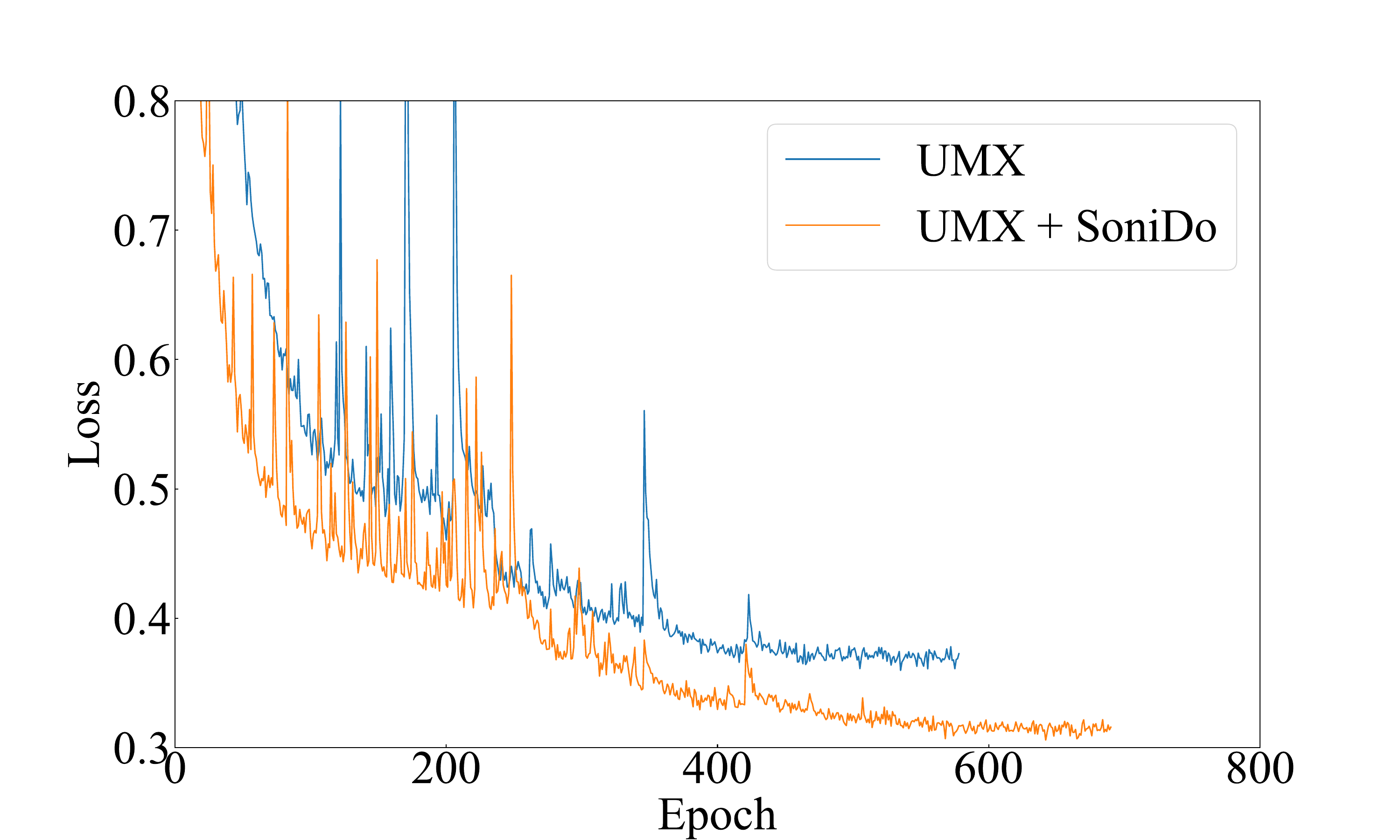} }}
    \subfloat[\centering Validation loss]{{\includegraphics[width=.45\textwidth]{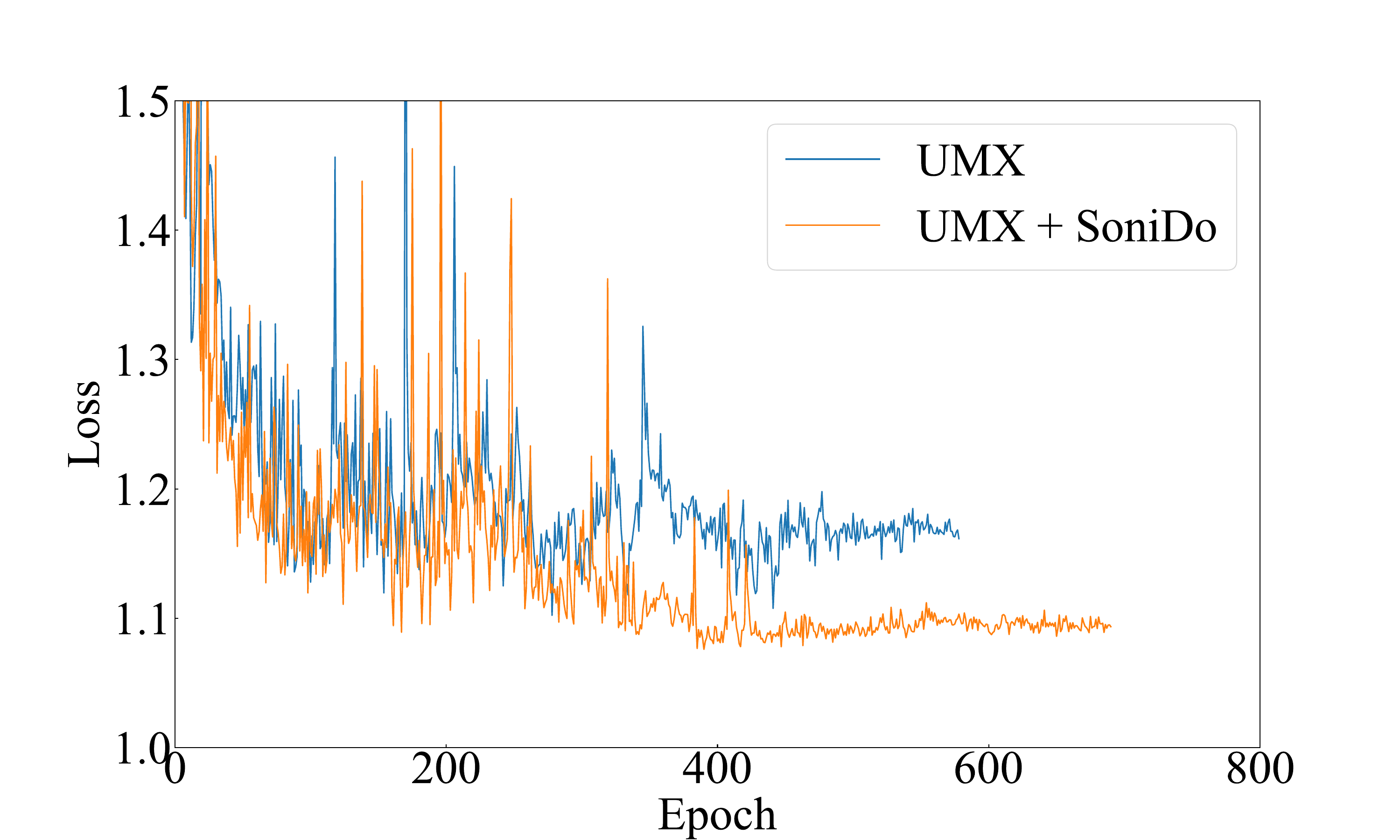} }}    
\caption{Training and validation curves for UMX experiments in music separation task}
\label{fig:umx_trainvalid_curves}
\end{figure}
Figure~\ref{fig:umx_trainvalid_curves} shows the training and validation curves for UMX and UMX with \mfm (UF, EI).
The loss curve of UMX with \mfm tends to be lower than that of UMX in both training and validation. \textit{i.e.}, the number of epochs required to reach target loss was smaller when UMX was trained with \mfm.
The converging loss of UMX with \mfm was also lower than that of the original UMX, which reveals that the benefit from \mfm can be consistently observed in the training phase, and the performance improvement is significant.

\subsection{Experiments: HTDemucs with \mfm}
\label{appendix:HTDemucsMFM}

\begin{figure}
\centering
    \subfloat[\centering Training loss]{{\includegraphics[width=.3125\textwidth]{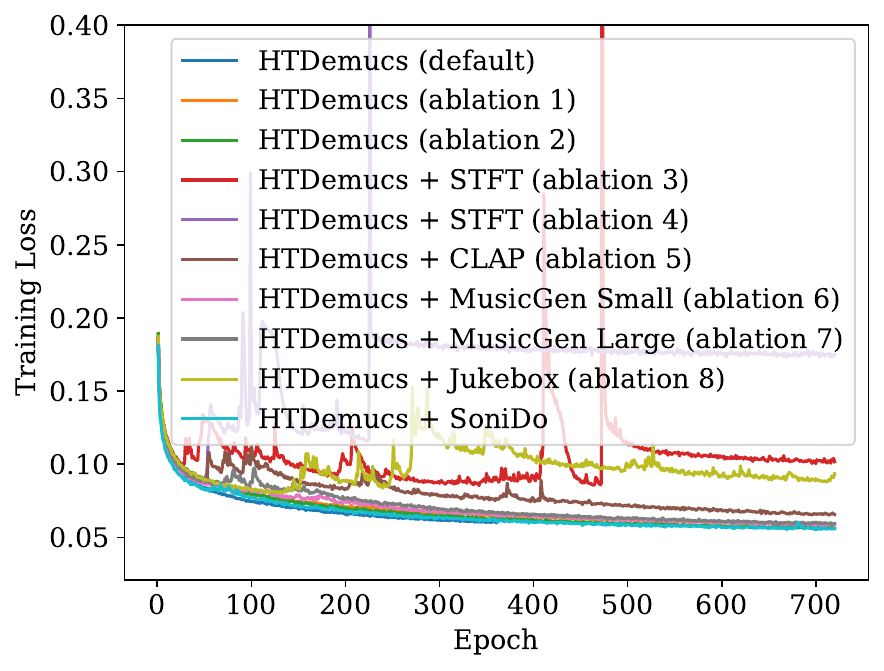} }}
    \subfloat[\centering Validation loss]{{\includegraphics[width=.3125\textwidth]{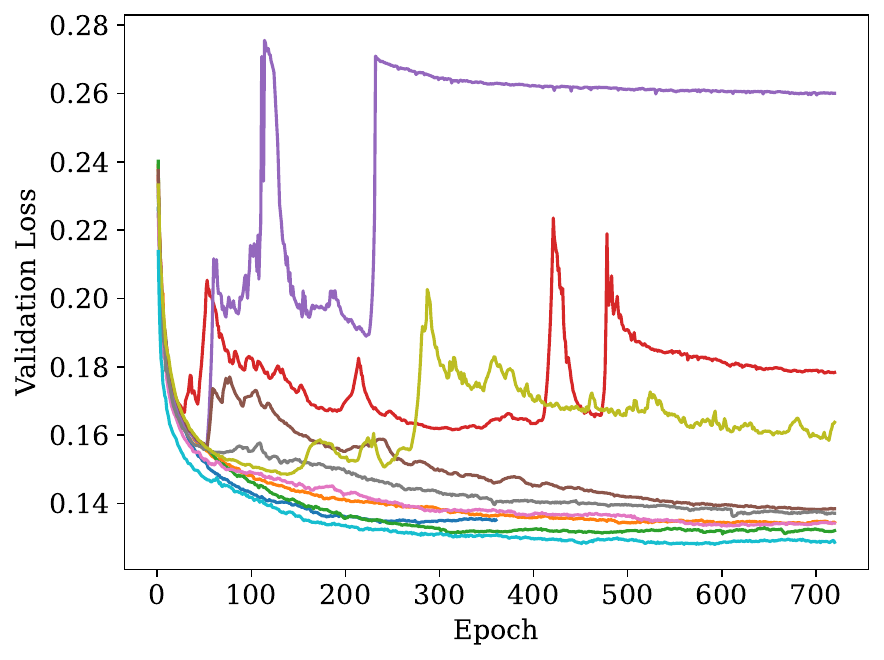} }}  
    \subfloat[\centering Validation SDR (dB)]{{\includegraphics[width=.3\textwidth]{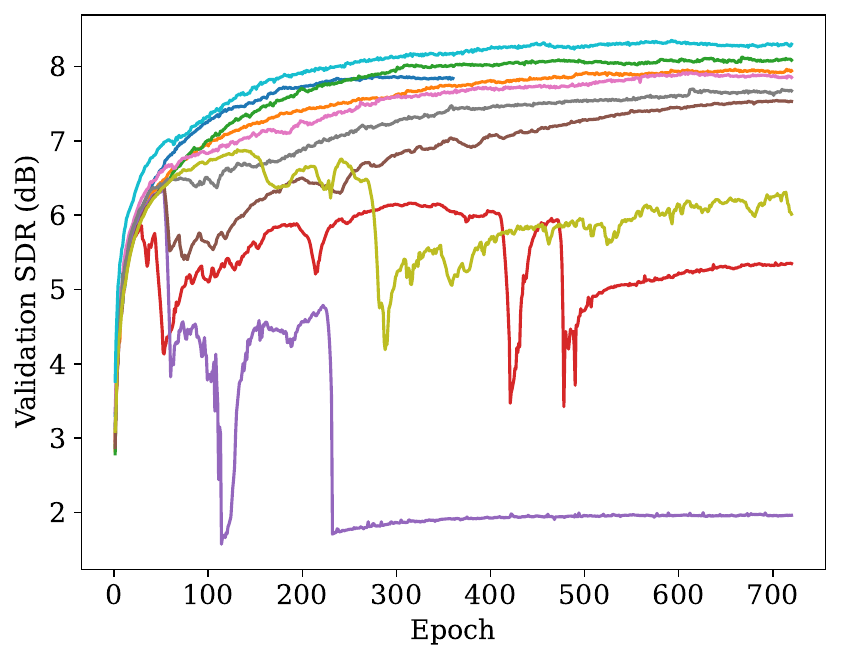} }}  
\caption{Training and validation curves for HTDemucs experiments in music source separation task. We conditioned HTDemucs with four different features (STFT, CLAP, MusicGen and \mfm). We achieved highest separation scores when injecting features of \mfm. Interestingly, we observed instabilities when injecting STFT and CLAP features but not when injecting \mfm features.}
\label{fig:htdemucs_trainvalid_curves}
\end{figure}

We provide more details for music source separation with HTDemucs~\citep{rouard2023hybrid} and \mfm from Section~\ref{sssec:music_editing_mss}. To investigate the effect of the \mfm features, we trained several models on MUSDB18~\citep{rafii2017musdb18}. We compare the following models:
\begin{itemize}
    \item \emph{HTDemucs (default)}: Model with default settings (Dora\footnote{\url{https://github.com/facebookresearch/dora}} signature `955717e8'). The default batch size is $32$ which corresponds to $4$ samples per GPU as we trained in parallel on $8$ GPUs. The default number of training epochs is $360$.
    \item \emph{HTDemucs + \mfm}: Combination of HTDemucs with \mfm. We introduced two new cross-domain transformer encoders into HTDemucs which are inserted directly after the original cross-domain transformer encoder. The first encoder facilitates information exchange between the spectral and \mfm feature sequence, while the second one enables interaction between the waveform and \mfm feature sequence. The \mfm features are computed from the monaural downmix of the mixture using the top prior. Hence, the \mfm feature sequence has a dimension of $(T = 2688) \times (C = 4800)$, reflecting the characteristics of training samples with a duration of $7.8$~s. Subsequently, these features undergo normalization through a LayerNorm and further projected from their original dimension of $4800$ to $2048$ using a linear layer, giving them the same size as the features in the sequences from both the spectral and waveform branches of HTDemucs. Each additional cross-domain transformer encoder has a depth of $3$, where we set \verb|cross_first=True|. Due to the additional \mfm model, we needed to reduce the batch size to 16 (corresponding to 2 samples per GPU as we trained on 8 GPUs). To keep the number of samples that the models seen during training the same, the number of training epochs was increased to $720$. Additionally, to match the random remixing augmentation of the default HTDemucs model, we added $860$ random mixes, generated from the $86$ training songs in MUSDB18. It is important to note that this was done to ensure the same augmentation and does not introduce new songs to the training set; we still exclusively trained on the train split of MUSDB18.
    \item \emph{HTDemucs (ablation 1)}: Training with default settings `955717e8` where we also reduced the batch size to $16$ and increased the number of training epochs to $720$, together with the additional $860$ random mixes used for \emph{HTDemucs + \mfm}.
    \item \emph{HTDemucs (ablation 2)}: Training as \emph{HTDemucs (ablation 1)} but where the number of layers in the cross-domain transformer was increased from $5$ to $11$, matching the $2\cdot 3 = 6$ additional transformer layers of \emph{HTDemucs + \mfm}. 
    \item \emph{HTDemucs + STFT-2048}: Same training settings as \emph{HTDemucs + \mfm} but where we used STFT features instead of \mfm features. We computed the STFT with a Hann window of $2048$ samples and hop size of $512$ from the monaural downmix of the mixture.
    \item \emph{HTDemucs + STFT-4096}: Same training settings as \emph{HTDemucs + \mfm} but where we used STFT features instead of \mfm features. We computed the STFT with a Hann window of $4096$ samples and hop size of $256$ from the monaural downmix of the mixture.
    \item \emph{HTDemucs + CLAP}: Same training settings as \emph{HTDemucs + \mfm} but where we used the embeddings from CLAP~\citep{laionclap2023} instead of the \mfm features. More specifically, we used the `fine-grained embeddings` before the final AP in CLAP. This experiment enabled us to compare features obtained with two trained models (as opposed to the STFT)\footnote{Note that \mfm and CLAP were trained on different datasets.}.
    \item \emph{HTDemucs + MusicGen Small}: Same training settings as \emph{HTDemucs + \mfm} but where we used intermediate features from MusicGen~\citep{copet2023musicgen} instead of the \mfm features. More specifically, we used the `musicgen-small` activations at the output of the \nth{12} layer. This experiment enabled us to compare features obtained with two music foundation models\footnote{Note that \mfm and MusicGen were trained on different datasets.}.
    \item \emph{HTDemucs + MusicGen Large}: Same training settings as \emph{HTDemucs + MusicGen} but where we used the `musicgen-large` activations at the output of the \nth{24} layer. This enabled us to compare features obtained with two music foundation models with the same order of magnitude of number of learnable parameters. Note that, to reduce instability during training, this experiment used AdamW~\citep{adamw} as the optimizer, instead of Adam.
    \item \textcolor{c_revise}{\emph{HTDemucs + Jukebox}: Same training settings as \emph{HTDemucs + \mfm} but where we used intermediate features from Jukebox~\citep{dhariwal2020jukebox} instead of the \mfm features. More specifically, we used the activations at the output of the \nth{36} layer of the model `5b`\footnote{Note that \mfm and Jukebox were trained on different datasets.}.}
\end{itemize}
Figure~\ref{fig:htdemucs_trainvalid_curves} displays the training and validation curves for these models.

\section{Music Mixing}
\label{sec:app_mixing}

Music mixing is a crucial task in music production and typically conducted using audio processors or audio effects, which are signal processing systems that alter specific characteristics of the input signal. Several signal processing and machine learning methods have been investigated to automatize this task~\citep{steinmetz2022automix}, with the goal of simplifying the process for less experienced content creators and enhancing the workflow capabilities of professionals~\citep{moffat2019approaches}.

Data-driven deep learning approaches for automatic music mixing have focused on two fundamental frameworks: direct transformation networks, in which the model executes the mixing in a black-box manner, and parameter estimation networks, in which the mixing is carried out via differentiable audio processors. ~\citet{martinez2021deep} proposed Mix-Wave-U-Net, a modified Wave-U-Net for drum mixing as a direct transformation system, while \citet{steinmetz2021automatic} introduced a differentiable mixing console in which neural proxies act as a parameter estimation network. Both systems have been acknowledged as having limited performance due to the scarcity of available training data, failing to meet professional audio engineering standards. Such systems require unprocessed or dry multitrack recordings and their corresponding mixtures, and large datasets are not readily accessible. To address this limitation, \citet{martinez2022automatic} proposed an Fx-normalization preprocessing method that enables the training of direct transformation automatic mixing systems using processed or wet multi-track audio datasets, akin to the datasets used in source separation. Building on this approach, \citet{koo2022music} introduced a contrastive learning approach that allows a direct transformation network to execute mixing style transfer. \citet{vanka2024diff} proposed a differentiable mixing style transfer system that predicts console parameters from raw tracks and a reference mix, advancing both parameter estimation and style transfer approaches.

While \citet{koo2022music} used SSL embeddings from a reference mixture to guide the mixing style transfer task, to the best of our knowledge, our approach is the first data-driven automatic mixing approach that incorporates SSL features from the input stems or high-level information related to genre, instrumentation, or mood to enhance automatic mixing performance.

\subsection{Details of Mix-Wave-U-Net and CRAFx2 with \mfm}
\label{appendix:mixing_waveunet}

Mix-Wave-U-Net extends the U-Net architecture for audio signal processing tasks. We used FiLM  layers~\citep{perez2018film} to incorporate the \mfm features into each up-sampling 1D convolutional block and the bottleneck 1D 
convolutional block in the Mix-Wave-U-Net. Following~\citet{meseguer2019conditioned}, we positioned FiLM layers after the normalization layer and before the LeakyReLU activation function.

CRAFx2~\citep{martinez2022automatic} comprises (1) an adaptive front-end that learns a filter bank, (2) latent-space mixer that learns a mixing mask functioning as equalizer, dynamic range compression, and reverberation transformations, and (3) a synthesis back-end that, through adaptive gains, implements loudness and panning transformations for each filter-bank channel. To enhance the mixing performance of the network, we also incorporated the \mfm features into the relevant layers of the network. Following a similar approach to Mix-Wave-U-Net, we use FiLM layers to condition both the latent-space mixer and synthesis back-end.
The latent-space mixer was first constructed on the basis of a temporal dilated convolutional network (TCN) followed by stacked bidirectional long short-term memory (BLSTM) layers, and FiLM layers were inserted within the TCN block and before the BLSTM layers.
For the TCN part, the FiLM layers were placed after the depthwise convolution operation and before the second nonlinear activation function. 
Before the input of the BLSTM layers, we introduced a FiLM layer.
Finally, the synthesis back-end incorporated a squeeze-and-excitation block (SE)~\citep{hu2018squeeze}, which scales channel-wise information by applying adaptive gains. A GroupNorm and FiLM layer are added after the second linear layer of the SE and before the sigmoid function.

\subsection{Experiments: Mix-Wave-U-Net and CRAFx2 with \mfm}
\label{appendix:mixing_experiments}

We used Mix-Wave-U-Net and CRAFx2 to explore the effect of the \mfm features on the music mixing task. The input to all networks is the Fx-normalized~\citep{martinez2022automatic} stereo stems; vocals, bass, drums, and other, with the output being the stereo mixture. Each input stem consists of $2$ channels of $10$-s audio frames at $44.1$ kHz. The \mfm features are computed from the summation of the Fx-normalized input stems using the top prior and unconditional extraction. Therefore, the \mfm feature sequence has a dimension of $(T=3446) \times (C=4800)$. 

Although music mixing is a time-varying task, we hypothesize that high-level concepts such as genre, instrumentation, and mood, might improve mixing performance. Therefore, following the pre-processing for time-invariant retrieval described in Section~\ref{ssec:dt_pp_1}, we carried out AP over the time dimension, and the tokens were time-averaged to a dimension of $(T=1) \times (C=4800)$, subsequently normalized via a non-trainable LayerNorm layer, and dimensionally reduced to $512$ through a linear layer. We found that reducing the dimension too much (\textit{e.g.}, $128$) decreased performance. The linear layer is followed by a dropout layer with a probability of $0.25$ to avoid overfitting. All networks underwent the same token-processing steps.

From MUSDB18, $86$ songs were used for training, and $14$ and $50$ for validation and testing, respectively. Since this dataset is significantly smaller than that used by~\citet{martinez2022automatic}, the number of batches per epoch was reduced from $1600$ to $320$, and all models were trained for $520$ epochs. For the default networks, we used the suggested initial learning rate of 1e$^{-3}$. For models involving \mfm, to ensure stability, we set the initial learning rate to 1e$^{-4}$.

The loss function corresponds to the stereo-invariant loss that \citet{martinez2022automatic} reported as the best-performing, which they referred to as $L_{\text{b}}$, and consists of A-weighting pre-emphasis and low-pass finite impulse response filters, L2-norm on the spectral magnitude, and the L1-norm on the spectral log-magnitude.

To investigate the effect of the \mfm features, we trained various models using features from both MusicGen Small and MusicGen Large. The training and feature injection settings are identical to those of Mix-Wave-U-Net + \mfm and CRAFx2 +\mfm, respectively, and the MusicGen features were extracted in the same manner as described in Appendix~\ref{appendix:HTDemucsMFM}/

To objectively evaluate the performance of all mixing systems, we used audio features related to the main audio characteristics that audio engineers manipulate during the mixing process, as shown in several works~\citep{colonel2021reverse, steinmetz2022automix, martinez2022automatic, vanka2024diff}. We computed the following audio features, spectral: centroid, bandwidth, contrast, roll-off, and flatness~\citep{peeters2004large}; panning: the panning root mean square (PRMS) for total panning, low, mid, and high frequencies~\citep{tzanetakis2007stereo}; dynamic: RMS level, dynamic spread, and crest factor~\citep{ma2015intelligent}; and loudness units full scale (LUFS) level and peak loudness~\citep{itu2011itu}. All features were computed using a running mean of $0.5$-s~\citep{tzanetakis2007stereo}. The objective evaluation test sets are identical to~\citet{martinez2022automatic}. These include MDXDB21-dry, an 18-song dry test set from MDXDB21, and the MUSDB18 test set, comprising 50 wet multi-track songs.

\subsection{Results: Mix-Wave-U-Net and CRAFx2 with \mfm}
\label{appendix:mixing_results}

\begin{figure}
\centering\subfloat[Mix-Wave-U-Net: Training loss]{\includegraphics[width=0.45\textwidth]{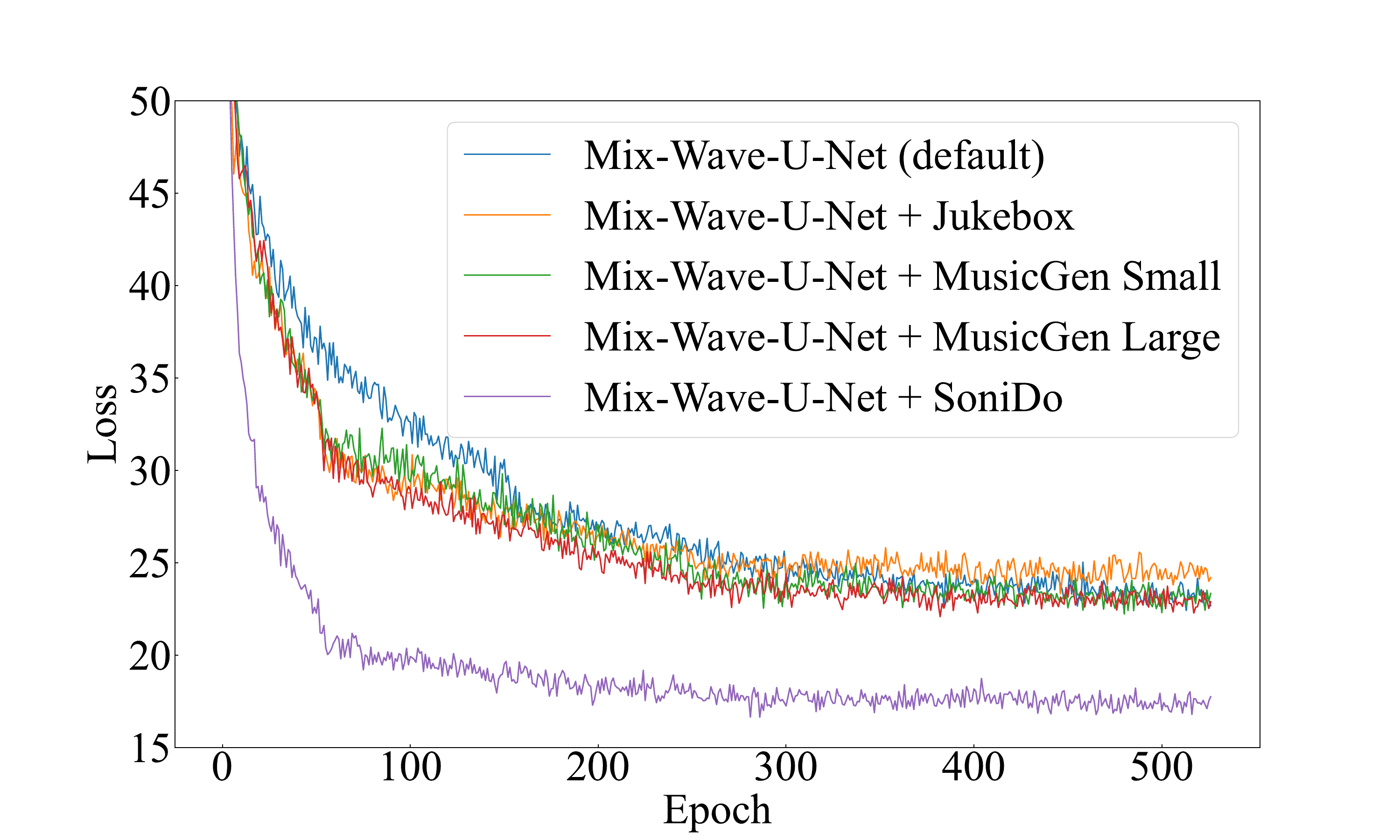}}
\centering\subfloat[Mix-Wave-U-Net: Validation loss]{\includegraphics[width=0.45\textwidth]{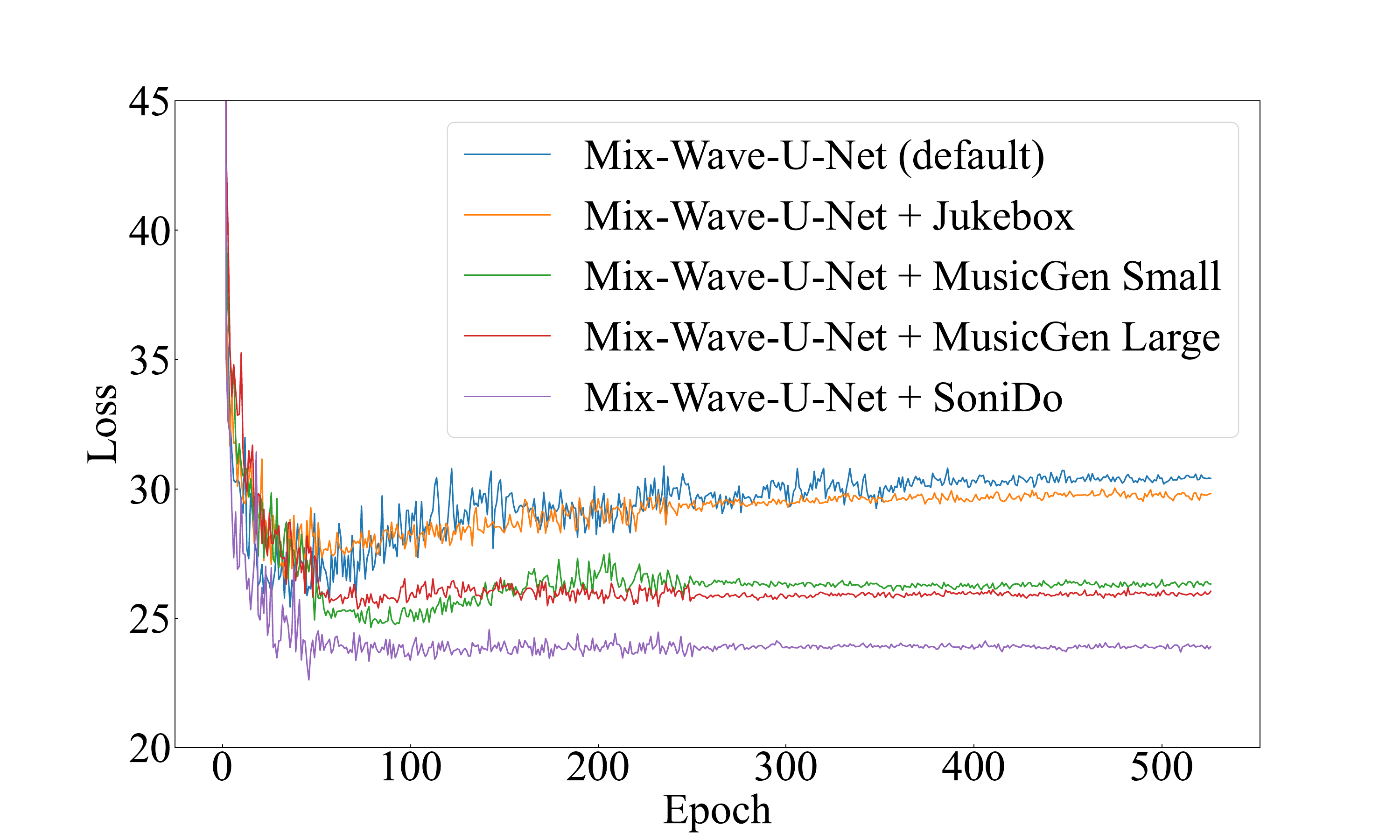}}\\
\centering\subfloat[CRAFx2: Training loss]{\includegraphics[width=0.45\textwidth]{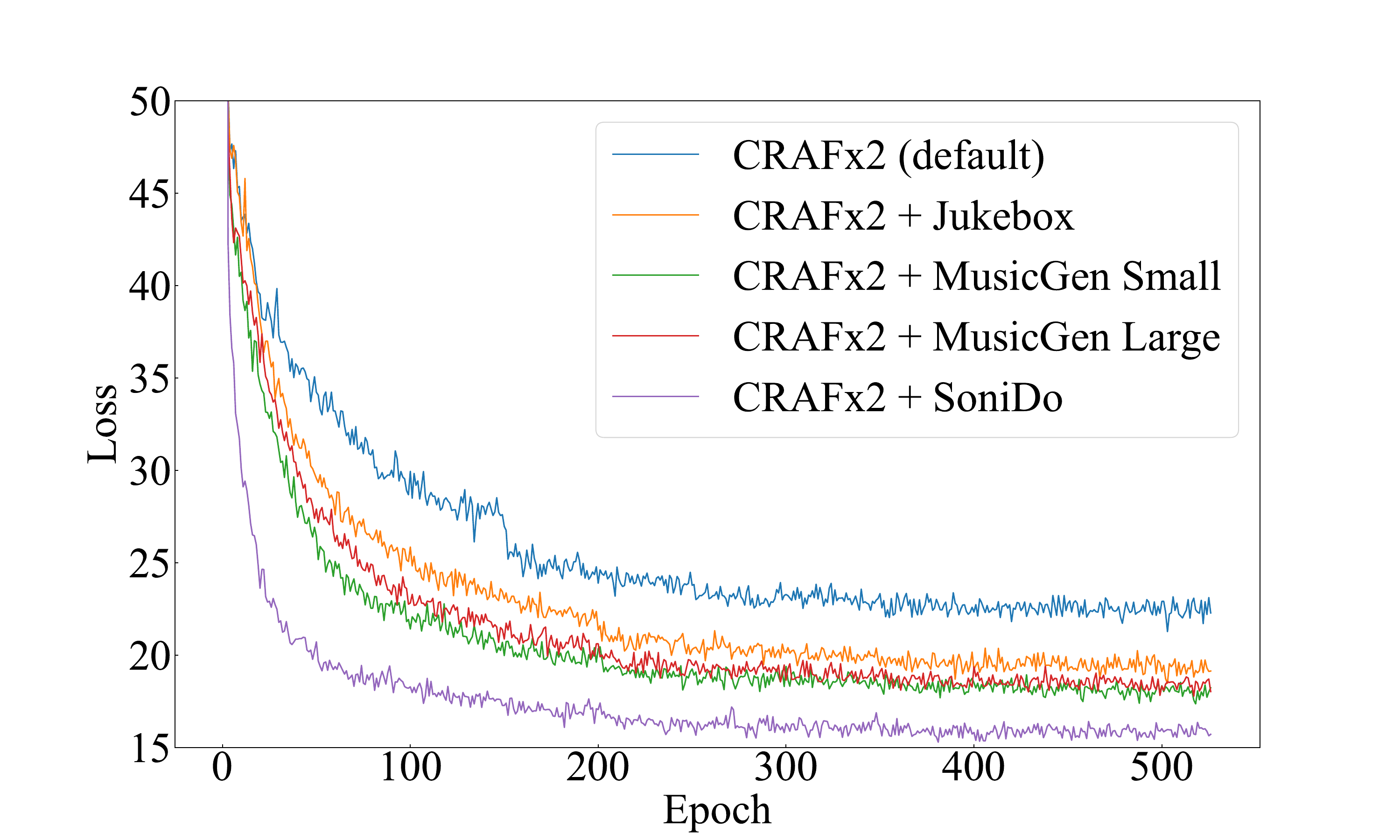}}
\centering\subfloat[CRAFx2: Validation loss]{\includegraphics[width=0.45\textwidth]{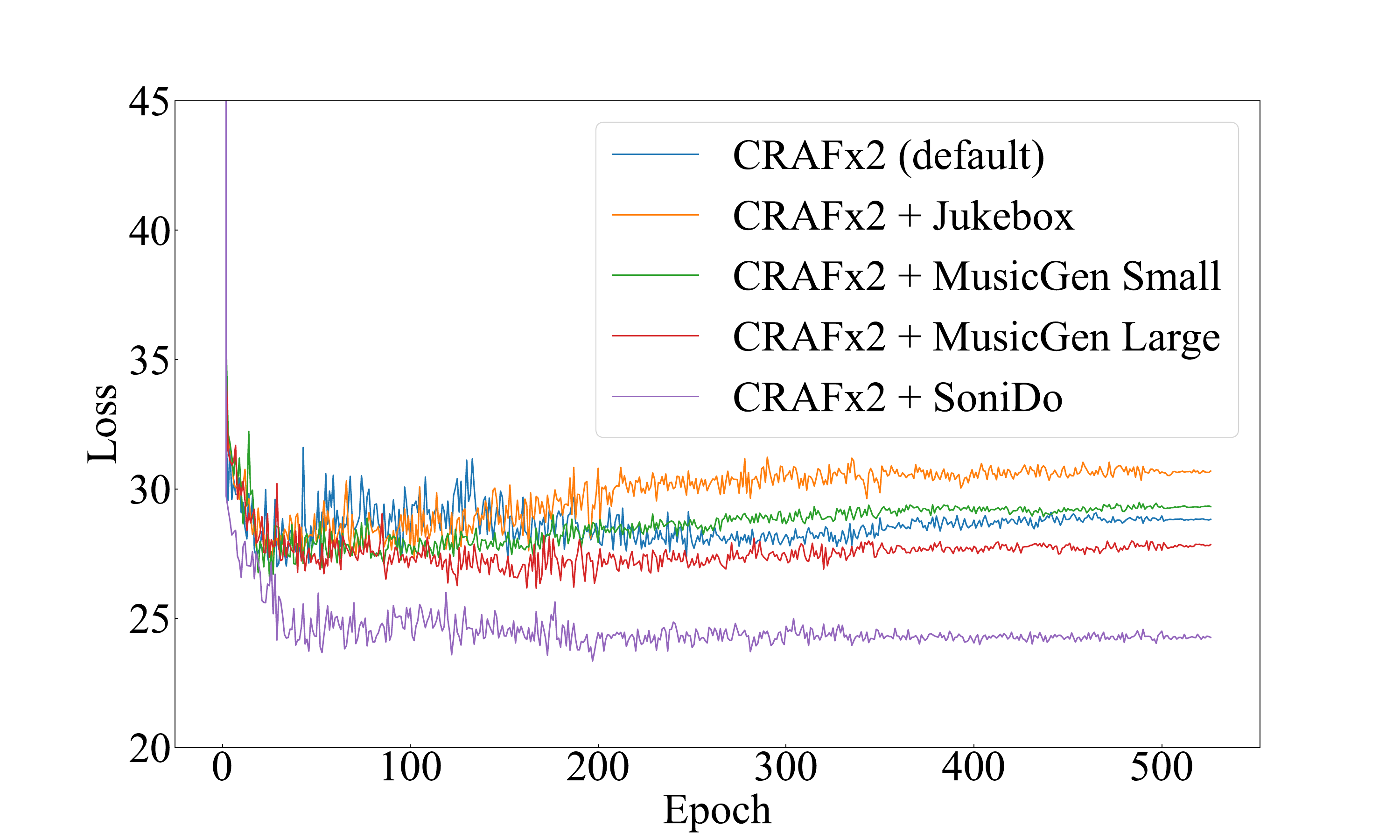}}
\caption{Training and validation curves for Mix-Wave-U-Net and CRAFx2 in music mixing task}
\label{fig:mixing_trainvalid_curves}
\end{figure}

\begin{table}[t]
\centering
\caption{Objective metrics correspond to mean absolute percentage error per low-level audio feature. Results are presented for MDXDB21-dry test set and MUSDB18 test set.}
\setlength\tabcolsep{3pt}
\footnotesize
\resizebox{\textwidth}{!}{
\begin{tabular}{@{}p{4.7cm}cccccccccccccc@{}}
\toprule
\multirow{3}{*}{\textbf{Test set / Model}}  & \multicolumn{5}{c}{\textbf{Spectral}} & \multicolumn{4}{c}{\textbf{Panning}} & \multicolumn{3}{c}{\textbf{Dynamic}} & \multicolumn{2}{c}{\textbf{Loudness}} \\
\cmidrule(r){2-6} \cmidrule(r){7-10} \cmidrule(r){11-13} \cmidrule(r){14-15}
& \multirow{2}{*}{centroid} & band- & \multirow{2}{*}{contrast} & \multirow{2}{*}{roll-off} & \multirow{2}{*}{flatness} & \multirow{2}{*}{PRMS} & \multirow{2}{*}{PRMS} & \multirow{2}{*}{PRMS} & \multirow{2}{*}{PRMS} & \multirow{2}{*}{RMS} & \multirow{2}{*}{spread} & \multirow{2}{*}{crest} & \multirow{2}{*}{LUFS} & \multirow{2}{*}{peak} \\

&  & width &  &  &  & $_{\text{total}}$ & $_{\text{low}}$ & $_{\text{mid}}$ & $_{\text{high}}$ &  &  &  &  &  \\

\midrule
\textit{\textbf{MDXDB21-dry}} \\
Mix-Wave-U-Net (default)  & 0.250 & 0.178 & 0.286 & 0.312 & 0.143  & 0.217 & 0.296 & 0.101 & 0.246  & 0.077 & 0.047 & 0.096 & 0.094 & 0.243 \\

\textcolor{c_revise}{Mix-Wave-U-Net $+$ Jukebox} & \textcolor{c_revise}{0.236} & \textcolor{c_revise}{0.176} & \textcolor{c_revise}{0.300} & \textcolor{c_revise}{0.322} & \textcolor{c_revise}{\textbf{0.140}} & \textcolor{c_revise}{0.278} & \textcolor{c_revise}{0.236} & \textcolor{c_revise}{0.084} & \textcolor{c_revise}{0.325} & \textcolor{c_revise}{0.078} & \textcolor{c_revise}{0.047} & \textcolor{c_revise}{0.100} & \textcolor{c_revise}{0.086} & \textcolor{c_revise}{0.222} \\

Mix-Wave-U-Net $+$ MusicGen Small    & 0.265 & 0.182 & \textbf{0.273} & 0.324 & 0.147  & 0.192 & 0.266 & 0.119 & 0.211  & \textbf{0.065} & 0.044 & \textbf{0.082} & 0.087 & 0.206 \\

Mix-Wave-U-Net $+$ MusicGen Large    & 0.264 & 0.191 & 0.286 & 0.323 & 0.149  & 0.262 & \textbf{0.243} & 0.130 & 0.288  & \textbf{0.065} & 0.044 & 0.089 & \textbf{0.083} & 0.207 \\

Mix-Wave-U-Net $+$ \mfm     & \textbf{0.228} &\textbf{ 0.159} & 0.295 & \textbf{0.290} & 0.158  & \textbf{0.183} & \textbf{0.243} & \textbf{0.088} & \textbf{0.205}  & 0.071 & \textbf{0.041} & 0.089 & 0.088 & \textbf{0.173} \\ \\

CRAFx2 (default)      & \textbf{0.216} & 0.165 & 0.196 & \textbf{0.264} & \textbf{0.123}  & 0.168 & \textbf{0.273} & \textbf{0.078} & 0.197  & 0.072 & 0.049 & 0.087 & 0.086 & 0.218 \\

\textcolor{c_revise}{CRAFx2 $+$ Jukebox} & \textcolor{c_revise}{0.282} & \textcolor{c_revise}{0.206} & \textcolor{c_revise}{0.176} & \textcolor{c_revise}{0.339} & \textcolor{c_revise}{0.151} & \textcolor{c_revise}{0.266} & \textcolor{c_revise}{0.333} & \textcolor{c_revise}{0.082} & \textcolor{c_revise}{0.315} & \textcolor{c_revise}{0.082} & \textcolor{c_revise}{0.046} & \textcolor{c_revise}{0.098} & \textcolor{c_revise}{0.084} & \textcolor{c_revise}{0.204} \\

CRAFx2 $+$ MusicGen Small         & 0.262 & 0.195 & 0.172 & 0.338 & 0.144  & 0.265 & 0.312 & 0.079 & 0.318  & 0.081 & 0.047 & 0.089 & 0.090 & 0.209 \\

CRAFx2 $+$ MusicGen Large         & 0.274 & 0.215 & \textbf{0.170} & 0.334 & 0.147  & 0.230 & 0.292 & 0.086 & 0.267  & 0.077 & 0.047 & 0.094 & \textbf{0.075} & \textbf{0.190} \\

CRAFx2 $+$ \mfm         & 0.226 & \textbf{0.157} & 0.273 & 0.283 & 0.169  & \textbf{0.145} & 0.307 & 0.085 & \textbf{0.162}  & \textbf{0.068} & \textbf{0.044} & \textbf{0.084} & 0.109 & 0.232 \\

\midrule
\textit{\textbf{MUSDB18}} \\

Mix-Wave-U-Net (default)  & 0.260 & 0.173 & 0.170 & 0.291 & 0.109  & 0.156 & 0.222 & 0.104 & 0.172  & 0.087 & 0.069 & 0.098 & 0.094 & 0.241 \\

\textcolor{c_revise}{Mix-Wave-U-Net $+$ Jukebox} & \textcolor{c_revise}{0.273} & \textcolor{c_revise}{0.179} & \textcolor{c_revise}{0.169} & \textcolor{c_revise}{0.302} & \textcolor{c_revise}{0.108} & \textcolor{c_revise}{0.211} & \textcolor{c_revise}{0.204} & \textcolor{c_revise}{0.094} & \textcolor{c_revise}{0.241} & \textcolor{c_revise}{0.089} & \textcolor{c_revise}{0.059} & \textcolor{c_revise}{0.098} & \textcolor{c_revise}{\textbf{0.089}} & \textcolor{c_revise}{\textbf{0.225}} \\

Mix-Wave-U-Net $+$ MusicGen Small   & 0.294 & 0.192 & \textbf{0.162} & 0.301 & 0.119  & \textbf{0.145} & \textbf{0.211} & 0.116 & \textbf{0.160}  & 0.088 & 0.056 & 0.093 & 0.097 & 0.230 \\

Mix-Wave-U-Net $+$ MusicGen Large   & 0.276 & 0.176 & 0.170 & 0.299 & \textbf{0.107}  & 0.203 & 0.232 & 0.108 & 0.225  & 0.079 & 0.057 & 0.090 & 0.096 & 0.237 \\

Mix-Wave-U-Net $+$ \mfm     & \textbf{0.240} & \textbf{0.152} & \textbf{0.162} & \textbf{0.265} & 0.110  & 0.179 & 0.217 & \textbf{0.103} & 0.200  & \textbf{0.068} & \textbf{0.050} & \textbf{0.069} & \textbf{0.089} & 0.269 \\ \\

CRAFx2 (default)      & 0.253 & 0.173 & 0.152 & \textbf{0.274} & \textbf{0.111}  & \textbf{0.132} & 0.253 & \textbf{0.086} & 0.147  & 0.097 & 0.047 & 0.098 & \textbf{0.089} & 0.241 \\

\textcolor{c_revise}{CRAFx2 $+$ Jukebox} & \textcolor{c_revise}{0.284} & \textcolor{c_revise}{0.210} & \textcolor{c_revise}{0.154} & \textcolor{c_revise}{0.303} & \textcolor{c_revise}{0.129} & \textcolor{c_revise}{0.238} & \textcolor{c_revise}{0.264} & \textcolor{c_revise}{0.104} & \textcolor{c_revise}{0.275} & \textcolor{c_revise}{0.097} & \textcolor{c_revise}{0.047} & \textcolor{c_revise}{0.103} & \textcolor{c_revise}{\textbf{0.089}} & \textcolor{c_revise}{0.241} \\

CRAFx2 $+$ MusicGen Small     & 0.279 & 0.190 & 0.157 & 0.303 & 0.125  & 0.223 & \textbf{0.242} & 0.091 & 0.258  & 0.098 & 0.047 & 0.105 & 0.101 & 0.255 \\

CRAFx2 $+$ MusicGen Large     & 0.293 & 0.212 & 0.163 & 0.317 & 0.136  & 0.209 & 0.277 & 0.099 & 0.237  & \textbf{0.093} & 0.048 & 0.098 & 0.099 & 0.251 \\

CRAFx2 $+$ \mfm         & \textbf{0.243} & \textbf{0.157} & \textbf{0.148} & 0.276 & 0.113  & \textbf{0.132} & 0.249 & 0.088 & \textbf{0.146}  & \textbf{0.093} & \textbf{0.046} & \textbf{0.089} & 0.097 & \textbf{0.240} \\
\bottomrule
\end{tabular}
}
\label{tab:features_mixing_combined}
\end{table}

From the training and validation curves in Figure~\ref{fig:mixing_trainvalid_curves}, we can see that both Mix-Wave-U-Net + \mfm and CRAFx2 + \mfm exhibited a significant improvement during training and enhanced generalization during validation. These results are reflected in Table~\ref{tab:mixing}, where the models that included \mfm tokens consistently displayed improved performance of the stereo-invariant loss. Regarding the audio effect-related features, Table~\ref{tab:features_mixing_combined} shows that across most feature categories, Mix-Wave-U-Net + \mfm outperformed the default and Mix-Wave-U-Net. 

Incorporating Jukebox \textcolor{c_revise}{and MusicGen} features generally leads to improvement over the default network, although not as much as \mfm. \musicgens performed slightly better than the large model, confirming the trend observed in previous downstream tasks. \textcolor{c_revise}{However, for both \musicgen and Jukebox, improvement over the default network is not always the case with CRAFx2}. In the MDXDB21-dry test set, the default model performed better than \musicgen in terms of spectral and panning features as well as stereo-invariant loss. \textcolor{c_revise}{Yet, \musicgen did enhance the dynamic and loudness features, while Jukebox only improved loudness and generally led to overfitting}. The performance gap of \musicgen in spectral-related metrics, and generally when compared with \mfm, might stem from \musicgen being trained on $32$-kHz audio, while both our evaluation datasets and training dataset of \mfm correspond to $44.1$-kHz audio. 

This discrepancy in training data sampling rates leads us to assume that the relatively lower performance of \musicgen may be attributed to this difference. Further investigation is required to confirm this hypothesis and explore the impact of sampling rates on music foundation models as boosters for music mixing tasks.

It is worth noting that the inherent differences between the MDXDB21-dry and MUSDB18 test sets are reflected in these scores. This is expected, given that the MDXDB21-dry data consists of dry multi-tracks, presenting a more realistic mixing scenario than the wet multi-tracks of MUSDB18. Therefore, we attribute the substantial difference in reported stereo-invariant values in Table~\ref{tab:mixing} to the fact that the target mixture has been processed with different audio processors, and the timbral characteristics cannot always be matched by the trained networks.

The mismatch in performance between the stereo-invariant loss and low-level features related to audio effects reflects the challenge of objectively evaluating automatic mixing systems, which remains an ongoing and open research direction~\citep{IMPbook19,steinmetz2022automix}. Thus, for a more in-depth analysis of such systems, a listening test is required. However, for the scope of this paper, the reported objective results support our hypothesis that incorporating the \mfm features, i.e. high-level knowledge of the input stems to be mixed, overall improves training and generalization when these features are properly used.

\end{document}